\documentclass[
 reprint,
 showpacs, preprintnumbers,
 amsmath, amssymb,
 aps,
 pra,
 floatfix,
 superscriptaddress
]{revtex4-1}

\usepackage{graphicx}
\usepackage{tikz}
\usepackage{xcolor}
\usetikzlibrary{arrows}

\makeatletter
\newcommand{\colorcaption}[2][]{%
  \begingroup%
  \renewcommand{\@caption@fignum@sep}{ (Color online). }%
  \caption[#1]{#2}%
  \endgroup%
}
\makeatother

\usepackage{dcolumn}
\usepackage{bm}
\usepackage{bbm}
\usepackage{upgreek}
\usepackage{hyperref}
\usepackage{blindtext}

\usepackage{changes}

\usepackage{txfonts}
\usepackage{dsfont}

\newcommand{\im}{\mathrm{i}}
\makeatletter
\def\maketag@@@#1{\hbox{\m@th\normalfont\normalsize#1}}
\makeatother

\definecolor{darkgreen}{RGB}{0 100 0}

\usepackage{pdfpages}
\makeatletter
\AtBeginDocument{\let\LS@rot\@undefined}
\makeatother

\begin{document}

    %% ================================================================================== %%
    %% ================================================================================== %%

	%% Header: Title, Authors, Institutions, Abstract %%
	\title{Anomalous Josephson Hall effect charge and transverse spin currents \\ in superconductor/ferromagnetic~insulator/superconductor junctions}
	
	\author{Andreas Costa}%
	\email[E-Mail: ]{andreas.costa@physik.uni-regensburg.de}
 	\affiliation{Institute for Theoretical Physics, University of Regensburg, 93040 Regensburg, Germany}
 	
	\author{Jaroslav Fabian}%
 	\affiliation{Institute for Theoretical Physics, University of Regensburg, 93040 Regensburg, Germany}

	\date{\today}
    
    \begin{abstract}
    	Interfacial spin-orbit coupling in Josephson junctions offers an intriguing way to combine anomalous Hall and Josephson physics in a single device. %%
    	We study theoretically how the superposition of both effects impacts superconductor/ferromagnetic insulator/superconductor junctions' transport properties. %%
    	Transverse momentum-dependent skew tunneling of Cooper pairs through the spin-active ferromagnetic insulator interface creates \emph{sizable} transverse Hall supercurrents, to which we refer as \emph{anomalous Josephson Hall effect currents}. %%
    	We generalize the Furusaki--Tsukada formula, which got initially established to quantify usual (tunneling) Josephson current flows, to evaluate the transverse current components and demonstrate that their amplitudes are \emph{widely adjustable} by means of the spin-orbit coupling strengths or the superconducting phase difference across the junction. %%
    	As a clear spectroscopic fingerprint of Josephson junctions, well-localized subgap bound states form around the interface. %%
    	By analyzing the spectral properties of these states, we unravel an unambiguous correlation between spin-orbit coupling-induced asymmetries in their energies and the transverse current response, founding the currents' \emph{microscopic} origin. %%
    	Moreover, skew tunneling simultaneously acts like a transverse spin filter for spin-triplet Cooper pairs and complements the discussed charge current phenomena by their spin current counterparts. %%
    	The junctions' universal spin--charge current cross ratios provide valuable possibilities to experimentally detect and characterize interfacial spin-orbit coupling. %%
    \end{abstract}

    \maketitle

    %% ================================================================================== %%
    %% ================================================================================== %%

    %% Introduction %%
    %%\paragraph*{Introduction.}%
    \section{Introduction}
    Superconducting~junctions offer unique possibilities to generate and control charge and spin~supercurrents, and provide the key~ingredients for spintronics~applications~\cite{Eschrig2011,Linder2015}. Particularly rich physics occurs when superconductivity is brought together with the antagonistic ferromagnetic~phase. 
    Prominent~examples cover magnetic~Josephson~junctions~\cite{Bulaevskii1977,*Bulaevskii1977alt,Buzdin1982,*Buzdin1982alt,Andreev1991,Demler1997,Golubov2004,Buzdin2005,Bergeret2005,AnnunziatA2011,Campagnano2015,Gingrich2016,Minutillo2018}, in which the combination of superconductivity and ferromagnetism can add intrinsic phase~shifts to the junctions' characteristic current-phase~relation and reverse the Josephson~currents'~directions. 
    
    The interplay of magnetism and superconductivity gets even more fascinating in the presence of Rashba~\cite{Bychkov1984} and/or Dresselhaus~\cite{Dresselhaus1955} spin-orbit~coupling~(SOC)~\cite{Fabian2004,Fabian2007}, which induces spin-triplet~correlations~\cite{Bergeret2001,Volkov2003,Keizer2006,Halterman2007,Eschrig2008,Eschrig2011,Sun2015}, triggers long-range proximity~effects~\cite{Duckheim2011,Bergeret2013,Bergeret2014,Jacobsen2015}, and is furthermore expected to host Majorana~states in proximitized superconducting regions~\cite{Nilsson2008,Duckheim2011,Lee2012,Nadj-Perge2014,Dumitrescu2015,Pawlak2016,Ruby2017,Livanas2019}. Tunneling~barriers invariably introduce interfacial SOC into various types of (superconducting) tunnel~junctions. Earlier theoretical studies concluded that skew~tunneling of \emph{spin-polarized} electrons through such barriers gives rise to (extrinsic) tunneling~anomalous~Hall~effects~(TAHEs)~\cite{Vedyayev2013,Vedyayev2013a,MatosAbiague2015,HuongDang2015,Mironov2017,HuongDang2018}. Although first experiments carried out on granular~nanojunctions~\cite{Zhuravlev2018} essentially confirmed the theoretical~expectations, the effect is typically weak in normal-state~junctions. More sizable TAHE~conductances, coming along with a spontaneous transverse supercurrent~response, were predicted for superconducting~junctions~\cite{Costa2019}, opening several novel perspectives, e.g., the possibility to experimentally verify superconducting magnetoelectric~effects~\cite{Edelstein1995,Edelstein2003}. 
    
    From that viewpoint, integrating TAHEs into Josephson~junctions could likewise attract considerable interest. The resulting \emph{dissipationless} transverse supercurrent~flows might be efficiently tuned by means of the phase~difference between the superconducting junction~electrodes, becoming exploitable for a variety of spintronics~applications~\cite{Eschrig2011,Linder2015}. However, already one of the initial works into that direction~\cite{Malshukov2008} demonstrated that the fundamental time-reversal~(electron--hole)~symmetry in stationary Josephson~junctions acts against the spontaneous flow of (spin)~Hall~supercurrents. To overcome this obstacle, one could either apply a finite bias~voltage to the system~\cite{Malshukov2011} or modify the considered junction~geometry. Several proposals suggested to focus on intricate magnetic~multilayer~configurations~\cite{Asano2005,Asano2006,Lu2009,Malshukov2010,Brydon2011,*Brydon2012,Wang2011,Wang2011a,Bujnowski2012,Ren2013,Alidoust2015,Wakamura2015,Yokoyama2015,Bergeret2016,Linder2017,Mironov2017,Risinggard2019}, which break time-reversal symmetry and simultaneously facilitate a mixture of spin-singlet and spin-triplet~correlations~(caused, e.g., by strong~SOC), eventually leading not only to nonzero charge~Hall~supercurrents~\cite{Wang2011a,Yokoyama2011,Yokoyama2015,Mironov2017}, but also to their spin~counterparts~\cite{Lu2009,Malshukov2010,Wang2011,Ren2013,Wakamura2015,Bergeret2016,Linder2017,Ouassou2017,Risinggard2019}. 
    
    In this paper, we consider a ballistic superconductor~(S)/ferromagnetic~insulator~(F-I)/S~Josephson~junction, whose magnetic (F-I)~tunneling~barrier introduces strong interfacial SOC into the system. We demonstrate that Cooper~pairs \emph{skew~tunnel} through the spin-active~interface and spontaneously generate charge~Hall~supercurrents along the transverse directions~(i.e., parallel to the interface), to which we refer as \emph{anomalous~Josephson~Hall~effect~(AJHE)~currents}~\footnote{In an earlier study~\cite{Yokoyama2015}, the term \emph{AJHE} refers to the anomalous~Hall~conductances appearing in the nonsuperconducting electrode of magnet/triplet~S~junctions. Although we use the same terminology, it shall be noted that the physics is different in our case.}. When compared to most of the previously predicted geometries, our system brings along the great advantage that its physical~properties can be much better controlled in experiments. Generalizing the Green's~function-based~\cite{McMillan1968} Furusaki--Tsukada~method~\cite{Furusaki1991}, we quantify the AJHE~currents for representative junction~parameters and discuss their characteristic dependence on the F-I's magnetization~orientation and the phase~difference across the junction. 
    \pagebreak
    
    A clear spectroscopic fingerprint of Josephson~junctions is the formation of subgap~bound~states, which are strongly localized around the nonsuperconducting link. In~fact, two distinct types of bound~states play a major role in S/F-I/S~junctions~\cite{Costa2018,Rouco2019}: the Andreev~bound~states~(ABS)~\cite{Andreev1964,*Andreev1964alt,Andreev1966,*Andreev1966alt} and the Yu--Shiba--Rusinov~(YSR)~\cite{Yu1965,Shiba1968,Shiba1969,Rusinov1968,*Rusinov1968alt}~states. Up to now, it remained unclear whether one can draw connections between these states' features and the  Josephson~Hall~effects. To answer this question, we identify our junction's ABS and YSR~states, together with their respective energies, and formulate an alternative approach that allows us to compute the AJHE~currents directly from the bound~state wave~functions. The additional calculations offer not only an essential cross-check for the Furusaki--Tsukada~method, but enable us to resolve the single current~contributions that originate from the ABS and the YSR~states. We identify SOC-induced transverse~momentum-dependent asymmetries in the bound~state~energies, most clearly apparent in the YSR~branch of the spectrum, as the \emph{microscopic} origin of the~AJHE. 
    
    The spin-active F-I~barrier simultaneously induces interfacial spin~flips and converts some of the spin-singlet Cooper~pairs into triplet pairs. We extend the Cooper~pair skew~tunneling~picture to these \emph{spin-polarized} triplet~pairs and develop a qualitative physical understanding to unravel the most essential features of the resulting transverse \emph{spin~current}~flows. We evaluate the spin~current~amplitudes once from an extended Furusaki--Tsukada~\emph{spin~current}~formula and once from the bound~state wave~functions, comment on their distinct magnetization~angle~dependence when compared to their AJHE~charge~current~counterparts, and eventually deduce that the \emph{magnetization-independent} spin--charge~current~cross~ratios could be exploited to classify the interfacial~SOC. 
    
    We structured the paper in the following way. In~Sec.~\ref{SecTheory}, we formulate the theoretical~model used to investigate our junction. After working out the qualitative skew~tunneling~picture, justifying the existence of nonzero AJHE~currents and bringing along valuable physical insight, in~Sec.~\ref{SecSkewAR}, we compute the current~components for realistic parameter~configurations and discuss their generic properties~(see~Sec.~\ref{SecAJHECurrents}). Section~\ref{SecBoundStates} is dedicated to a thorough analysis of the connections between the bound~states that form around the junction's F-I~barrier and the emergent AJHE. Finally, we are concerned with the charge~currents' spin~counterparts in~Sec.~\ref{SecJSCurrents}, before closing with a short summary~(Sec.~\ref{SecSummary}). The Appendices contain the most important technical~details of our calculations.

    %% ================================================================================== %%
    %% ================================================================================== %%

    %% Theoretical modeling %%
    %%\paragraph*{Theoretical~modeling.}%
    \section{Theoretical~modeling       \label{SecTheory}}
\begin{figure}
    	    \includegraphics[width=0.45\textwidth]{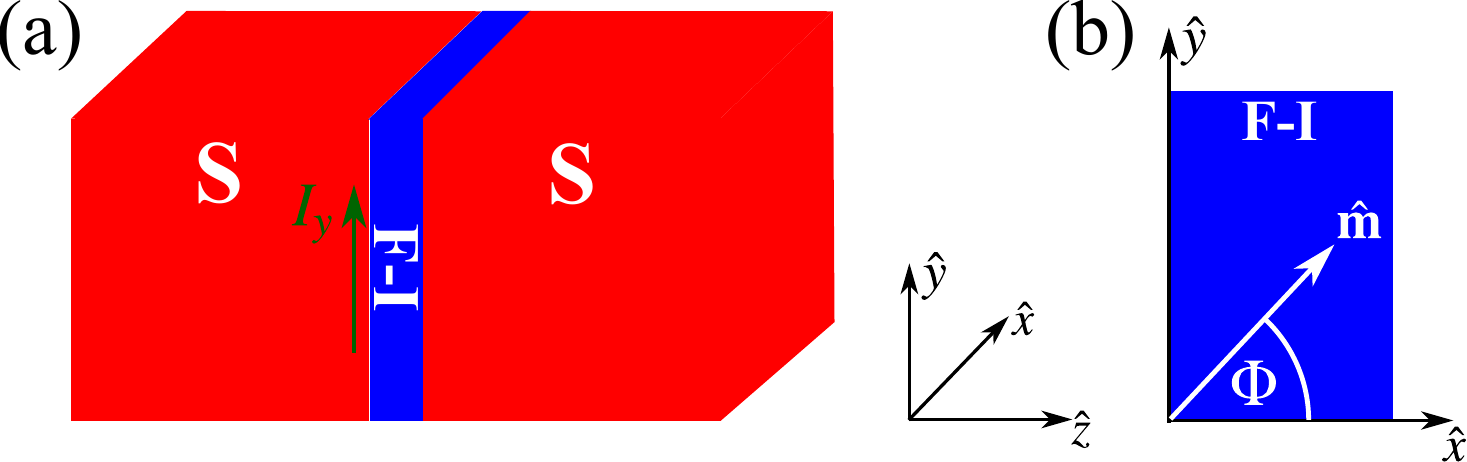}
        	\colorcaption{(a)~Sketch of the regarded S/F-I/S~junction, using the $ C_{2v} $~principal crystallographic orientations~$ \hat{x} \parallel [110] $, $ \hat{y} \parallel [\overline{1}10] $, and $ \hat{z} \parallel [001] $; Cooper~pair~tunneling generates (tunneling)~Josephson~currents along~$ \hat{z} $, while the AJHE~currents flow transversely along~$ \hat{x} $ and~$ \hat{y} $~($ I_y $ is exemplarily illustrated by the green~arrow). (b)~The direction of the magnetization~vector inside the F-I, $ \hat{\mathbf{m}} $, is determined by the angle~$ \Phi $.%%
    	       \label{FigSystem}}
        \end{figure}

        We consider a ballistic three-dimensional S/F-I/S~junction grown along the $ \hat{z} $-direction, in which the two semi-infinite S~regions are separated by an ultrathin F-I~(could, e.g., be a thin layer of~$ \mathrm{EuS} $~\cite{Moodera1988}, $ \mathrm{EuO} $~\cite{Tedrow1986}, a $ \mathrm{GaAs} $/$ \mathrm{Fe} $~slab~\cite{Hupfauer2015}, or another thin semiconducting layer proximitized by a ferromagnet); see~Fig.~\ref{FigSystem}(a). %%
        %% Fig. 1 %%
        %% %%
        The barrier itself introduces potential~scattering and, owing to the broken space~inversion~symmetry, simultaneously additional strong interfacial Rashba~\cite{Bychkov1984} and, for $ C_{2v} $-symmetrical~interfaces, Dresselhaus~\cite{Dresselhaus1955} SOC~\cite{Fabian2004,Fabian2007}. %%
        Our system is modeled by means of the stationary Bogoljubov--de~Gennes~(BdG)~Hamiltonian~\cite{DeGennes1989},
        \begin{equation}
            \hat{\mathcal{H}}_\mathrm{BdG} = \left[ \begin{matrix} \hat{\mathcal{H}}_\mathrm{e} & \hat{\Delta}_\mathrm{S}(z) \\ \hat{\Delta}_\mathrm{S}^\dagger(z) & \hat{\mathcal{H}}_\mathrm{h} \end{matrix} \right]
            ,
        \end{equation}
        with $ \hat{\mathcal{H}}_\mathrm{e} = [-\hbar^2 / (2m) \, \boldsymbol{\nabla}^2 - \mu] \, \hat{\sigma}_0 + \hat{\mathcal{H}}_\text{F-I} $ representing the single-electron~Hamiltonian and $ \hat{\mathcal{H}}_\mathrm{h} = -\hat{\sigma}_y \, \hat{\mathcal{H}}_\mathrm{e}^* \, \hat{\sigma}_y $ its holelike counterpart~($ \hat{\sigma}_0 $ and $ \hat{\sigma}_i $ indicate the two-by-two~identity and the $ i $th~Pauli~matrix). %%
        Analogously to previous studies~\cite{DeJong1995,Zutic1999,Zutic2000,Costa2017,Costa2018,Costa2019}, the ultrathin F-I~region is included into our model as an effective potential- and SOC-dependent deltalike barrier, 
        \begin{multline}
            \hat{\mathcal{H}}_\text{F-I} = \Big[ \lambda_\mathrm{SC} \, \hat{\sigma}_0 + \lambda_\mathrm{MA} \, (\hat{\mathbf{m}} \cdot \hat{\boldsymbol{\sigma}}) %%
            \\%%
            + \alpha \, (k_y \, \hat{\sigma}_x - k_x \, \hat{\sigma}_y) - \beta \, (k_y \, \hat{\sigma}_x + k_x \, \hat{\sigma}_y) \Big] \, \delta(z)
            ,
            \label{EqBarrierHamiltonian}
        \end{multline}
        where the first two parts describe scalar and magnetic tunneling with amplitudes~$ \lambda_\mathrm{SC} $ and $ \lambda_\mathrm{MA} $, respectively. The unit~vector along the magnetization~direction in the F-I, $ \hat{\mathbf{m}} = [ \cos \Phi, \, \sin \Phi, \, 0 \big]^\top $, is determined with respect to the $ \hat{x} \parallel [110] $-reference~direction~[see~Fig.~\ref{FigSystem}(b)], while the vector~$ \hat{\boldsymbol{\sigma}} = [ \hat{\sigma}_x , \, \hat{\sigma}_y , \, \hat{\sigma}_z ]^\top $ comprises the Pauli~spin~matrices. %%
        Finally, the remaining contributions resemble the interfacial Rashba and (linearized) Dresselhaus~SOC with the effective strengths~$ \alpha $ in the first and $ \beta $ in the second case; the SOC~Hamiltonian is given with respect to the $ C_{2v} $~principal crystallographic~axes~$ \hat{x} \parallel [110] $ and~$ \hat{y} \parallel [\overline{1}10] $. %%
        Inside the S~electrodes, the $ s $-wave superconducting~pairing~potential, $ \hat{\Delta}_\mathrm{S}(z) = |\Delta_\mathrm{S}| \, [ \Theta(-z) + \mathrm{e}^{\mathrm{i} \phi_\mathrm{S}} \Theta(z) ] $ ($ |\Delta_\mathrm{S}| $ is the superconductors' isotropic energy gap, which is taken to be the same in both electrodes, and~$ \phi_\mathrm{S} $ the phase~difference across the junction) couples the BdG~Hamiltonian's electron and hole~blocks. Writing $ \hat{\Delta}_\mathrm{S} $ in that way is a rigid approximation as it fully neglects proximity~effects. Nevertheless, this approach drastically simplifies the subsequent theoretical analyses, while still yielding reliable results for common transport~calculations~\cite{Likharev1979,Beenakker1997}. %%
        For further simplification and without losing generality, we additionally consider equal effective~carrier~masses, $ m $, and the same Fermi~level, $ \mu = (\hbar^2 q_\mathrm{F}^2) / (2m) $~($ q_\mathrm{F} $ is the associated Fermi~wave~vector), in all junction~constituents.
        \pagebreak
        
        Assuming translational invariance parallel to the F-I~interface, the solutions of the BdG~equation, $ \hat{\mathcal{H}}_\mathrm{BdG} \, \Psi(\mathbf{r}) = E \, \Psi(\mathbf{r}) $, can be factorized into~$ \Psi(\mathbf{r}) = \psi(z) \, \mathrm{e}^{\mathrm{i} (\mathbf{k_\parallel} \cdot \mathbf{r_\parallel})} $, where $ \mathbf{k_\parallel} = [k_x, \, k_y, \, 0]^\top $~($ \mathbf{r_\parallel} = [x, \, y, \, 0]^\top $) is the transverse momentum~(position) vector and $ \psi(z) $ the BdG~equation's individual solution for the effective one-dimensional scattering~problem along~$ \hat{z} $. %%
        The latter distinguishes between the involved quasiparticle scattering~processes at the interface. Quasiparticles incident from one~S may, for~instance, either undergo Andreev~reflection~(AR) or specular~reflection~(SR), or may be transmitted into the second~S. %%
        The AR~process contains all the information concerning the transfer of Cooper~pairs across the barrier and is therefore \emph{the} process on which we need to focus subsequently to understand the physical origin of transverse supercurrent~flows. Putting the scattering~picture on a mathematical~ground is rather technical and can be partly found in~Appendix~\ref{AppA} and in all details in the Supplemental~Material~(SM)~\footnote{See the attached Supplemental~Material, which includes~Refs.~\cite{Martinez2018,Fabian2007,MatosAbiague2009,Moser2007,Bychkov1984,Dresselhaus1955,Fabian2004,DeJong1995,Zutic1999,Zutic2000,Costa2017,Costa2018,Costa2019,DeGennes1989,McMillan1968,Furusaki1991,Carbotte1990,Rouco2019,Andreev1964,*Andreev1964alt,Andreev1966,*Andreev1966alt,Yu1965,Shiba1968,Shiba1969,Rusinov1968,*Rusinov1968alt,Blonder1982,Furusaki1999,Dyakonov1971,Dyakonov1971b,*Dyakonov1971a,MatosAbiague2015,Asano2005,Hoegl2015,*Hogl2015,Beenakker1991,Golubov2004,Bulaevskii1977,*Bulaevskii1977alt}, for more details.}.

    %% Quasiparticle picture---Skew Andreev tunneling %%
    \section{Quasiparticle~picture---Skew~AR        \label{SecSkewAR}}   
        
        \begin{figure}
    	    \includegraphics[width=0.45\textwidth]{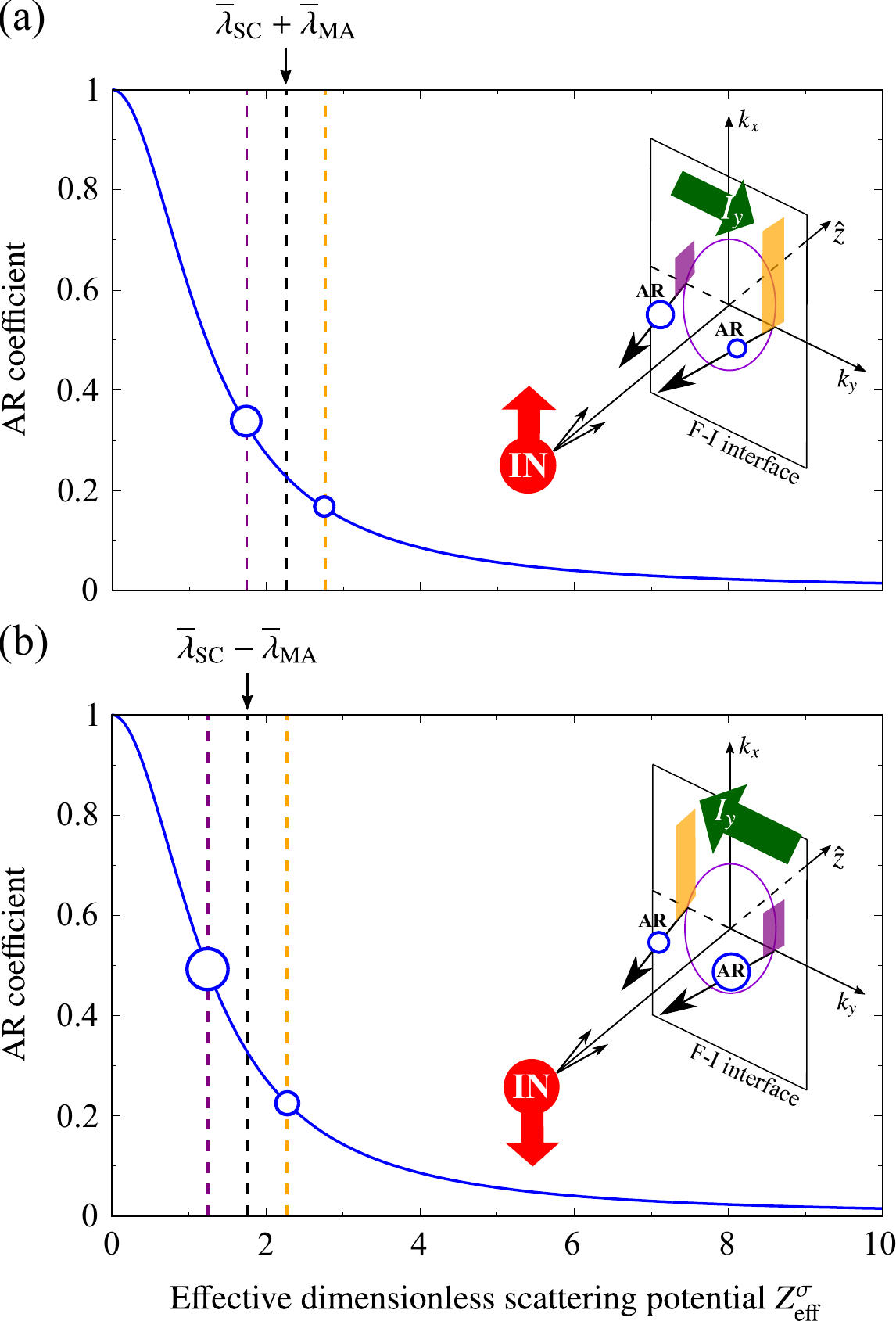}
            \colorcaption{(a)~Calculated (zero-energy) AR~coefficient~(determining the AR~probability) for spin~up electrons~(IN) incident on the F-I~interface and as a function of~$ Z_\mathrm{eff}^\sigma = (2mV_\mathrm{eff}^\sigma)/(\hbar^2 q_\mathrm{F}) $, essentially modeling the effective scattering~potential in~Eq.~(\ref{EqPotentialEffective}). The dashed black line indicates the tunneling~parameters~$ \overline{\lambda}_\mathrm{SC} = (2m\lambda_\mathrm{SC})/(\hbar^2 q_\mathrm{F}) = 2 $ and~$ \overline{\lambda}_\mathrm{MA} = (2m\lambda_\mathrm{MA})/(\hbar^2 q_\mathrm{F}) = 0.25 $, which combine to~$ \overline{\lambda}_\mathrm{SC} + \overline{\lambda}_\mathrm{MA} $ for up-spin~electrons. Assuming the Rashba~SOC~strength~$ \overline{\lambda}_\mathrm{R} = (2m\alpha) / \hbar^2 = 1 $, incoming electrons with~$ k_y > 0 $ are exposed to a raised~(dashed orange line) and those with~$ k_y < 0 $ to a lowered~(dashed violet line) effective scattering~potential. AR becomes suppressed at positive~$ k_y $ and favorable at negative~$ k_y $, highlighted by the different size of the (blue) Andreev~reflected~holes. This skew~AR generates a net transverse current along~$ \hat{y} $~(the direction of the current is usually defined \emph{oppositely} to the electron~flow~direction; the latter points along~$ -\hat{y} $), which flows as a dissipationless AJHE~current, $ I_y $, in the superconductors. (b)~Same as in~(a), but for incident spin~down~electrons. Skew~AR causes now an AJHE~current along~$ -\hat{y} $. Since the effective tunneling~strength~(without~SOC) for down-spin~electrons is~$ \overline{\lambda}_\mathrm{SC} - \overline{\lambda}_\mathrm{MA} $, the skew~AR~coefficients for spin~down are always slightly greater than for spin~up so that the AJHE~currents originating from both processes do not completely compensate. 
                \label{FigModel}}
        \end{figure}
        On the quasiparticle~level, the supercurrent generating exchange of Cooper~pairs between the superconductors is mediated by the peculiar AR~process. An (unpaired) electronlike~quasiparticle incident on the F-I~barrier from one~S gets transmitted into the second~S, pairs with another correlated electronlike~quasiparticle, and effectively transfers a Cooper~pair across the barrier. Formally, the transmission of two correlated electronlike~quasiparticles is modeled by having the incident electronlike~quasiparticle Andreev~reflected as a holelike~quasiparticle with opposite spin. As long as more Cooper~pairs enter the right~S than the left~one~(or vice~versa), net (tunneling)~Josephson~currents start to flow. In the following, we will simply refer to electronlike~(holelike)~quasiparticles as electrons~(holes). Electrons incident on the F-I~barrier are exposed to an effective scattering~potential that combines the scalar and (spin-dependent) magnetic potential~terms with an additional transverse~momentum- and spin-dependent contribution originating from the interfacial SOC. Assuming, for~simplicity, that only Rashba~SOC is present~($ \alpha > 0 $ and~$ \beta = 0 $), the F-I's magnetization points along~$ \hat{x} $~(meaning~$ \Phi = 0 $), and~$ k_x = 0 $, the effective scattering~potential takes the form
        \begin{equation}
            V_\mathrm{eff}^\sigma = \lambda_\mathrm{SC} + \sigma \lambda_\mathrm{MA} + \sigma \alpha k_y ,
            \label{EqPotentialEffective}
        \end{equation}
        where $ \sigma = +(-) 1 $ indicates a spin~parallel~(antiparallel) to~$ \hat{x} $; we will equivalently use the terms spin~up~(spin~down). \emph{How does $ V_{eff}^\sigma $ impact the peculiar AR~process at the F-I~barrier?} To address this central question, Fig.~\ref{FigModel} illustrates the dependence of the AR~coefficient~\cite{Note2} on the strength of~$ V_\mathrm{eff}^\sigma $~[represented by the dimensionless~parameter~$ Z_\mathrm{eff}^\sigma = (2mV_\mathrm{eff}^\sigma) / (\hbar^2 q_\mathrm{F}) $]. We just focus on (spin-conserving)~AR since this scattering~process essentially drives the supercurrents we are predominantly interested in. Earlier studies~\cite{Costa2019} showed that the contributions of spin-flip~AR, i.e., the triplet Cooper~pair~currents are small within the considered limit and can be neglected when formulating a qualitative picture.
        
        Following Eq.~\eqref{EqPotentialEffective}, incident up-spin~electrons with~$ k_y > 0 $ experience a raised effective scattering~potential, while~$ V_\mathrm{eff}^\sigma $ gets lowered for incoming $ k_y < 0 $-electrons. Since the probability to undergo~AR typically decreases with increasing~$ V_\mathrm{eff}^\sigma $, up-spin~electrons get predominantly Andreev~reflected for negative~$ k_y $. In that way, this \emph{skew~AR} generates a transverse AJHE~quasiparticle~current along the $ \hat{y} $-direction. Although we are solely dealing with quasiparticle~currents at the moment, skew~AR effectively cycles Cooper~pairs across the F-I~interface and triggers a supercurrent~response~\cite{Costa2019}. Therefore, the transverse AJHE~quasiparticle~currents building up at the interface are immediately converted into transverse AJHE~supercurrents inside the two superconducting~electrodes~(basically generated by skew~tunneling Cooper~pairs). 
        Flipping the incident electrons' spin reverses the skew~AR~picture. It is now the positive range of~$ k_y $ that causes preferential~ARs, leading to an AJHE~current that flows along~$ -\hat{y} $. If the F-I~barrier would be nonmagnetic, the net AJHE~current~amplitudes stemming from skew~ARs of incoming up-spin and down-spin~electrons would become equal and, as they flow along reversed directions, no net AJHE~currents are expected. Already a weak exchange~splitting in the F-I, however, is sufficient that skew~ARs happen more likely for incoming down-spin than for up-spin~electrons~(see our explanations to~Fig.~\ref{FigModel}). The individual AJHE~currents in the (weakly) magnetic~junction do then not completely cancel and nonzero AJHE~currents build up.

    %% AJHE currents %%
    \section{AJHE~currents      \label{SecAJHECurrents}}
    
        Measuring a finite AJHE~supercurrent~response is an unambiguous experimental evidence for skew~ARs at the spin-active F-I~interface. To mathematically access the \emph{interfacial} AJHE~currents in our junction~(we refer to them as~$ I_\eta $ flowing along the $ \hat{\eta} \in \{ \hat{x} ; \hat{y} \} $-directions), we generalize the quasiparticle-based Furusaki--Tsukada~approach~\cite{Furusaki1991} and end up with~\cite{Note2,Costa2019}
        \begin{align}
            I_\eta &\approx \frac{e k_\mathrm{B} T}{2\hbar} |\Delta_\mathrm{S}(0)| \tanh\left( 1.74 \sqrt{\frac{T_\mathrm{C}}{T}-1} \right) \nonumber \\
            &\hspace{10 pt} \times \frac{A}{(2\pi)^2} \int \mathrm{d}^2 \mathbf{k}_\parallel \sum_{\omega_n} \frac{k_\eta}{\sqrt{q_\mathrm{F}^2 - \mathbf{k}_\parallel^2}} \nonumber \\
            &\hspace{10 pt} \times \left[ \frac{\mathcal{C}^{(1)}(\im \omega_n) + \mathcal{D}^{(2)}(\im \omega_n) + \mathcal{A}^{(3)}(\im \omega_n) + \mathcal{B}^{(4)}(\im \omega_n)}{\sqrt{\omega_n^2 + |\Delta_\mathrm{S}(0)|^2 \tanh^2 \left( 1.74 \sqrt{T_\mathrm{C}/T - 1} \right)}} \right] ,
                \label{EqCurrentFurusakiTsukada}
        \end{align}
        where $ e $ denotes the (positive) elementary~charge, $ k_\mathrm{B} $ Boltzmann's~constant, and $ \omega_n = (2n+1) \pi k_\mathrm{B} T $, with integer~$ n $, indicates the fermionic Matsubara~frequencies~(at temperature~$ T $ and given in units of~$ 1/\hbar $); for~simplicity, we assume that the tunneling and Hall~contact areas are equal and determined by~$ A $. All information necessary to evaluate the AJHE~current~components enters via the \emph{spin-conserving} AR~coefficients for incoming~(from the left) up-spin~(down-spin) electronlike~quasiparticles, $ \mathcal{C}^{(1)}(\im \omega_n) $~[$ \mathcal{D}^{(2)}(\im \omega_n) $], as well as the ones belonging to incident up-spin~(down-spin) holelike~quasiparticles, $ \mathcal{A}^{(3)}(\im \omega_n) $~[$ \mathcal{B}^{(4)}(\im \omega_n) $]. The latter are required to properly capture the AJHE~currents originating from skew~ARs of electrons incident on the F-I~interface from the right. Further details on the methodology are included into Appendix~\ref{AppA} and the~SM~\cite{Note2}. 
        
\begin{figure}
    	    \includegraphics[width=0.45\textwidth]{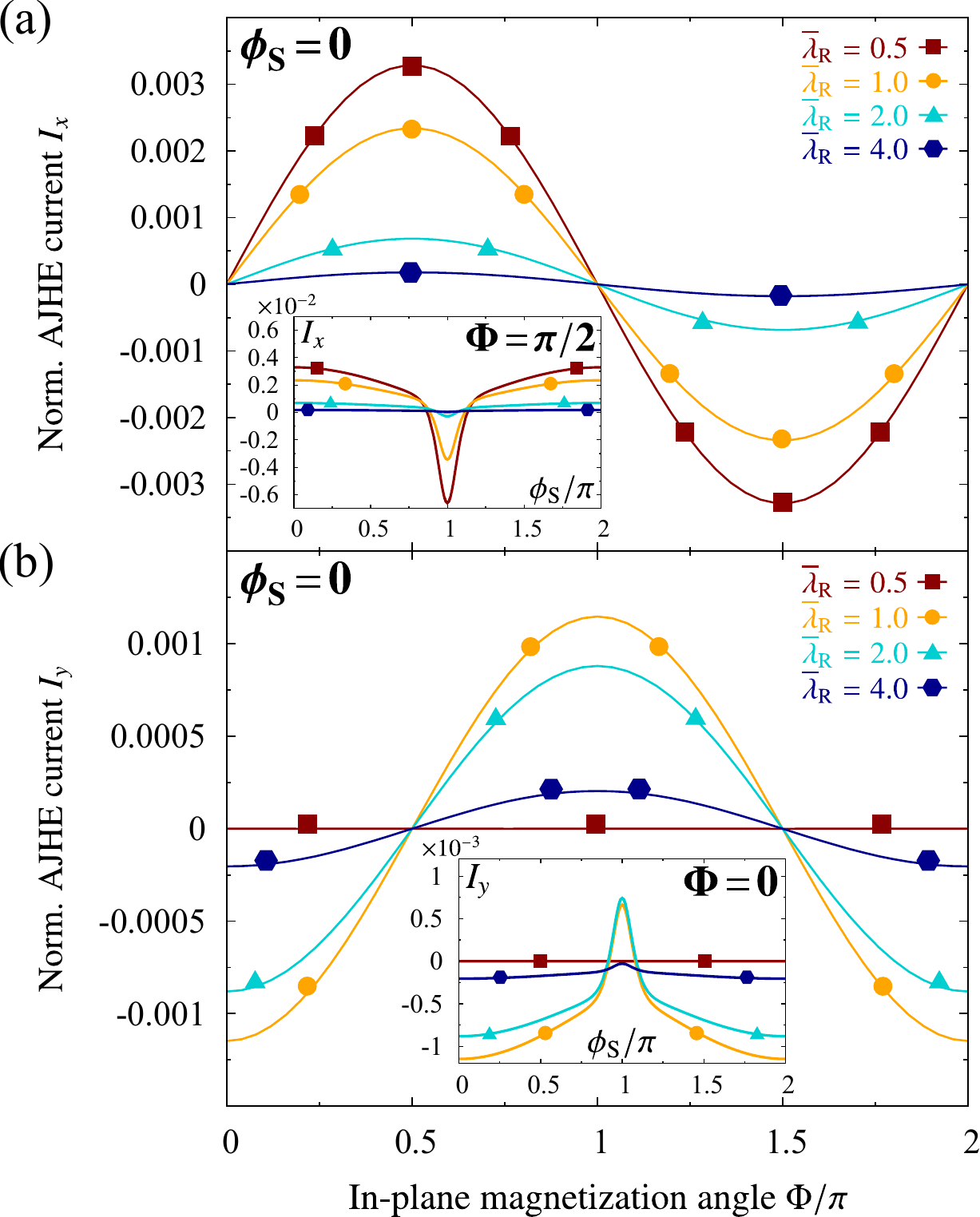}
            \colorcaption{(a)~Calculated dependence of the AJHE~current along~$ \hat{x} $, $ I_x $, normalized according to~$ (I_x e) / [G_\mathrm{S} \pi |\Delta_\mathrm{S}(0)|] $~[$ e $ is the (positive)~elementary~charge and $ G_\mathrm{S} $ represents Sharvin's~conductance of a three-dimensional point~contact], on the F-I's in-plane~magnetization~angle, $ \Phi $, and for various indicated (dimensionless) Rashba~SOC~strengths, $ \overline{\lambda}_\mathrm{R} = (2m\alpha) / \hbar^2 $. The remaining parameters are~$ \overline{\lambda}_\mathrm{SC} = (2m\lambda_\mathrm{SC}) / (\hbar^2 q_\mathrm{F}) = 1 $, $ \overline{\lambda}_\mathrm{MA} = (2m\lambda_\mathrm{MA}) / (\hbar^2 q_\mathrm{F}) = 0.005 $, and $ \overline{\lambda}_\mathrm{D} = (2m\beta) / \hbar^2 = 0.5 $. The temperature is chosen such that~$ T / T_\mathrm{C} = 0.1 $, where~$ T_\mathrm{C} \approx 16 \, \mathrm{K} $ abbreviates the superconductors'~critical~temperature. The inset shows the \emph{maximal}~$ I_x $~(i.e., for $ \Phi = \pi / 2 $) as a function of the superconducting~phase~difference, $ \phi_\mathrm{S} $. (b)~Similar calculations as in~(a) for the AJHE~current along~$ \hat{y} $, $ I_y $. 
                        \label{FigTransverseCurrent}}
        \end{figure}

        In~Fig.~\ref{FigTransverseCurrent}, we show the numerically extracted AJHE~currents, $ I_x $ and~$ I_y $, for one representative S/F-I/S~junction. 
        For the superconducting~materials' zero-temperature~gap and their critical~temperature, we substituted realistic values for $ s $-wave~superconductors~\cite{Carbotte1990}, $ |\Delta_\mathrm{S}(0)| \approx 2.5 \, \mathrm{meV} $ and~$ T_\mathrm{C} \approx 16 \, \mathrm{K} $. The F-I~parameters refer, e.g., to a weakly magnetic barrier~(exchange~couplings in the $ \mathrm{meV} $-range) with a height of about~$ 0.75 \, \mathrm{eV} $ and a width of about~$ 0.40 \, \mathrm{nm} $~(assuming $ q_\mathrm{F} \approx 8 \times 10^7 \, \mathrm{cm}^{-1} $ as a typical Fermi~wave~vector~\cite{Martinez2018}); the chosen Dresselhaus~SOC, $ \overline{\lambda}_\mathrm{D} = (2m\beta) / \hbar^2 = 0.5 $, corresponds to typical~Dresselhaus~SOC~strengths of~$ \beta \approx 1.9 \, \mathrm{eV} \, \text{\AA}^2 $~(for~example, $ \mathrm{AlP} $~barriers with the considered height and width would have~$ \beta \approx 1.7 \, \mathrm{eV} \, \text{\AA}^2 $~\cite{Fabian2007,Note2}), while the dimensionless Rashba~measure got varied between~$ \overline{\lambda}_\mathrm{R} = (2m\alpha) / \hbar^2 = 0.5 $ and~$ \overline{\lambda}_\mathrm{R} = 4.0 $, indicating bare Rashba~SOC~strengths between~$ \alpha \approx 1.9 \, \mathrm{eV} \, \text{\AA}^2 $ and~$ \alpha \approx 15.2 \, \mathrm{eV} \, \text{\AA}^2 $, respectively. A recent study~\cite{Martinez2018} concluded that the Rashba~SOC arising at $ \mathrm{Fe} $/$ \mathrm{MgO} $/$ \mathrm{V} $~junctions' interfaces can reach values up to~$ \alpha \approx 4.6 \, \mathrm{eV} \, \text{\AA}^2 $~(for a $ 1.7 \, \mathrm{nm} $~thick $ \mathrm{MgO} $~barrier), which lies well within the range we considered. Even larger Rashba~couplings were furthermore predicted to appear at~$ \mathrm{BiTeBr} $~interfaces~\cite{Ideue2017}.
        
        Let us first discuss the dependence of the AJHE~currents on the in-plane~magnetization~angle, $ \Phi $, and at zero~superconducting~phase~difference~($ \phi_\mathrm{S}=0 $). The apparent sinelike~(cosinelike)~variations of~$ I_x $~($ I_y $) with respect to~$ \Phi $ are a direct consequence of the intriguing interplay of ferromagnetism and the interfacial~SOC~\cite{Costa2019}, and a distinct (experimental)~fingerprint for the junction's magnetoanisotropic charge~transport~properties~\cite{MatosAbiague2015}. To be more specific, we deduced~$ I_x \sim -(\alpha+\beta) \sin \Phi $ and~$ I_y \sim (\alpha - \beta) \cos \Phi $ in an earlier work~\cite{Costa2019}. The latter explains the vanishing~$ I_y $ for~$ \alpha \sim \overline{\lambda}_\mathrm{R} = 0.5 $~(equals the considered~Dresselhaus~SOC, $ \beta \sim \overline{\lambda}_\mathrm{D} = 0.5 $), illustrated by the dark red~curve in~Fig.~\ref{FigTransverseCurrent}(b). In~fact, inspecting the SOC~part of the single-particle barrier~Hamiltonian in~Eq.~\eqref{EqBarrierHamiltonian} suggests that~$ \alpha = \beta $ completely suppresses the skew~AR~mechanism along~$ \hat{y} $, which we identified as the physical origin of nonzero AJHE~currents, and thus simultaneously~$ I_y $. Already a slight change of the Rashba~SOC~strength~(while keeping all remaining parameters fixed) typically significantly alters the AJHE~currents'~amplitudes and offers hence an efficient experimental way to control skew~ARs. The real interplay of all system~parameters is rather intricate. This can be observed, e.g., in our simulations for~$ I_y $. Contrary to~$ I_x $, whose amplitudes get continuously damped with increasing Rashba~SOC, stronger~Rashba~SOC reverses $ I_y $'s~direction~(sign) and initially even enhances its absolute amplitudes. In the limit of strong~SOC, both currents are heavily damped since strong interfacial~SOC acts like large~(additional) scattering~potentials; see~Eq.~\eqref{EqPotentialEffective}. Similar features, especially the reversal of the AJHE~current with enlarging~$ \overline{\lambda}_\mathrm{R} $, can also appear for~$ I_x $. Reversing the AJHE~currents requires a reversal of the skew~AR~mechanism, depicted in~Fig.~\ref{FigModel}, with respect to~$ \mathbf{k}_\parallel $'s sign. This may be most conveniently achieved by varying either the scalar~tunneling~strength, $ \lambda_\mathrm{SC} $, or the Rashba~SOC~strength, $ \alpha $, both governing the effective scattering~potential in~Eq.~\eqref{EqPotentialEffective} responsible for skew~ARs, in an appropriate way~\cite{Costa2019,Note2}. Overall, when compared to conventional anomalous~Hall~effects~\cite{MatosAbiague2015,Rylkov2017,Zhuravlev2018,Costa2019}, the AJHE~currents are sizable.
        
        Next, we analyze the influence of the superconducting~phase~difference, $ \phi_\mathrm{S} $, on the \emph{maximal} AJHE~currents; see the insets in~Fig.~\ref{FigTransverseCurrent}. While the junction's (tunneling)~Josephson~current always follows the well-established sinusoidal current-phase~relation~(not explicitly shown; see~Ref.~\cite{Costa2018}), the transverse AJHE~currents vary with~$ \phi_\mathrm{S} $ in a remarkably different way. The greatest AJHE~currents flow at those phase~differences at which the (tunneling)~Josephson~current itself vanishes, i.e., at~$ \phi_\mathrm{S} = 0 \, \, (\mathrm{mod} \, \pi) $. To develop a simple understanding of the AJHE~currents' phase~dependence, we may look once again into our Cooper~pair skew~tunneling~picture~(mediated by the skew~ARs as outlined in the explanations to~Fig.~\ref{FigModel}). 
        \begin{figure}
    	    \includegraphics[width=0.425\textwidth]{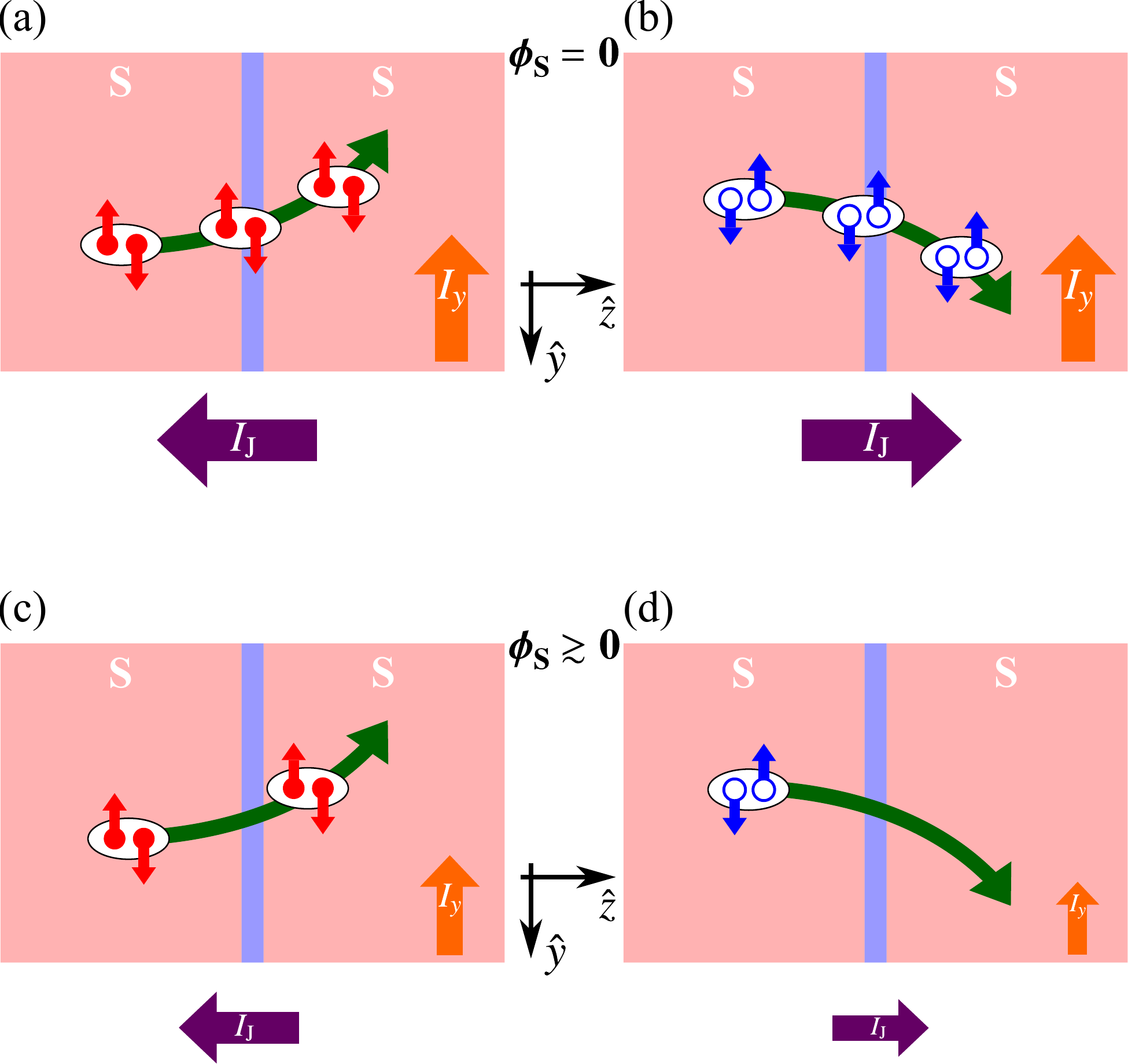}
            \colorcaption{(a)~Illustration of the \emph{electron} Cooper~pair tunneling from the left into the right~S across the F-I~barrier~(light blue), generating the (tunneling)~Josephson~current, $ I_\mathrm{J} $, and, owing to the skew~tunneling~mechanism~(illustrated by the green~arrows), the transverse AJHE~current, $ I_y $; the superconducting~phase~difference is~$ \phi_\mathrm{S} = 0 $ and the current~amplitudes are proportional to the size of the violet and orange arrows. (b)~Same as in~(a), but for the tunneling of \emph{hole} Cooper~pairs from the left into the right~S, essentially modeling electron Cooper~pair~tunneling from right to left. At~$ \phi_\mathrm{S} = 0 $, $ I_\mathrm{J} $'s amplitude is the same as in~(a), but the current flows along the opposite~direction~(recall that hole~currents enter with opposite signs). The overall (tunneling)~Josephson~current vanishes. The transverse AJHE~currents~(both have again the same magnitude), contrarily, flow along the same direction and become maximal. (c),~(d)~Same as in~(a) and~(b), but at~$ \phi_\mathrm{S} \gtrsim 0 $. Finite phase introduces a ``bias'' so that more electron Cooper~pairs tunnel from left to right than vice~versa and the overall (tunneling)~Josephson~current slowly starts to increase~(the contributions no longer completely compensate, though they are both smaller than at~$ \phi_\mathrm{S} = 0 $). The decrease of the (tunneling)~Cooper~pair currents simultaneously damps their contributions to the generated AJHE~current.
                        \label{FigIllustrationSpinJosephsonCurrentGenerationPaper}}
        \end{figure}
        All supercurrent~flows through the junction are essentially generated by the tunneling of Cooper~pairs from one into the other~S, each happening with certain probabilities. At zero superconducting~phase~difference~($ \phi_\mathrm{S} = 0 $), tunnelings of Cooper~pairs from the left into the right~S and vice~versa become equally likely. All Cooper~pairs leaving one~S are therefore fully compensated by others entering this~S and no net (tunneling)~Josephson~currents flow; see Figs.~\ref{FigIllustrationSpinJosephsonCurrentGenerationPaper}(a)--(b) for illustration~(the tunneling of Cooper~pairs from right to left is modeled in terms of hole~Cooper~pairs that tunnel from left to right). Increasing~$ \phi_\mathrm{S} $ acts now as an effective ``bias''. While the probability for  \emph{forward~tunneling}~(meaning from the left into the right~S) is only barely affected, \emph{backward~tunneling}~(meaning from the right into the left~S) becomes much less likely. In the end, more (electron)~Cooper~pairs are transferred into the right~S than leave, giving rise to a \emph{finite} (tunneling)~Josephson~current. The imbalance~(``bias'') between forward and backward~tunnelings gets more distinct with further enhancing~$ \phi_\mathrm{S} $ so that simultaneously the (tunneling)~Josephson~current rises. Owing to the tunneling~probabilities' periodicity, the situation eventually reverses at~$ \phi_\mathrm{S} \approx \pi / 2 $~(assuming ideal or dirty~junctions; otherwise the reversal happens at other values of~$ \phi_\mathrm{S} $) and the Josephson~current decreases again, finally resembling the typical sinusoidal Josephson~current-phase~relation. 
        
        In sharp contrast, the AJHE~current~contributions stemming from forward and backward~tunneling of Cooper~pairs flow along the \emph{same} direction and thus add up. As a consequence, the largest AJHE~currents appear whenever forward and backward~tunnelings become maximal~(and equal in magnitudes), i.e., precisely at~$ \phi_\mathrm{S} = 0 \, \, (\mathrm{mod} \, \pi) $, as calculated in~Fig.~\ref{FigTransverseCurrent}. Increasing~$ \phi_\mathrm{S} $ then primarily suppresses backward~tunneling and simultaneously the total AJHE~currents; see~Figs.~\ref{FigIllustrationSpinJosephsonCurrentGenerationPaper}(c) and~\ref{FigIllustrationSpinJosephsonCurrentGenerationPaper}(d) for illustration.

    %% Bound state picture---SOC asymmetries %%
    \section{Bound~state picture---SOC~asymmetries      \label{SecBoundStates}}

        The formation of \emph{interfacial} subgap~bound~states counts to the most distinct spectroscopic characteristics of Josephson~junctions. Particularly interesting is the case in which the junctions additionally comprise magnetic~components and the bound~state~spectrum splits into ABS and YSR~branches. The latter turned out to possess unique spectral~properties~\cite{Vecino2003,Kawabata2012,Costa2018,Rouco2019} already in one-dimensional point~contacts.
       
        Those states are especially relevant to our study since all electrical~current inside the F-I~barrier is essentially carried by single~electrons, which initially formed Cooper~pairs in one of the superconductors, and now tunnel through the barrier via the available bound~states. Each bound state \emph{occupied} by an electron characteristically contributes to the (tunneling)~Josephson and the AJHE~currents. Instead of dealing with the Furusaki--Tsukada~approach~(see~Sec.~\ref{SecAJHECurrents}), one can equivalently access the current~components via the bound~state wave~functions. The full calculations are rather cumbersome and can be looked up in~Appendix~\ref{AppB} and the~SM~\cite{Note2}. The resulting \emph{interfacial} AJHE~currents, $ I_\eta $, read as
        \begin{align}
             I_\eta &= -e \sum_{E_\mathrm{B}} \frac{|\Delta_\mathrm{S}(0)| \tanh \left( 1.74 \sqrt{T_\mathrm{C}/T-1} \right)}{2E_\mathrm{B}} \nonumber \\
             &\hspace{10 pt} \times \frac{A}{(2\pi)^2} \int \mathrm{d}^2 \mathbf{k}_\parallel \, \frac{\hbar k_\eta}{m} \left[ \big| e(\mathbf{k}_\parallel ; E_\mathrm{B}) \big|^2 + \big| f(\mathbf{k}_\parallel ; E_\mathrm{B}) \big|^2 \right. \nonumber \\
             &\hspace{20 pt} \left. + \big| g(\mathbf{k}_\parallel ; E_\mathrm{B}) \big|^2 + \big| h(\mathbf{k}_\parallel ; E_\mathrm{B}) \big|^2 \right] \times \tanh \left( \frac{E_\mathrm{B}}{2k_\mathrm{B}T} \right)
             %%&\hspace{10 pt} \times \tanh \left( \frac{E_\mathrm{B}}{2k_\mathrm{B}T} \right)
             \label{EqCurrentBoundStates}
             ,
        \end{align}
       %%\small
       %%\begin{widetext}
        %%    \begin{multline}
        %%        I_\eta = -e \sum_{E_\mathrm{B}} \frac{|\Delta_\mathrm{S}(0)| \tanh \left( 1.74 \sqrt{T_\mathrm{C}/T-1} \right)}{2E_\mathrm{B}} 
        %%        \frac{A}{(2\pi)^2} \int \mathrm{d}^2 \mathbf{k}_\parallel \, \frac{\hbar k_\eta}{m} \left[ \big| e(\mathbf{k}_\parallel ; E_\mathrm{B}) \big|^2 + \big| f(\mathbf{k}_\parallel ; E_\mathrm{B}) \big|^2 \right. 
        %%        \left. + \big| g(\mathbf{k}_\parallel ; E_\mathrm{B}) \big|^2 + \big| h(\mathbf{k}_\parallel ; E_\mathrm{B}) \big|^2 \right] \tanh \left( \frac{E_\mathrm{B}}{2k_\mathrm{B}T} \right) ,
        %%            \label{EqCurrentBoundStates}
        %%    \end{multline}
        %%\end{widetext}
        %%\normalsize
        where $ E_\mathrm{B} $ refers to the bound~states' energies~(ABS \emph{and} YSR~states), while $ e(\mathbf{k}_\parallel ; E_\mathrm{B}) $, $ f(\mathbf{k}_\parallel ; E_\mathrm{B}) $, $ g(\mathbf{k}_\parallel ; E_\mathrm{B}) $, and $ h(\mathbf{k}_\parallel ; E_\mathrm{B}) $ represent the electronlike and holelike coefficients of the underlying bound~state wave~function~(see~Appendix~\ref{AppB} and the~SM~\cite{Note2} for details). The thermal occupation~factor, $ \tanh [E_\mathrm{B} / (2k_\mathrm{B}T)] $, ensures that only occupied states are counted to the current. Simply speaking, the AJHE~currents are given by the electrons' transverse~velocities, $ v_\eta = (\hbar k_\eta) / m $, multiplied by their charge, $ -e $, and a ``weighting~factor'', which is mostly determined by the bound~state~energy~(via the wave~function~coefficients). 

        \begin{figure}
    	    \includegraphics[width=0.435\textwidth]{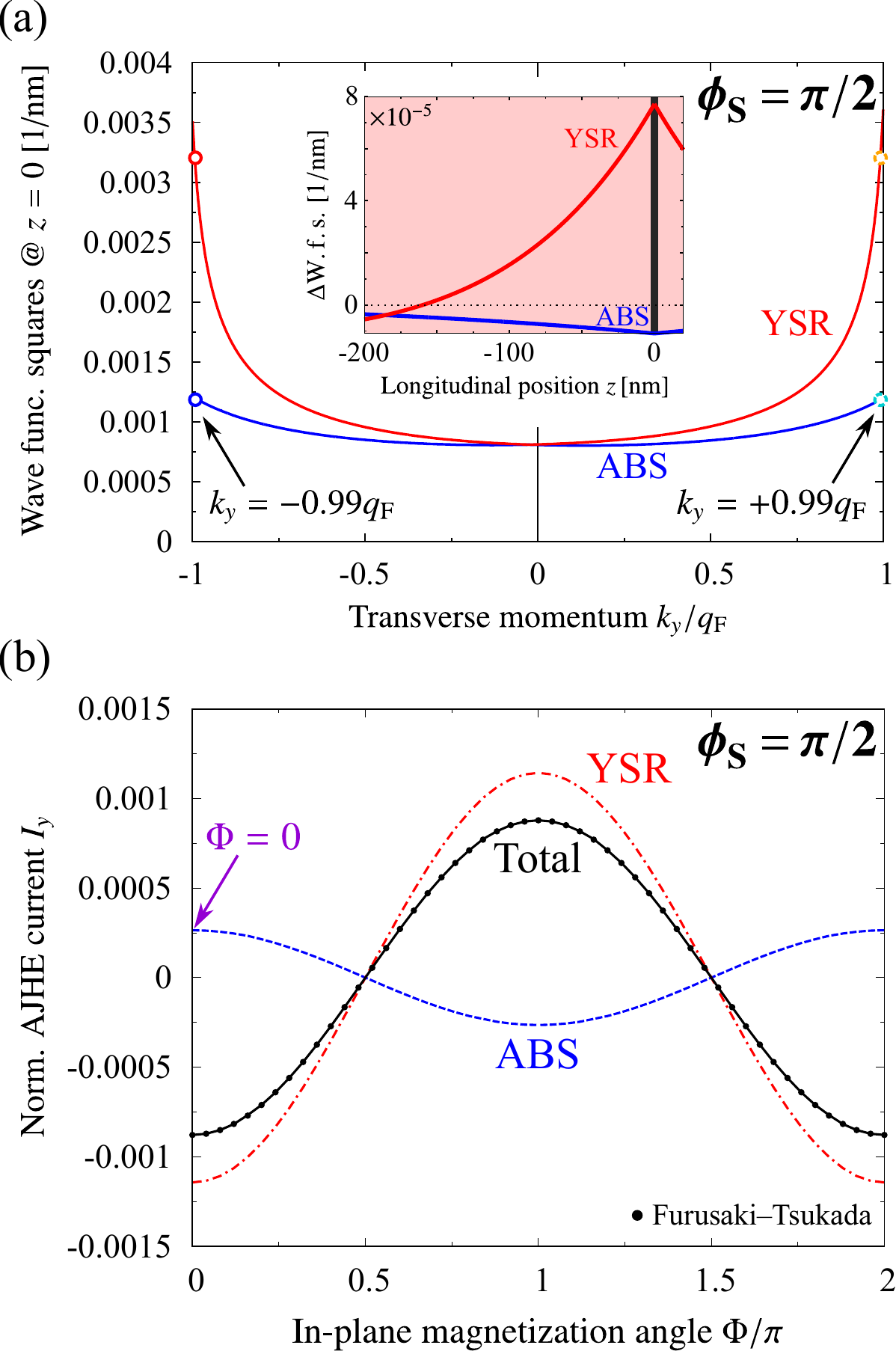}
            \colorcaption{(a)~Calculated absolute~squares of the bound~state wave~functions at the F-I~interface~($ z=0 $) as a function of the transverse~momentum~$ k_y $~(normalized to the Fermi~wave~vector, $ q_\mathrm{F} $) and for the superconducting~phase~difference~$ \phi_\mathrm{S}=\pi/2 $; for~simplicity, we set~$ k_x = 0 $ and~$ \Phi = 0 $. The Rashba~SOC~strength is~$ \overline{\lambda}_\mathrm{R} = (2m\alpha)/\hbar^2 = 1 $ and all other parameters are the same as in~Fig.~\ref{FigTransverseCurrent}. The blue~curve corresponds to ABS and the red~curve to YSR~states. The inset shows the \emph{spatial~dependence} of the bound~state wave~functions' absolute~square~\emph{differences}, exemplarily in the left~S and for~$ k_y = \pm 0.99q_\mathrm{F} $, as a deeper analysis~\cite{Note2} suggests that the dominant current~contributions stem from states with~$ |\mathbf{k}_\parallel| \to q_\mathrm{F} $. The positive YSR~tail indicates that the wave~function~squares at~$ k_y = 0.99q_\mathrm{F} $ exceed those at~$ k_y = -0.99q_\mathrm{F} $~(and vice~versa for the ABS). Though being small (as expected from the small AJHE~currents), the $ k_y $-asymmetry explained in the text becomes clearly evident. (b)~Dependence of~$ I_y $ on~$ \Phi $~[same normalization as in~Fig.~\ref{FigTransverseCurrent} and for~$ \overline{\lambda}_\mathrm{R} = (2m\alpha) / \hbar^2 = 1 $], calculated from the bound~state~spectrum. The contributions of ABS and YSR~states are separately resolved; all other parameters are the same as in~Fig.~\ref{FigTransverseCurrent}, except~$ \phi_\mathrm{S} = \pi/2 $. As a cross-check, the dots show the total AJHE~current evaluated from the Furusaki--Tsukada~approach.
                \label{FigAnisotropy}}
        \end{figure}

        As long as the interfacial~SOC remains absent, the junction's bound~state~spectrum is symmetric with respect to a reversal of~$ \mathbf{k}_\parallel $. To each electron with transverse velocity~$ v_\eta = (\hbar k_\eta) / m $, being transferred through the F-I via a bound~state at energy~$ E_\mathrm{B} $, one finds a second electron with opposite~velocity~($ -v_\eta $), occupying a bound~state with precisely the same~energy. Consequently, two occupied states always carry the same amount of current along opposite~directions so that the overall AJHE~currents vanish. Since SOC scales \emph{linearly} with the components of~$ \mathbf{k}_\parallel = [ k_x, \, k_y , \, 0 ]^\top $, nonzero SOC causes an asymmetry of the bound~state~energies with respect to $ \mathbf{k}_\parallel $'s sign. Depending on the chosen SOC~strength and the magnetic tunneling~parameter, the energies of the bound~states getting occupied by the propagating~(with transverse velocity~$ v_\eta $) and its counterpropagating~(with transverse velocity~$ -v_\eta $) electron are no longer identical and may noticeably differ. In contrast to the case without SOC, the current~contributions stemming from the propagating and counterpropagating states cannot fully compensate~[as the energy-dependent ``weighting~factors'' entering~Eq.~\eqref{EqCurrentBoundStates} differ once the $ E_\mathrm{B} $'s of the propagating and counterpropagating~states are no longer equal], and finite AJHE~currents start to flow. Such SOC-controlled $ \mathbf{k}_\parallel $-asymmetries in the bound~state~energies are thus the \emph{microscopic} physical manifestation of the~AJHE.

        Figure~\ref{FigAnisotropy}(a) illustrates this asymmetry for~$ k_y $~(keeping~$ k_x = 0 $ fixed) and the same parameters as considered in~Fig.~\ref{FigTransverseCurrent}, except that we additionally assume~$ \phi_\mathrm{S} = \pi/2 $ to stress that our explanations are general and not restricted to zero phase~difference. Since the SOC~asymmetry of the bound~state~energies is rather small and hard to visualize~(owing to the small~$ \lambda_\mathrm{MA} $ used for our calculations), we focus on the absolute~squares of the bound~state wave~functions~(see the~SM~\cite{Note2} for details). Apparently, the $ k_y $-asymmetry is more pronounced for the YSR than for the ABS~branch of the spectrum. Furthermore, the SOC~asymmetry impacts the ABS and the YSR~states in the opposite way. While the YSR~states' wave~function~squares are raised at~$ k_y > 0 $, those belonging to ABS decrease there. Translating both observations into current~flows, we expect that the single current~contributions stemming from the two bound~state~bands must flow along opposite directions and the YSR~part must be the dominant one. This is also the deeper reason why sizable AJHE~currents require not only interfacial~SOC, but also (at least weak) ferromagnetism. If the latter would not be there, the bound~state~bands simply merge into the usual ABS and the $ \mathbf{k}_\parallel $-asymmetry (and simultaneously the~AJHE) immediately disappear. 
        
        Evaluating the AJHE~currents from~Eq.~\eqref{EqCurrentBoundStates}~[see~Fig.~\ref{FigAnisotropy}(b)] essentially confirms all predicted features. The AJHE~currents obtained from the bound~state~spectrum coincide with the results extracted from the Furusaki--Tsukada~approach. Although the first method is computationally more challenging and less general, it establishes an important cross-check for the second technique and brings along more physical insight. For~example, the spatial~dependence of the bound~state wave~function~squares~[see~Fig.~\ref{FigAnisotropy}(a)] allows us to deduce the AJHE~currents' spatial~dependence, which was not covered by the Furusaki--Tsukada~formula~(we computed the currents at the interface there). Since the squares of the wave~function~coefficients directly enter the bound~state current~formula~[see~Eq.~\eqref{EqCurrentBoundStates}], the AJHE~currents decay in exactly the same way with increasing distance from the interface, i.e., exponentially over the characteristic decay~length~$ \kappa = 1/ \{ 2 \mathrm{Im} [q_{z,\mathrm{e}} (E_\mathrm{B})] \} $, where~$ q_{z,\mathrm{e}} (E_\mathrm{B}) = q_\mathrm{F} [ 1 + \im (|\Delta_\mathrm{S}|^2 - E_\mathrm{B}^2)^{1/2} / \mu - \mathbf{k}_\parallel^2 / q_\mathrm{F}^2 ]^{1/2} $ indicates the electronlike wave~vector inside the superconductors. We provide a more comprehensive discussion of the SOC-induced $ \mathbf{k}_\parallel $-asymmetries, with special attention on the bound~state~spectra and their correlation to the AJHE~currents, in the~SM~\cite{Note2}.
        %%\pagebreak

    %% Transverse spin currents %%
    \section{Transverse spin~currents      \label{SecJSCurrents}}

        Apart from the AJHE~charge~currents, also their spin~current~counterparts might provide indispensable ingredients for spintronics~applications. When tunneling through the spin-active F-I~barrier, some of the spin-singlet~Cooper~pairs' electrons undergo spin~flips and generate \emph{spin-polarized} triplet~pairs~\cite{Ouassou2017}. Those pairs' spin wave~functions may be composed of all possible triplet~pairings, $ | {\uparrow \uparrow} \rangle $, $ | {\downarrow \downarrow} \rangle $, and $ (| {\uparrow \downarrow} \rangle + | {\downarrow \uparrow} \rangle ) / \sqrt{2} $, where $ | {\uparrow} \rangle $~($ | {\downarrow} \rangle $) denotes a single~electron up-spin~(down-spin)~state with respect to the $ \hat{z} $-spin~quantization~axis~(inside the superconductors). The~$ (| {\uparrow \downarrow} \rangle + | {\downarrow \uparrow} \rangle ) / \sqrt{2} $-contribution is usually neglected since it decays rapidly inside real tunneling~barriers~\cite{Ouassou2017}. The remaining $ | {\uparrow \uparrow} \rangle $- and $ | {\downarrow \downarrow} \rangle $-pairs, however, are also subject to the proposed skew~tunneling~mechanism and may separate along the transverse directions. %%~(i.e., if $ | {\downarrow \downarrow} \rangle $-pairs are mostly transmitted at~$ \eta > 0 $, $ | {\uparrow \uparrow} \rangle $-pairs tunnel predominantly at~$ -\eta $).
        From that point~of~view, skew~tunneling acts like a \emph{transverse Cooper~pair spin~filter} and generates nonzero transverse~spin~\emph{super}current~flows, combining the advantages of the conventional spin~Hall~effect~(referring to pure transverse spin~currents in the absence of charge~currents)~\cite{Dyakonov1971,Dyakonov1971b,*Dyakonov1971a} with the dissipationless character of supercurrents.

        Anyhow, earlier studies~\cite{Malshukov2008} demonstrated that superconductors' fundamental time-reversal~(electron--hole)~symmetry suppresses the spin~Hall~effect. The recent prediction of sizable tunneling~spin~Hall~currents in metal/insulator/metal~junctions~\cite{MatosAbiague2015}, essentially triggered by interfacial skew~tunneling just as in our study, boosted new hopes to efficiently integrate the spin~Hall~effect into superconducting tunnel~junction~geometries. Nonetheless, replacing one of the junction's normal-conducting~electrodes by a S will dramatically impact the underlying physics. The resulting strong competition between skew~ARs and skew~SRs~(being another consequence of the electron--hole~symmetry) will again heavily suppress the tunneling~spin~Hall~currents~\cite{Note2}.

        \begin{figure}
    	    \includegraphics[width=0.475\textwidth]{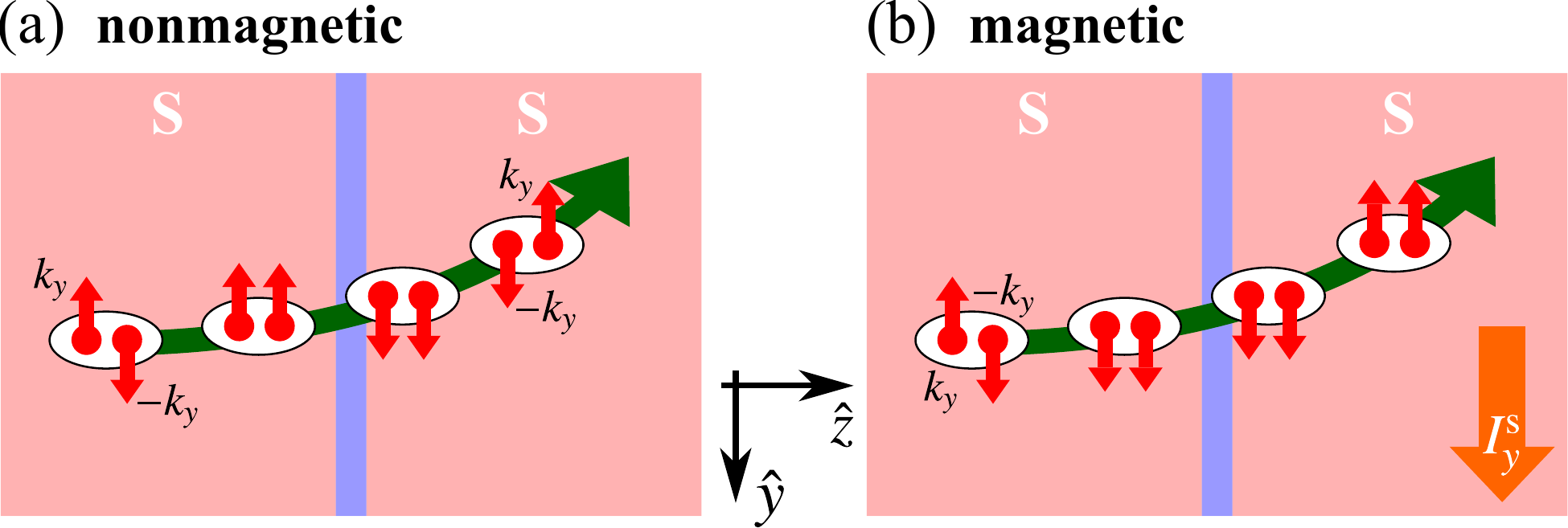}
            \colorcaption{(a)~Illustration of the Cooper~pair skew~tunneling from the left into the right~S across the F-I~barrier~(light blue). Each Cooper~pair initially consists of one up-spin~electron with transverse momentum~$ k_y > 0 $ and one down-spin~electron with~$ -k_y $~(assuming, for~simplicity, $ k_x=0 $). When tunneling through the spin-active~interface, at which the present SOC gives rise to nonzero \emph{spin-flip~probabilities}, some Cooper~pair~electrons flip their spins, converting spin-unpolarized singlet into spin-polarized triplet~pairs. In the absence of exchange~coupling~($ \lambda_\mathrm{MA}=0 $), interfacial spin~flips generate, on average, the same amount of polarized~$ | {\uparrow \uparrow} \rangle $- and~$ | {\downarrow \downarrow} \rangle $-Cooper~pairs (per transverse channel) so that eventually the overall transverse spin~current vanishes. (b)~If exchange~coupling is present~($ \lambda_\mathrm{MA} \neq 0 $), interfacial spin~flips cause an excess of either $ | {\uparrow \uparrow} \rangle $- or $ | {\downarrow \downarrow} \rangle $-pairs in the skew~tunneling~channel along~$ -\hat{y} $~(and vice~versa along~$ \hat{y} $). The result is a finite transverse spin~\emph{super}current, denoted by~$ I_y^\mathrm{s} $ and highlighted by the orange arrow.
                \label{FigIllustrationZeroSpinCurrent}}
        \end{figure}

        Before we evaluate the transverse spin~current~components that flow through our Josephson~junction, we therefore need to understand the connections between the triplet pair skew~tunneling and the generated transverse~spin~currents. Both superconductors act as reservoirs for spin-singlet~Cooper~pairs, each consisting of two electrons with opposite spin and antiparallel momenta~(recall that~$ \mathbf{k}_\parallel = [k_x, \, k_y, \, 0]^\top $). To be more specific, the allowed spin and transverse~momenta configurations of the Cooper~pairs are~$ ( {\mathbf{k}_\parallel,\uparrow} ; {-\mathbf{k}_\parallel,\downarrow} ) $, $ ( {-\mathbf{k}_\parallel,\downarrow} ; {\mathbf{k}_\parallel,\uparrow} ) $, $ ( {\mathbf{k}_\parallel,\downarrow} ; {-\mathbf{k}_\parallel,\uparrow} ) $, and~$ ( {-\mathbf{k}_\parallel,\uparrow} ; {\mathbf{k}_\parallel,\downarrow} ) $; the two parts always indicate the transverse momentum and spin of the two electrons forming a singlet~pair. 
        %%Starting from the quasiparticle~level, all pairings are mediated by one related AR~process of incoming electronlike~(holelike)~quasiparticles. 
        Approaching the barrier, the Cooper~pairs are exposed to the aforementioned skew~tunneling~mechanism. As a consequence, they are spatially separated along the transverse $ \hat{\eta} \in \{ \hat{x} ; \hat{y} \} $-directions, i.e., if the~$ ( {\mathbf{k}_\parallel,\uparrow} ; {-\mathbf{k}_\parallel,\downarrow} ) $- and~$ ( {-\mathbf{k}_\parallel,\downarrow} ; {\mathbf{k}_\parallel,\uparrow} ) $-pairs are predominantly transmitted at~$ \eta < 0 $, the remaining pairs tunnel mostly at positive~$ \eta $. For a further characterization, we distinguish between \emph{nonmagnetic} and \emph{magnetic} junctions. 
        
        \paragraph*{Nonmagnetic junctions.}
        As long as the barrier is \emph{nonmagnetic}, the numbers of Cooper~pairs involved in the skew~tunneling~processes at~$ \eta < 0 $ and~$ \eta > 0 $ are always equal. Therefore, both channels generate the same charge~current~flows along reversed directions and no net transverse charge~currents build up. %%Recall that we also argued within our quasiparticle~picture that ferromagnetism is necessary to avoid compensation of the AJHE~currents originating from skew~ARs of incident up-spin and down-spin~electrons~(now we are concerned with the Cooper~pair~picture in which both spin~channels are already simultaneously incorporated). 
        Close to the barrier, the interfacial SOC gives additionally rise to nonzero \emph{spin-flip~probabilities}, determined by the respective spin-flip~potential, $ V_\mathrm{flip} $. In the nonmagnetic~junction~(and assuming~$ \beta=0 $, as well as~$ k_x = 0 $, to further simplify our considerations), we deduce~$ V_\mathrm{flip} \sim \alpha k_y \sigma $, where $ k_y $ and $ \sigma $ denote one Cooper~pair~electron's $ \hat{y} $-component of~$ \mathbf{k}_\parallel $ and its spin~[note the close analogy with~Eq.~\eqref{EqPotentialEffective}]. In our case, this means that an up-spin~electron with~$ k_y > 0 $ flips its spin with the same probability as a down-spin~electron with~$ -k_y $. On~average, each transverse skew~tunneling~channel~(along~$ \pm \hat{y} $) contains then the same amount of~$ | {\uparrow \uparrow} \rangle $- and $ | {\downarrow \downarrow} \rangle $-triplet~pairs, and the overall transverse spin~current~components must vanish~[see~Fig.~\ref{FigIllustrationZeroSpinCurrent}(a) for illustration]. To get the full picture, one would also need to include the electron~Cooper~pairs tunneling from right to left~(or hole~pairs tunneling from left to right). Since similar arguments apply to hole Cooper~pairs, this would still not lead to finite transverse spin~currents.
        
        \paragraph*{Magnetic junctions.}
        The situation starts to change if the barrier becomes (at least weakly) \emph{magnetic}. The Cooper~pair~electrons' spin-flip~probabilities are then governed by the spin-flip~potential~$ V_\mathrm{flip} \sim (\lambda_\mathrm{MA} \sin \Phi) \sigma  + \alpha k_y \sigma $, and become asymmetric with respect to the electrons' spins. A $ k_y $-electron with spin~up flips its spin now with a different probability than a~$ (-k_y) $-spin~down electron. Therefore, the skew~tunneling~channel along~$ -\hat{y} $ comprises an excess of either $ | {\uparrow \uparrow} \rangle $- or $ | {\downarrow \downarrow} \rangle $-pairs and the channel along~$ \hat{y} $ either more $ | {\downarrow \downarrow} \rangle $- or $ | {\uparrow \uparrow} \rangle $-pairs. The result is a nonzero transverse spin~current; see~Fig.~\ref{FigIllustrationZeroSpinCurrent}(b). Note that, aside from the configuration involving magnetic~barriers, one could achieve similar effects, e.g., by replacing one of the superconducting~electrodes by a two-dimensional~S with strong bulk Rashba~SOC~\cite{Yang2012}. Furthermore, our qualitative explanations suggest that a reversal of~$ \lambda_\mathrm{MA} $'s sign must be sufficient to reverse the direction of the spin~current~(since this simultaneously reverses the sign of the spin-dependent magnetization~part of~$ V_\mathrm{flip} $).
        
        \begin{widetext}
        To access and quantify the \emph{particle}~\footnote{We compute \emph{particle}~spin~currents, which only distinguish between spin~up and spin~down, but do not take care of electrons' and holes' opposite charge. In the literature, some authors prefer to rather calculate \emph{charge}~spin~currents, additionally accounting for the electron and hole~charges.}~spin~currents in our junction, we can either generalize the Furusaki--Tsukada~technique or our bound~state~approach. Within an extended Furusaki--Tsukada~formulation~\cite{Asano2005}, the \emph{interfacial} $ \hat{\sigma}_z $-spin~currents along the $ \hat{\eta} $-direction are given by
        \begin{equation}
            \small
            I_{\eta,\hat{z}}^\mathrm{s} \approx \frac{k_\mathrm{B} T}{4} |\Delta_\mathrm{S}(0)| \tanh \left( 1.74 \sqrt{\frac{T_\mathrm{C}}{T} - 1} \right)
            \frac{A}{(2\pi)^2} \int \mathrm{d}^2 \mathbf{k}_\parallel \, \sum_{\omega_n} \frac{k_\eta}{\sqrt{q_\mathrm{F}^2-\mathbf{k}_\parallel^2}}
            \left[ \frac{\mathcal{C}^{(1)}(\im \omega_n) - \mathcal{D}^{(2)}(\im \omega_n) - \mathcal{A}^{(3)}(\im \omega_n) + \mathcal{B}^{(4)}(\im \omega_n)}{\sqrt{\omega_n^2 + |\Delta_\mathrm{S}(0)|^2 \tanh^2 \left( 1.74 \sqrt{T_\mathrm{C}/T - 1} \right)}} \right] ,
                \label{EqSpinCurrentFurusakiTsukada}
        \end{equation}
        \normalsize
        while the bound~state~modeling yields
        \small
        %%\begin{widetext}
            \begin{multline}
                I_{\eta,\hat{z}}^\mathrm{s} = \frac{\hbar}{2} \sum_{E_\mathrm{B}} \frac{|\Delta_\mathrm{S}(0)| \tanh \left( 1.74 \sqrt{T_\mathrm{C}/T-1} \right)}{2E_\mathrm{B}} 
                \frac{A}{(2\pi)^2} \int \mathrm{d}^2 \mathbf{k}_\parallel \, \frac{\hbar k_\eta}{m} \left[ \big| e(\mathbf{k}_\parallel ; E_\mathrm{B}) \big|^2 - \big| f(\mathbf{k}_\parallel ; E_\mathrm{B}) \big|^2 - \big| g(\mathbf{k}_\parallel ; E_\mathrm{B}) \big|^2 + \big| h(\mathbf{k}_\parallel ; E_\mathrm{B}) \big|^2 \right] \tanh \left( \frac{E_\mathrm{B}}{2k_\mathrm{B}T} \right) .
                    \label{EqSpinCurrentBoundStates}
            \end{multline}
        \normalsize
        Reasoning for the two formulas is given in~Appendix~\ref{AppC} and the SM~\cite{Note2}. 
        \end{widetext}

        \begin{figure}
    	    \includegraphics[width=0.45\textwidth]{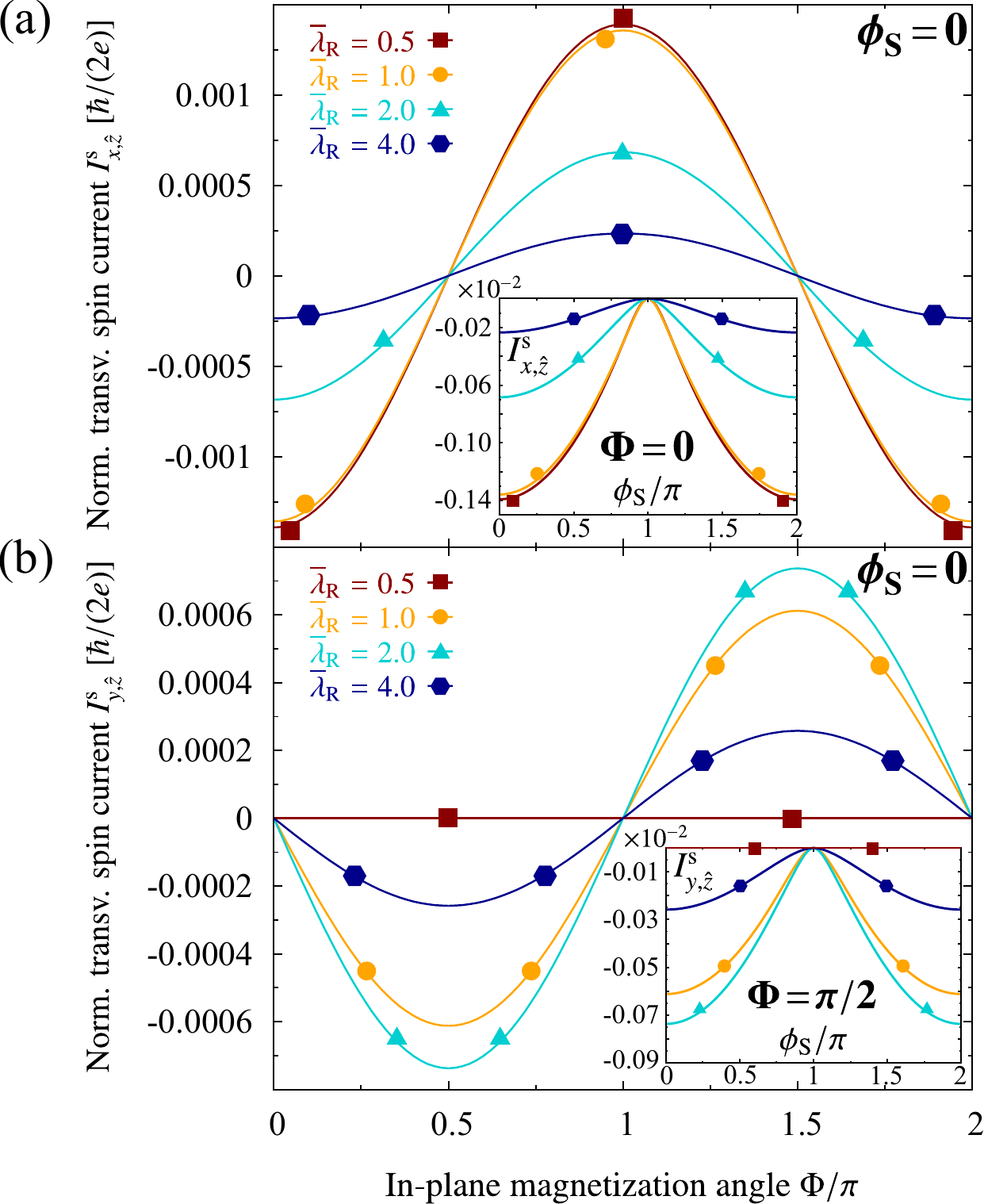}
            \colorcaption{(a)~Calculated dependence of the $ \hat{\sigma}_z $-spin~current along~$ \hat{x} $, $ I_{x,\hat{z}}^\mathrm{s} $, given in units of~$ \hbar / (2e) $ and normalized according to~$ (I_{x,\hat{z}}^\mathrm{s} e) / [G_\mathrm{S} \pi |\Delta_\mathrm{S}(0)|] $, on the F-I's in-plane~magnetization~angle, $ \Phi $, and for the same parameters as considered in~Fig.~\ref{FigTransverseCurrent}. The inset shows the \emph{maximal}~$ I_{x,\hat{z}}^\mathrm{s} $~(i.e., for~$ \Phi=0 $) as a function of the superconducting~phase~difference, $ \phi_\mathrm{S} $. (b)~Similar calculations as in~(a) for the $ \hat{\sigma}_z $-spin~current along~$ \hat{y} $, $ I_{y,\hat{z}}^\mathrm{s} $.
                \label{FigTransverseSpinCurrents}}
        \end{figure}
        Figure~\ref{FigTransverseSpinCurrents} presents the numerically computed~[by means of~Eq.~\eqref{EqSpinCurrentFurusakiTsukada}] transverse spin~current~components, $ I_{x,\hat{z}}^\mathrm{s} $ and $ I_{y,\hat{z}}^\mathrm{s} $, for the same set of junction~parameters considered when evaluating the AJHE~charge~currents in~Fig.~\ref{FigTransverseCurrent}. As stated above, putting the F-I's magnetic~tunneling~parameter to zero~(which basically means that the barrier becomes nonmagnetic) would immediately lead to vanishing transverse spin~currents. 
        In contrast, already the weak magnetic~tunneling~strength assumed for our AJHE~charge~current~calculations is sufficient to trigger sizable transverse spin~current~responses.

        Regarding the spin~currents' dependence on the F-I's in-plane~magnetization~angle, $ \Phi $, we observe an experimentally promising trend. While the charge~currents scale according to~$ I_x \sim \sin \Phi $ and~$ I_y \sim \cos \Phi $, the spin~currents obey~$ I_{x,\hat{z}}^\mathrm{s} \sim \cos \Phi $ and~$ I_{y,\hat{z}}^\mathrm{s} \sim \sin \Phi $. These well-distinct $ \Phi $-variations come along with another particularly auspicious property. The spin~current~components become maximal precisely at those magnetization~angles at which the AJHE~charge~current~counterparts simultaneously vanish. As a result, tuning the magnetization~angle allows for an experimental switch between the pure AJHE~charge~current and the pure transverse~spin~current~regimes. Owing to its analogy with conventional spin~Hall~effects, the latter phenomenon could be termed \emph{anomalous~Josephson~spin~Hall~effect}; \emph{anomalous} stresses that our junction needs to be weakly magnetic, in contrast to the conventional spin~Hall~effect which occurs already in nonmagnetic systems. Altering~$ \Phi $ essentially modulates the spin-flip~potential, controlling the spin-flip~probabilities of Cooper~pair~electrons and thereby the generation~rate of triplet~pairs. Particularly at~$ \Phi=\pi/2 $, the negative amplitudes of $ I_{y,\hat{z}}^\mathrm{s} $ indicate that each transverse skew~tunneling~channel along~$ \hat{y} $ involves an excess of $ | {\downarrow \downarrow} \rangle $-pairs. Moreover, the spin-flip~potential does not depend on the superconducting~phase~difference, $ \phi_\mathrm{S} $. Thus, varying~$ \phi_\mathrm{S} $ does not qualitatively impact the spin~current~flow~(i.e., not reverse its direction, in sharp contrast to the AJHE~charge~currents), but simply changes its overall amplitudes by introducing the ``bias'' between the mutually enhancing electron and hole Cooper~pairs we encountered when analyzing the AJHE~currents. At~$ \phi_\mathrm{S} = \pi $, maximal AJHE~charge~currents come again along with vanishing transverse spin~currents, which might offer another interesting parameter~configuration for following experiments. As claimed earlier when investigating the generic form of the spin-flip~potential, switching the magnetic~tunneling~parameter's sign would reverse the directions of the transverse spin~currents.

        \begin{figure}
    	    \includegraphics[width=0.475\textwidth]{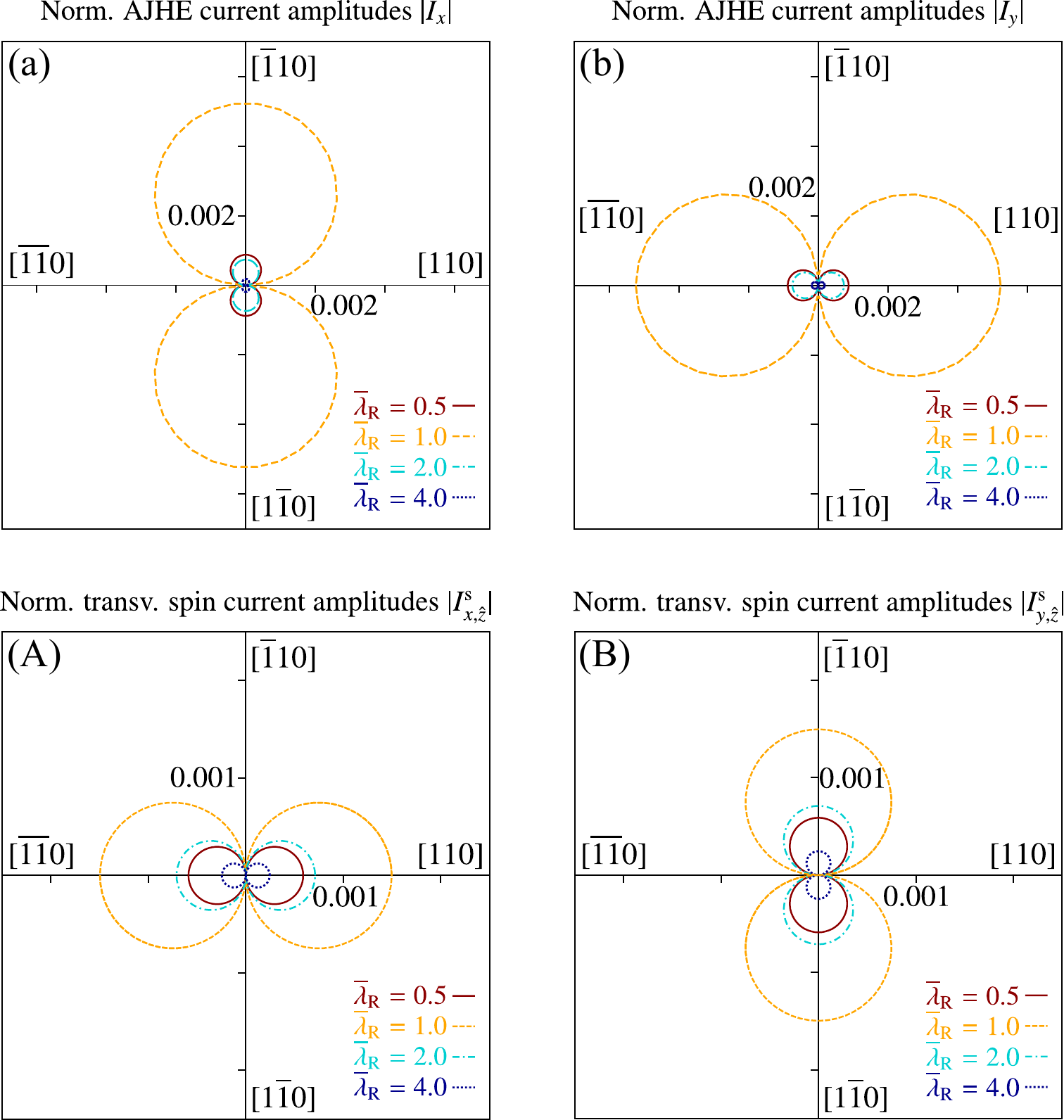}
            \colorcaption{(a)~Calculated angular dependence of the AJHE~charge~current~amplitudes along~$ \hat{x} $, $ I_x $, on the F-I's in-plane~magnetization~angle, $ \Phi $. All parameters and the normalization are the same as in~Fig.~\ref{FigTransverseCurrent}, except that we assume~$ \overline{\lambda}_\mathrm{D} = (2m\beta) / \hbar^2 = 0 $ now. (b)~Similar calculations as in~(a) for the AJHE~charge~current~amplitudes along~$ \hat{y} $, $ I_y $. (A),~(B)~Similar calculations as in~(a) and~(b), but for the transverse $ \hat{\sigma}_z $-spin~current~amplitudes, $ I_{x,\hat{z}}^\mathrm{s} $ and~$ I_{y,\hat{z}}^\mathrm{s} $, given in units of~$ \hbar / (2e) $ and normalized as in~Fig.~\ref{FigTransverseSpinCurrents}.
                \label{FigCurrentsAngularDependence}}
        \end{figure}
        We also computed all AJHE~charge and transverse spin~current~parts assuming that just Rashba~SOC is present and Dresselhaus~SOC is absent~($ \beta \sim \overline{\lambda}_\mathrm{D} = 0 $); all remaining parameters were not changed. This situation might often be the experimentally more realistic one since tunneling~barriers inevitably introduce interfacial Rashba~SOC due to the broken space~inversion~symmetry, whereas only those additionally lacking bulk~inversion~symmetry give rise to nonzero Dresselhaus~SOC. The results of our calculations are summarized in~Fig.~\ref{FigCurrentsAngularDependence}. Contrary to the tunneling~Josephson~(charge)~current, whose magnetoanisotropy disappears if only either interfacial Rashba or Dresselhaus~SOC is considered, the AJHE~charge and spin~currents' still clearly reveal their unique and well-distinct scaling with respect to the magnetization~angle we mentioned in the previous paragraph. Since~$ I_x \sim -(\alpha + \beta) \sin \Phi $ and~$ I_y \sim (\alpha - \beta) \cos \Phi $~(and adapted relations hold for the spin~currents), the maximal amplitudes of the $ \hat{x} $- and $ \hat{y} $-current~components become exactly equal once Dresselhaus~SOC is no longer there~(i.e., when setting~$ \beta = 0 $). For appropriately chosen Rashba~SOC~strengths, the current~amplitudes can now even overcome those we extracted in the simultaneous presence of Rashba and Dresselhaus~SOC. Measuring the currents' angular~dependencies for concrete junction~geometries and fitting the results to our modeling might provide valuable insight into the characteristics of the system's interfacial~SOC.

        \begin{figure}
    	    \includegraphics[width=0.445\textwidth]{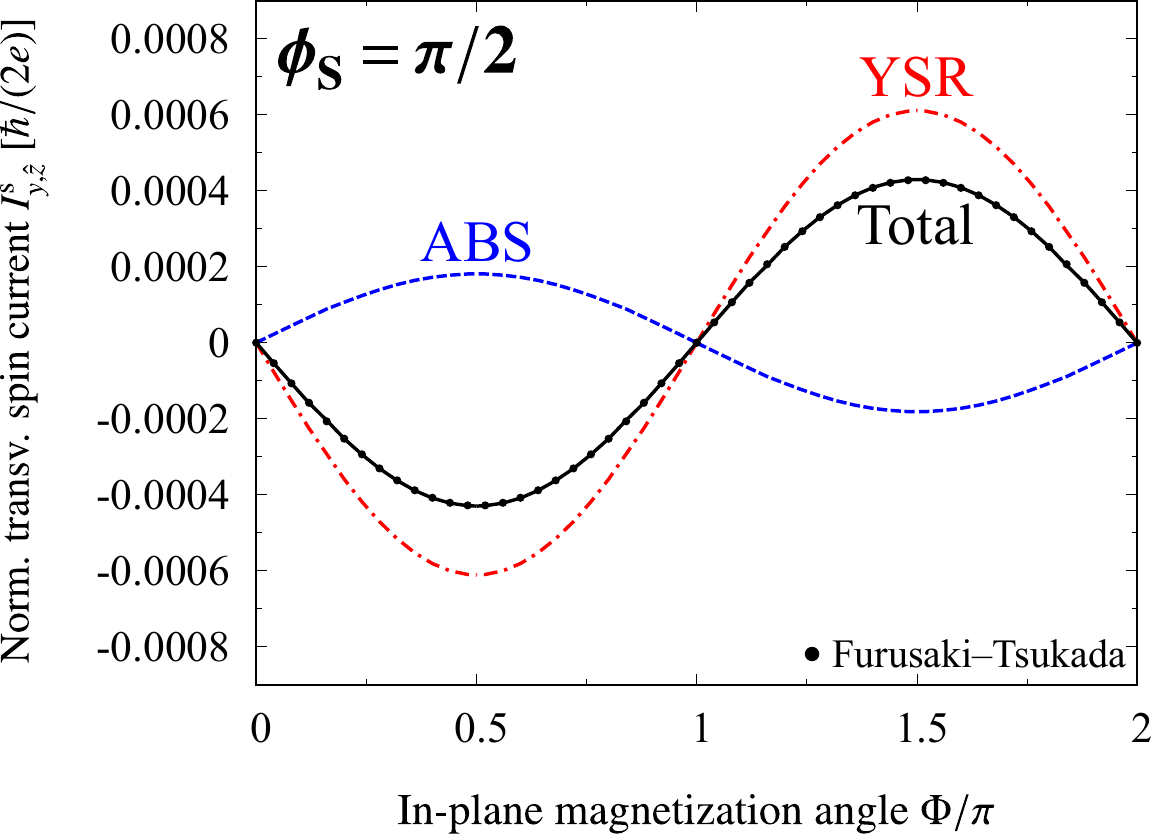}
            \colorcaption{Calculated~(from the bound~state~spectrum) dependence of the $ \hat{\sigma}_z $-spin~current along~$ \hat{y} $, $ I_{y,\hat{z}}^\mathrm{s} $, given in units of~$ \hbar/(2e) $ and normalized as in~Fig.~\ref{FigTransverseSpinCurrents}, on the F-I's in-plane~magnetization~angle, $ \Phi $, for the Rashba~SOC~parameter~$ \overline{\lambda}_\mathrm{R} = (2m\alpha) / \hbar^2 = 1 $ and the superconducting~phase~difference~$ \phi_\mathrm{S} = \pi / 2 $; all other parameters are the same as in~Fig.~\ref{FigTransverseCurrent}. The individual contributions of ABS and YSR~states are separately resolved. As a cross-check, the dots represent the total spin~current extracted from the Furusaki--Tsukada~formula.
                \label{FigTransverseSpinCurrentsResolved}}
        \end{figure}
        Similarly to our analyses of the AJHE~charge~currents, we finally evaluate the transverse spin~currents from the junction's bound~state~spectrum~[by means of~Eq.~\eqref{EqSpinCurrentBoundStates}]. Figure~\ref{FigTransverseSpinCurrentsResolved} illustrates the total spin~current along~$ \hat{y} $, $ I_{y,\hat{z}}^\mathrm{s} $, together with its individual contributions stemming from the junction's ABS and YSR~states, and, for comparison, the related~$ I_{y,\hat{z}}^\mathrm{s} $ obtained from the Furusaki--Tsukada~method~[using~Eq.~\eqref{EqSpinCurrentFurusakiTsukada}]. We regarded the same junction~parameters as in~Fig.~\ref{FigTransverseSpinCurrents}~(i.e., Rashba and Dresselhaus~SOC are both nonzero), except that we keep the superconducting~phase~difference at~$ \phi_\mathrm{S} = \pi / 2 $~(as in~Fig.~\ref{FigAnisotropy} to stress that the trends are general). Analogously to the AJHE~charge~currents, the transverse spin~currents are also mostly dominated by the YSR~states, which contribute again with an opposite sign to the overall spin~current compared to the ABS. The negative~(positive) sign of the YSR~states~(ABS)~parts~(at~$ 0 < \Phi < \pi $) actually entails that down-spin~(up-spin)~electrons with transverse momenta~$ \mathbf{k}_\parallel = [k_x > 0 , \, k_y > 0, \, 0]^\top $ tunnel predominantly through the F-I~interface via the available YSR~states~(ABS). This observation has its physical origin in the peculiar spin~characteristics associated with ABS and YSR~states in magnetic Josephson~junctions~\cite{Costa2018}. For the considered parameters, the YSR~states~(at fixed~$ \mathbf{k}_\parallel = [k_x > 0, \, k_y > 0, \, 0]^\top $) correspond to down-spin~states~(through which the down-spin Cooper~pair~electrons tunnel) and the ABS to up-spin~states~(through which the up-spin Cooper~pair~electrons tunnel); see the comprehensive analysis of the states' spin~characteristics provided in~Ref.~\cite{Costa2018}. An excess of down-spin~electrons with momentum~$ \mathbf{k}_\parallel $ that skew~tunnel through the interface yields a negative spin~current~(essentially, this is then precisely the case for the YSR~states) and an excess of up-spin~electrons~(in the ABS) a positively counted spin~current~contribution. The perfect agreement of the bound~state and the Furusaki--Tsukada~approach persuades that our results are reliable.

        \paragraph*{Spin--charge~current~cross~ratios.}
        \begin{figure}
    	    \includegraphics[width=0.445\textwidth]{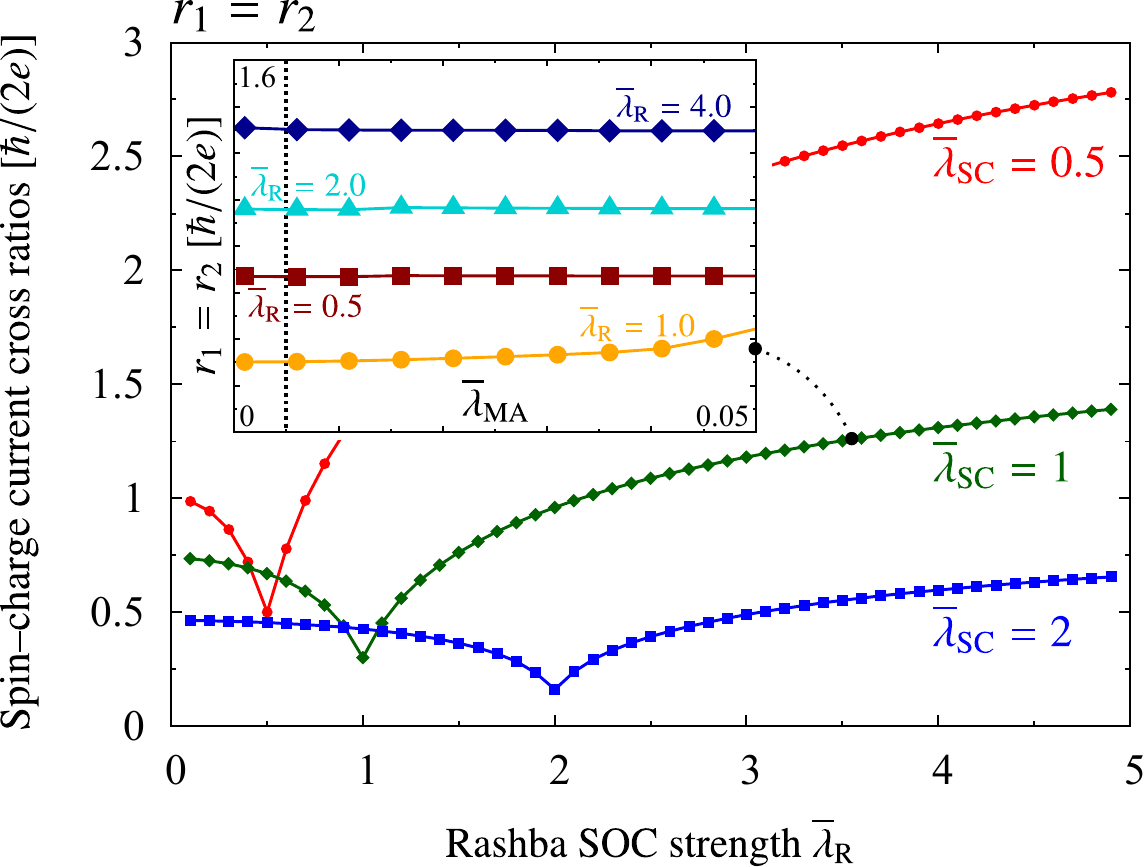}
            \colorcaption{Calculated dependence of the universal spin--charge~current~cross~ratios, $ r_1 $ and~$ r_2 $, given in units of~$ \hbar/(2e) $, on the Rashba~SOC~strength, $ \overline{\lambda}_\mathrm{R} = (2m\alpha) / \hbar^2 $. Since the Dresselhaus~SOC~parameter is~$ \overline{\lambda}_\mathrm{D} = (2m\beta) / \hbar^2 = 0 $, $ r_1 = r_2 $; all other parameters are the same as in~Fig.~\ref{FigTransverseCurrent}~(i.e., also~$ \phi_\mathrm{S} = 0 $), except that the scalar~tunneling gets gradually increased from~$ \overline{\lambda}_\mathrm{SC} = (2m\lambda_\mathrm{SC}) / (\hbar^2 q_\mathrm{F}) = 0.5 $~(red) to~$ \overline{\lambda}_\mathrm{SC} = 1 $~(dark green), and finally to~$ \overline{\lambda}_\mathrm{SC} = 2 $~(blue). The inset shows~$ r_1 $~($ = r_2 $) as a function of the magnetic~tunneling~parameter, $ \overline{\lambda}_\mathrm{MA} = (2m\lambda_\mathrm{MA}) / ( \hbar^2 q_\mathrm{F}) $, and for various Rashba~SOC~parameters, $ \overline{\lambda}_\mathrm{R} = (2m\alpha) / \hbar^2 $~(again assuming~$ \overline{\lambda}_\mathrm{D} = (2m\beta) / \hbar^2 = 0 $ for the Dresselhaus~SOC); $ \overline{\lambda}_\mathrm{SC} = (2m\lambda_\mathrm{SC}) / (\hbar^2 q_\mathrm{F}) = 1 $ is kept constant. The dotted vertical line indicates~$ \overline{\lambda}_\mathrm{MA} = 0.005 $, which we assumed for all previous calculations and for which the $ r $-ratios become indeed universal. 
                \label{FigSpinChargeRatio}}
        \end{figure}
        In \emph{weakly magnetic junctions}, both the AJHE~charge and transverse spin~currents increase \emph{linearly} with the magnetic~tunneling~parameter, $ \overline{\lambda}_\mathrm{MA} $. The \emph{spin--charge~current~cross~ratios}~\footnote{An alternative~(and probably more intuitive) definition of~$ r_1 $ and~$ r_2 $ might read as~$ r_1 := | I_{x,\hat{z}}^\mathrm{s}/I_x| $ and~$ r_2 := | I_{y,\hat{z}}^\mathrm{s}/I_y | $. However, owing to the distinct $ \Phi $-dependencies of~$ I_{x,\hat{z}}^\mathrm{s} $ and~$ I_x $~($ I_{y,\hat{z}}^\mathrm{s} $ and~$ I_y $), these ratios would \emph{not} become completely magnetization~independent, i.e., only the $ \overline{\lambda}_\mathrm{MA} $-dependence would drop out, but the $ \Phi $-dependence would remain.},
        \begin{equation}
            r_1 := \left| \frac{I_{x,\hat{z}}^\mathrm{s}}{I_y} \right| \quad \quad \text{and} \quad \quad r_2 := \left| \frac{I_{y,\hat{z}}^\mathrm{s}}{I_x} \right| ,
        \end{equation}
        turn then into universal, \emph{magnetization-independent}, measures, which are uniquely determined by the interfacial SOC~strengths~(keeping~$ \overline{\lambda}_\mathrm{SC} $ and~$ \phi_\mathrm{S} $ constant, and restricting ourselves to parameters for which all currents are nonzero). If only Rashba~SOC is present, both ratios become equal~($ r_1 = r_2 $), whereas the constructive~(destructive) interferences of finite Rashba and Dresselhaus~SOC impact the $ \hat{x} $- and~$ \hat{y} $-currents in a different manner so that generally~$ r_1 \neq r_2 $~(as~$ r_1 $ and~$ r_2 $ basically relate $ \hat{x} $- and~$ \hat{y} $-currents at the same time). Figure~\ref{FigSpinChargeRatio} illustrates the spin--charge~current~cross~ratios' characteristic scaling with respect to the Rashba~SOC~parameter, $ \overline{\lambda}_\mathrm{R} $, in the absence of Dresselhaus~SOC~($ \overline{\lambda}_\mathrm{D} = 0 $). Extracting~$ r_1 $ and~$ r_2 $ from experimental transport~data and fitting the results to our model provides one way to identify the SOC~parameters of the junction's F-I~interface, without having exact knowledge of~$ \overline{\lambda}_\mathrm{MA} $ or the magnetization~orientation. 
        \pagebreak
        
        As soon as~$ \overline{\lambda}_\mathrm{MA} $ overcomes some critical~value, the charge and spin~current~parts are additionally governed by \emph{nonlinear} $ \overline{\lambda}_\mathrm{MA} $-terms and the $ r $-ratios are no longer universal quantities of the system. To estimate the relevance of these nonlinearities, the inset of~Fig.~\ref{FigSpinChargeRatio} shows $ r_1 $~($ r_1 = r_2 $ since Dresselhaus~SOC is not present) as a function of~$ \overline{\lambda}_\mathrm{MA} $ and for various Rashba~SOC~strengths. Apparently, the spin--charge~current~cross~ratios remain indeed universal~(magnetization independent) for the small magnetic~tunneling~strengths considered in all previously discussed current~calculations~(i.e., for~$ \overline{\lambda}_\mathrm{MA} \approx 10^{-3} $) and can therefore be used to reliably quantify the present~SOC in experiments. Nonlinear~$ \overline{\lambda}_\mathrm{MA} $-terms do not affect the AJHE~charge and spin~currents unless~$ \overline{\lambda}_\mathrm{MA} $ gets further enhanced by at~least one order of magnitude.
        
        Another peculiar feature becomes visible once the Rashba~SOC~measure approaches the scalar~tunneling~strength, i.e., at $ \overline{\lambda}_\mathrm{R} \approx \overline{\lambda}_\mathrm{SC} $, as the spin--charge~current~cross~ratios' amplitudes always drop into a sharp dip there. To strengthen the generality of this observation, we considered three different $ \overline{\lambda}_\mathrm{SC} $-values in~Fig.~\ref{FigSpinChargeRatio}, essentially all causing the same behavior. Recalling our qualitative picture formulated in~Sec.~\ref{SecSkewAR}, the AJHE~charge~currents are generated by skew~ARs of incident up-spin and down-spin~electrons at the effective interfacial scattering~potential. The latter is stated in~Eq.~\eqref{EqPotentialEffective} for the limiting case of restricting ourselves to the current along~$ \hat{y} $, $ I_y $; similar arguments hold, nevertheless, also for the $ I_x $-current. Inspecting Eq.~\eqref{EqPotentialEffective}, we deduce that incoming down-spin~(up-spin)~electrons are exposed to the lowest~(largest) possible interfacial scattering~potential exactly when the Rashba~SOC and the scalar~tunneling~measures become equal. As a result, the down-spin~channel carries its maximal amount of AJHE~current, while the (oppositely oriented) contribution of the up-spin~channel becomes simultaneously minimal. The overall AJHE~current, $ I_y $, reaches its maximal value and even significantly overcomes the related spin~currents. Our numerical calculations discussed in~Figs.~\ref{FigCurrentsAngularDependence}(a)--\ref{FigCurrentsAngularDependence}(d) essentially confirm these characteristics. Note that Dresselhaus~SOC is not present; otherwise, the interference of Rashba and Dresselhaus~terms would give rise to more intricate features. Since the AJHE~charge~currents enter the spin--charge~current~cross~ratios' denominators, maximal $ I_y $~($ I_x $) eventually comes along with strongly suppressed $ r $-ratios, manifested by the $ r $-$ \overline{\lambda}_\mathrm{R} $~relations' sharp dips at~$ \overline{\lambda}_\mathrm{R} \approx \overline{\lambda}_\mathrm{SC} $. Moreover, an increase of~$ \overline{\lambda}_\mathrm{SC} $ notably damps the current~cross~ratios at large Rashba~SOC~($ \overline{\lambda}_\mathrm{R} > \overline{\lambda}_\mathrm{SC} $) since strong interfacial scalar~tunneling usually suppresses the generated spin~currents much faster than their charge~current~counterparts.

    %% Summary %%
    \section{Summary        \label{SecSummary}}

        To conclude, we investigated the intriguing interplay of SOC and ferromagnetism arising at the interface of S/F-I/S~Josephson~junctions. Starting from simplified qualitative arguments, we understood that skew~tunneling of Cooper~pairs through the spin-active interface can give rise to spontaneous transverse AJHE~charge~current~flows, which may become relevant to various superconducting spintronics applications, especially due to their dissipationless character and their wide tunability. We demonstrated the latter by evaluating the AJHE~current~amplitudes from a generalized Furusaki--Tsukada Green's~function technique and for a variety of realistic junction~parameters. The interfacial Rashba~SOC~strength, which is mostly determined by the material~composition of the system, and the magnetically adjustable phase~difference between the superconductors offer particularly auspicious possibilities to vary the AJHE~current~magnitudes over several orders of magnitude. Maximal AJHE~currents can reach a few percent of the (tunneling)~Josephson~current and thereby significantly exceed normal-state TAHE~conductances, which remain usually far below~$ 1 \, \% $ of the respective tunneling~conductances~\cite{MatosAbiague2015}. The AJHE~currents' unique sinelike~(cosinelike) variations with the magnetization~angle inside the F-I were identified as a clear evidence that all the fascinating physics really stems from the combination of SOC with ferromagnetism in one single junction. 
        
        To establish an alternative approach, which brings along more physical insight, we connected nonzero AJHE~currents to pronounced SOC-induced asymmetries in the junctions' ABS and YSR~bound~state energies, and elucidated that the AJHE on the one hand and these bound~state energy~asymmetries on the other hand are uniquely correlated. Resolving the individual states' current~contributions, we convinced ourselves that the huge AJHE~current~flows are predominantly maintained by the YSR~states, whose appearance counts to the most peculiar features of magnetic Josephson~junctions. 
        
        Finally, we outlined that SOC triggers interfacial spin~flips of Cooper~pair~electrons and produces \emph{spin-polarized} triplet pairs. Since these triplet~pairs are also subject to the skew~tunneling~mechanism, while carrying a net spin, we proposed that the AJHE~charge~current~phenomena come along with their transverse spin~current~counterparts. We qualitatively unraveled the spin~currents' general properties and computed their amplitudes once from Green's~functions and once exploiting the bound~state~asymmetries, again revealing a great tunability by means of the Rashba~SOC~parameter or the superconducting~phase~difference. We illustrated the spin~currents' well-distinct magnetization~angle~dependence when compared to the AJHE~charge~currents and characterized the universal (magnetization-independent) spin--charge~current~cross~ratios, which might provide a valuable experimental tool to probe interfacial SOC in superconducting tunnel~junctions.

    %% ================================================================================== %%
    %% ================================================================================== %%
    
    %% Acknowledgments %%
    \begin{acknowledgments}
        This work was supported by the International~Doctorate~Program Topological~Insulators of the Elite~Network of Bavaria and Deutsche~Forschungsgemeinschaft~(DFG, German~Research~Foundation)---Project-ID~314695032---SFB~1277~(Subproject~B07).
    \end{acknowledgments}

    %% ================================================================================== %%
    %% ================================================================================== %%

    %% APPENDIX %%
    \appendix
    
    %% Generalized Furusaki--Tsukada method %%
    \section{Generalized Furusaki--Tsukada~method
                \label{AppA}}
    
    Assuming translational invariance parallel to the F-I~interface, the solutions of the BdG~equation, $ \hat{\mathcal{H}}_\mathrm{BdG} \Psi(\mathbf{r}) = E \Psi(\mathbf{r}) $, describing quasiparticle~excitations of energy~$ E $, factorize into
    \begin{equation}
        \Psi(\mathbf{r}) = \psi(z) \mathrm{e}^{\mathrm{i} (\mathbf{k}_\parallel \cdot \mathbf{r}_\parallel)} ;
        \label{EqWFAnsatz}
    \end{equation}
    $ \mathbf{k}_\parallel = [k_x, \, k_y, \, 0]^\top $~($ \mathbf{r}_\parallel = [x, \, y, \, 0]^\top $) refers to the transverse wave~vector~(vector of transverse spatial~coordinates). Substituting Eq.~\eqref{EqWFAnsatz} into the BdG~equation, the most general solutions for the $ \hat{z} $-projected scattering~states inside the superconductors are found to read as
    \begin{multline}
        \psi^{(i)}(z<0) = \psi^{(i)}_\mathrm{incoming}(z<0) \\ %%
        + \mathcal{A}^{(i)} \left[ \begin{matrix} u \\ \phantom{a} 0 \phantom{a} \\ v \\ 0 \end{matrix} \right] \mathrm{e}^{-\mathrm{i} q_{z,\mathrm{e}} z} %%
        + \mathcal{B}^{(i)} \left[ \begin{matrix} \phantom{a} 0 \phantom{a} \\ u \\ 0 \\ v \end{matrix} \right] \mathrm{e}^{-\mathrm{i} q_{z,\mathrm{e}} z} \\ %%
        + \mathcal{C}^{(i)} \left[ \begin{matrix} v \\ \phantom{a} 0 \phantom{a} \\ u \\ 0 \end{matrix} \right] \mathrm{e}^{\mathrm{i} q_{z,\mathrm{h}} z} %%
        + \mathcal{D}^{(i)} \left[ \begin{matrix} \phantom{a} 0 \phantom{a} \\ v \\ 0 \\ u \end{matrix} \right] \mathrm{e}^{\mathrm{i} q_{z,\mathrm{h}} z} ,
        \label{EqStates1}
    \end{multline}
    as well as
    \begin{multline}
        \psi^{(i)}(z>0) = \mathcal{E}^{(i)} \left[ \begin{matrix} u \mathrm{e}^{\mathrm{i} \phi_\mathrm{S}} \\ 0 \\ v \\ 0 \end{matrix} \right] \mathrm{e}^{\mathrm{i} q_{z,\mathrm{e}} z} %%
        + \mathcal{F}^{(i)} \left[ \begin{matrix} 0 \\ u \mathrm{e}^{\mathrm{i} \phi_\mathrm{S}} \\ 0 \\ v \end{matrix} \right] \mathrm{e}^{\mathrm{i} q_{z,\mathrm{e}} z} \\ %%
        + \mathcal{G}^{(i)} \left[ \begin{matrix} v \mathrm{e}^{\mathrm{i} \phi_\mathrm{S}} \\ 0 \\ u \\ 0 \end{matrix} \right] \mathrm{e}^{-\mathrm{i} q_{z,\mathrm{h}} z} %%
        + \mathcal{H}^{(i)} \left[ \begin{matrix} 0 \\ v \mathrm{e}^{\mathrm{i} \phi_\mathrm{S}} \\ 0 \\ u \end{matrix} \right] \mathrm{e}^{-\mathrm{i} q_{z,\mathrm{h}} z} ,
        \label{EqStates2}
    \end{multline}
    where the electronlike and holelike wave~vectors' $ \hat{z} $-projections are given by
    \begin{align}
        q_{z,\mathrm{e}} &= q_{z,\mathrm{e}}(\mathbf{k}_\parallel ; E) = \sqrt{\frac{2m}{\hbar^2} \left[ \mu + \sqrt{E^2-|\Delta_\mathrm{S}|^2} \right] - \mathbf{k}_\parallel^2}
        \intertext{and}
        q_{z,\mathrm{h}} &= q_{z,\mathrm{h}}(\mathbf{k}_\parallel ; E) = \sqrt{\frac{2m}{\hbar^2} \left[ \mu - \sqrt{E^2-|\Delta_\mathrm{S}|^2} \right] - \mathbf{k}_\parallel^2} ,
    \end{align}
    and the coherence~factors, $ u=u(E) $ and $ v=v(E) $, need to satisfy
    \begin{equation}
        u(E) = \sqrt{\frac{1}{2} \left( 1+\sqrt{1-\frac{|\Delta_\mathrm{S}|^2}{E^2}} \right)} = \sqrt{1-v^2(E)} .
    \end{equation}
    The incoming waves, $ \psi_\mathrm{incoming}^{(i)} $, differentiate between (1)~up-spin electronlike, (2)~down-spin electronlike, (3)~up-spin holelike, and (4)~down-spin holelike quasiparticles incident on the F-I from the left superconductor. Formally, they can be written as
    \begin{align}
        \psi^{(1)}_\mathrm{incoming}(z<0) &= [ u, \, 0, \, v, \, 0 ]^\top \mathrm{e}^{\mathrm{i} q_{z,\mathrm{e}} z} , \\
        \psi^{(2)}_\mathrm{incoming}(z<0) &= [ 0, \, u, \, 0, \, v ]^\top \mathrm{e}^{\mathrm{i} q_{z,\mathrm{e}} z} , \\
        \psi^{(3)}_\mathrm{incoming}(z<0) &= [ v, \, 0, \, u, \, 0 ]^\top \mathrm{e}^{-\mathrm{i} q_{z,\mathrm{h}} z} ,
        \intertext{and}
        \psi^{(4)}_\mathrm{incoming}(z<0) &= [ 0, \, v, \, 0, \, u ]^\top \mathrm{e}^{-\mathrm{i } q_{z,\mathrm{h}} z} .
    \end{align}
    To attain the unknown reflection and transmission~coefficients entering the scattering~states, we apply the interfacial~($ z=0 $) boundary~conditions
    \begin{equation}
        \psi(z) \big|_{z=0_-} = \psi(z) \big|_{z=0_+} ,
        \label{EqBoundary1}
    \end{equation}
    as well as
    \begin{multline}
        \left \{ \left[ \frac{\hbar^2}{2m} \frac{\mathrm{d}}{\mathrm{d}z} + \lambda_\mathrm{SC} \right] \boldsymbol{\eta} + \lambda_\mathrm{MA} \boldsymbol{\omega} \right\} \psi(z) \big|_{z=0_-} \\
        + \left[ \begin{matrix} \boldsymbol{\Omega} \cdot \hat{\boldsymbol{\sigma}} & \mathbf{0} \\ \mathbf{0} & -(\boldsymbol{\Omega} \cdot \hat{\boldsymbol{\sigma}}) \end{matrix} \right] \psi(z) \big|_{z=0_-} = \frac{\hbar^2}{2m} \frac{\mathrm{d}}{\mathrm{d}z} \boldsymbol{\eta} \psi(z) \big|_{z=0_+} ,
        \label{EqBoundary2}
    \end{multline}
    with
    \begin{equation}
        \boldsymbol{\eta} = \left[ \begin{matrix} 1 & 0 & 0 & 0 \\ 0 & 1 & 0 & 0 \\ 0 & 0 & -1 & 0 \\ 0 & 0 & 0 & -1 \end{matrix} \right] \hspace{5 pt} \text{and} \hspace{5 pt} \boldsymbol{\omega} = \left[ \begin{matrix} 0 & \mathrm{e}^{-\im \Phi} & 0 & 0 \\ \mathrm{e}^{\im \Phi} & 0 & 0 & 0 \\ 0 & 0 & 0 & \mathrm{e}^{-\im \Phi} \\ 0 & 0 & \mathrm{e}^{\im \Phi} & 0 \end{matrix} \right]
        ,
    \end{equation}
    to the states and numerically solve the resulting linear systems of equations; $ \boldsymbol{\Omega} = \big[ ( \alpha - \beta ) k_y , \, - (\alpha + \beta ) k_x , \, 0 \big] $ contains the single-particle~Hamiltonians' Rashba and Dresselhaus~SOC~parts. 

    After identifying the AR~coefficients belonging to the four stated quasiparticle~injections, $ \mathcal{C}^{(1)} $, $ \mathcal{D}^{(2)} $, $ \mathcal{A}^{(3)} $, and~$ \mathcal{B}^{(4)} $, the interfacial AJHE~charge~currents can be evaluated from the extended Furusaki--Tsukada~formula~\cite{Furusaki1991}
    \begin{align}
        I_\eta &\approx \frac{e k_\mathrm{B} T}{2\hbar} |\Delta_\mathrm{S}(0)| \tanh\left( 1.74 \sqrt{\frac{T_\mathrm{C}}{T}-1} \right) \nonumber \\
        &\hspace{0 pt} \times \frac{A}{(2\pi)^2} \int \mathrm{d}^2 \mathbf{k}_\parallel \sum_{\omega_n} \frac{k_\eta}{\sqrt{q_\mathrm{F}^2 - \mathbf{k}_\parallel^2}} \nonumber \\
        &\hspace{0 pt} \times \left[ \frac{\mathcal{C}^{(1)}(\im \omega_n) + \mathcal{D}^{(2)}(\im \omega_n) + \mathcal{A}^{(3)}(\im \omega_n) + \mathcal{B}^{(4)}(\im \omega_n)}{\sqrt{\omega_n^2 + |\Delta_\mathrm{S}(0)|^2 \tanh^2 \left( 1.74 \sqrt{T_\mathrm{C}/T - 1} \right)}} \right] ,
        \label{EqCurrentFurusakiTsukadaAppendix}
    \end{align}
    where $ e $ indicates the (positive) elementary~charge, $ k_\mathrm{B} $ resembles Boltzmann's~constant, and $ \omega_n = (2n+1) \pi k_\mathrm{B} T $, where $ n $ is an integer, represents the fermionic Matsubara~frequencies~(at temperature~$ T $ and given in units of~$ 1/\hbar $). This current~formula is essentially given as~Eq.~\eqref{EqCurrentFurusakiTsukada} in~Sec.~\ref{SecAJHECurrents}. To simplify our considerations, we assumed that the junction's tunneling and Hall~contact~areas are equal and denoted by~$ A $. To account for temperature~effects, we substituted the Bardeen-Cooper-Schrieffer--type scaling of the superconducting~energy~gap, i.e., $ |\Delta_\mathrm{S}(T \neq 0)|  = |\Delta_\mathrm{S}(0)| \tanh (1.74 \sqrt{T_\mathrm{C}/T-1}) $, with $ |\Delta_\mathrm{S}(0)| $ referring to the gap at absolute zero and $ T_\mathrm{C} $ to the superconductors' critical~temperature. Further details can be looked up in the~SM~\cite{Note2}.

    %% Bound state technique %%
    \section{Bound state technique
                \label{AppB}}

        To access our junction's characteristic ABS and YSR~bound~state energies, we revisit the general ansatz for~$ \psi(z) $, Eqs.~\eqref{EqStates1}--\eqref{EqStates2}, \emph{without considering incoming waves}. Restricting ourselves to \emph{positive} bound~state~energies, $ E > 0 $, we can write
        \begin{widetext}
        \begin{multline}
            \psi(z<0 ; \mathbf{k}_\parallel ; E) = a(\mathbf{k}_\parallel ; E) \left[ \begin{matrix} u(E) \\ 0 \\ v(E) \\ 0 \end{matrix} \right] \mathrm{e}^{-\mathrm{i} q_{z,\mathrm{e}} (\mathbf{k}_\parallel ; E) z}  %%
            + b(\mathbf{k}_\parallel ; E) \left[ \begin{matrix} 0 \\ u(E) \\ 0 \\ v(E) \end{matrix} \right] \mathrm{e}^{-\mathrm{i} q_{z,\mathrm{e}}(\mathbf{k}_\parallel ; E) z} %%
            \\
            + c(\mathbf{k}_\parallel ; E) \left[ \begin{matrix} v(E) \\ 0 \\ u(E) \\ 0 \end{matrix} \right] \mathrm{e}^{\mathrm{i} q_{z,\mathrm{h}}(\mathbf{k}_\parallel ; E) z} %%
            + d(\mathbf{k}_\parallel ; E) \left[ \begin{matrix} 0 \\ v(E) \\ 0 \\ u(E) \end{matrix} \right] \mathrm{e}^{\mathrm{i} q_{z,\mathrm{h}}(\mathbf{k}_\parallel ; E) z}
        \end{multline}
        and likewise
        \begin{multline}
            \psi(z>0 ; \mathbf{k}_\parallel ; E) = e(\mathbf{k}_\parallel ; E) \left[ \begin{matrix} u(E) \mathrm{e}^{\mathrm{i} \phi_\mathrm{S}} \\ 0 \\ v(E) \\ 0 \end{matrix} \right] \mathrm{e}^{\mathrm{i} q_{z,\mathrm{e}}(\mathbf{k}_\parallel ; E) z} %%
            + f(\mathbf{k}_\parallel ; E) \left[ \begin{matrix} 0 \\ u(E) \mathrm{e}^{\mathrm{i} \phi_\mathrm{S}} \\ 0 \\ v(E) \end{matrix} \right] \mathrm{e}^{\mathrm{i} q_{z,\mathrm{e}}(\mathbf{k}_\parallel ; E) z} %%
            \\
            + g(\mathbf{k}_\parallel ; E) \left[ \begin{matrix} v(E) \mathrm{e}^{\mathrm{i} \phi_\mathrm{S}} \\ 0 \\ u(E) \\ 0 \end{matrix} \right] \mathrm{e}^{-\mathrm{i} q_{z,\mathrm{h}}(\mathbf{k}_\parallel ; E) z} %%
            + h(\mathbf{k}_\parallel ; E) \left[ \begin{matrix} 0 \\ v(E) \mathrm{e}^{\mathrm{i} \phi_\mathrm{S}} \\ 0 \\ u(E) \end{matrix} \right] \mathrm{e}^{-\mathrm{i} q_{z,\mathrm{h}}(\mathbf{k}_\parallel ; E) z}
            .
        \end{multline}
        \end{widetext}
    Requiring these states to satisfy the boundary~conditions in~Eqs.~\eqref{EqBoundary1} and~\eqref{EqBoundary2} yields a homogeneous system of equations, whose nontrivial solutions correspond to the bound~state~energies, $ E=E_\mathrm{B} $, we are looking for. Owing to the BdG~Hamiltonian's fundamental time-reversal~(electron--hole) symmetry, each of those states comes along with a second one located at energy~$ -E_\mathrm{B} $.
    
    After we identified \emph{all} bound~state~energies, we need to determine the unknown coefficients that appear in the bound~state wave~function ansatz. All those coefficients depend, in~general, on the transverse wave~vector, $ \mathbf{k}_\parallel $, and on the previously computed bound~state~energies, $ E=E_\mathrm{B} $. Properly normalizing the bound~state wave~functions according to
    \begin{equation}
        \int_{-\infty}^\infty \mathrm{d}z \, \big| \psi(z ; \mathbf{k}_\parallel ; E_\mathrm{B}) \big|^2 = 1
    \end{equation}
    leads to an equation which contains the (known) coherence~factors and wave~vectors, as well as the (unknown) absolute squares of all eight wave~function~coefficients. Making use of the boundary~conditions in~Eqs.~\eqref{EqBoundary1} and~\eqref{EqBoundary2} for another time, we can consecutively express seven coefficients in terms of the remaining eighth one and finally immediately invert the equation resulting from the wave~function normalization~condition to attain this coefficient. Afterwards, we go back with the same set of equations and determine all other coefficients. The obtained analytical expressions are rather cumbersome and can be found in the~SM~\cite{Note2}. 
    
    Inside our junction's F-I~layer~(i.e., at~$ z=0 $), all electrical~current is carried by \emph{single} particles that occupy the available bound~states. At a given temperature~$ T $, each \emph{occupied~state} of energy~$ E_\mathrm{B} $ contributes \emph{on average} an amount of
    \begin{multline}
        j_\eta (\mathbf{k}_\parallel ; E_\mathrm{B}) = \lim_{z \to 0_+} \Bigg\{ \bigg\langle \psi(z>0 ; \mathbf{k}_\parallel ; E_\mathrm{B}) \mathrm{e}^{\mathrm{i} (\mathbf{k}_\parallel \cdot \mathbf{r}_\parallel)} \bigg| \hat{j}_\eta \bigg| \\ %%
        \psi(z>0 ; \mathbf{k}_\parallel ; E_\mathrm{B}) \mathrm{e}^{\mathrm{i} (\mathbf{k}_\parallel \cdot \mathbf{r}_\parallel)} \bigg\rangle \tanh \left( \frac{E_\mathrm{B}}{2k_\mathrm{B}T} \right) \Bigg\}
        \label{EqCurrentDensityBoundStates}
    \end{multline}
    to the electrical current~density along the~$ \hat{\eta} $-direction~($ \hat{\eta} \in \{ \hat{x} ; \hat{y} \} $), with
    \begin{equation}
        \hat{j}_\eta = -e \left[ \begin{matrix} -\im \frac{\hbar}{m} \frac{\partial}{\partial \eta} & 0 & 0 & 0 \\ 0 & -\im \frac{\hbar}{m} \frac{\partial}{\partial \eta} & 0 & 0 \\ 0 & 0 & -\im \frac{\hbar}{m} \frac{\partial}{\partial \eta} & 0 \\ 0 & 0 & 0 & -\im \frac{\hbar}{m} \frac{\partial}{\partial \eta} \end{matrix} \right]
    \end{equation}
    corresponding to the respective electron~current~density~operator. As before, $ e $ represents the (positive) elementary~charge and $ k_\mathrm{B} $ stands for Boltzmann's~constant. Substituting the previously given bound~state wave~function~ansatz and evaluating~Eq.~\eqref{EqCurrentDensityBoundStates} provides an alternative way to derive the AJHE~current~components directly from the junction's bound~state~spectrum. After averaging over all transverse channels and the distinct bound~state~branches~(ABS and YSR~states), we eventually arrive at

    \begin{align}
        I_\eta &= -e \sum_{E_\mathrm{B}} \frac{|\Delta_\mathrm{S}(0)| \tanh \left( 1.74 \sqrt{T_\mathrm{C}/T-1} \right)}{2E_\mathrm{B}} \nonumber \\
        &\hspace{10 pt} \times \frac{A}{(2\pi)^2} \int \mathrm{d}^2 \mathbf{k}_\parallel \, \frac{\hbar k_\eta}{m} \left[ \big| e(\mathbf{k}_\parallel ; E_\mathrm{B}) \big|^2 + \big| f(\mathbf{k}_\parallel ; E_\mathrm{B}) \big|^2 \right. \nonumber \\
        &\hspace{20 pt} \left. + \big| g(\mathbf{k}_\parallel ; E_\mathrm{B}) \big|^2 + \big| h(\mathbf{k}_\parallel ; E_\mathrm{B}) \big|^2 \right] \times \tanh \left( \frac{E_\mathrm{B}}{2k_\mathrm{B}T} \right) ;
        \label{EqCurrentBSAppendix}
    \end{align}
    note that we approximated the Hall~contact~area again by the tunneling~contact~area, $ A $, and relied on the Bardeen-Cooper-Schrieffer--type scaling of the superconducting energy~gap. We stated this current~formula as~Eq.~\eqref{EqCurrentBoundStates} in~Sec.~\ref{SecBoundStates}. All ingredients required to evaluate the current, i.e., the bound~state~energies and the absolute~squares of the wave~function~coefficients, can be extracted from the previously outlined methodology. The bound~state~approach allows us to individually resolve the current~contributions stemming from ABS and YSR~states, as discussed when analyzing the results presented in~Fig.~\ref{FigAnisotropy}.

    %% Transverse spin current formulas %%
    \section{Transverse spin~current formulas
                \label{AppC}}
    
    In Sec.~\ref{SecJSCurrents}, we study the transverse (interfacial) $ \hat{\sigma}_z $-spin~(super)currents, $ I_{\eta,\hat{z}}^\mathrm{s} $, resulting from the skew~tunneling of triplet Cooper~pairs through the F-I~barrier. Inspecting the generic~form of the scattering~states inside the superconductors~[see, e.g., Eqs.~\eqref{EqStates1} and~\eqref{EqStates2}] suggests that the $ \hat{\sigma}_x $- and $ \hat{\sigma}_y $-spin~current projections must simultaneously vanish. 
    
    Simply speaking, we can obtain~$ I_{\eta,\hat{z}}^\mathrm{s} $ from the AJHE~charge~current Furusaki--Tsukada~formula in~Eq.~\eqref{EqCurrentFurusakiTsukadaAppendix} by replacing the electron~charge, $ -e $, in the equation's prefactor by~$ \hbar / (2e) $, and weighting all individual quasiparticle scattering~processes with proper signs depending on the quasiparticles' (transverse) propagation directions and their spins. Recall that we are calculating \emph{particle} spin~currents, which count up-spin and down-spin~particles' contributions with opposite signs, but do \emph{not} additionally differentiate between electrons' and holes' different charge. To give one example, let us consider the AR~coefficient in case of an incident up-spin electronlike~quasiparticle, $ \mathcal{C}^{(1)} $~[see~Eq.~\eqref{EqStates1}]. Although the retro-reflected hole has still the same spin~(as the incoming electron), it moves along the opposite transverse~direction and counts therefore \emph{negatively} to the \emph{particle} spin~current. In the same manner, we consistently identify the signs belonging to the spin~current~contributions caused by the remaining scattering~processes and end up with the extended Furusaki--Tsukada~spin~current~formula~\cite{Asano2005}
    \begin{align}
        I_{\eta,\hat{z}}^\mathrm{s} &\approx \frac{k_\mathrm{B} T}{4} |\Delta_\mathrm{S}(0)| \tanh\left( 1.74 \sqrt{\frac{T_\mathrm{C}}{T}-1} \right) \nonumber \\
        &\hspace{0 pt} \times \frac{A}{(2\pi)^2} \int \mathrm{d}^2 \mathbf{k}_\parallel \sum_{\omega_n} \frac{k_\eta}{\sqrt{q_\mathrm{F}^2 - \mathbf{k}_\parallel^2}} \nonumber \\
        &\hspace{0 pt} \times \left[ \frac{\mathcal{C}^{(1)}(\im \omega_n) - \mathcal{D}^{(2)}(\im \omega_n) - \mathcal{A}^{(3)}(\im \omega_n) + \mathcal{B}^{(4)}(\im \omega_n)}{\sqrt{\omega_n^2 + |\Delta_\mathrm{S}(0)|^2 \tanh^2 \left( 1.74 \sqrt{T_\mathrm{C}/T - 1} \right)}} \right] .
    \end{align}
    Alternatively, we could extract~$ I_{\eta,\hat{z}}^\mathrm{s} $ from the bound~state AJHE~current~formula in~Eq.~\eqref{EqCurrentBSAppendix}. Replacing the electron~charge, $ -e $, by~$ \hbar / (2e) $, and recognizing that the up-spin~(down-spin)~electronlike~parts, scaling with~$ |e(\mathbf{k}_\parallel ; E_\mathrm{B})|^2 $~[$ |f(\mathbf{k}_\parallel ; E_\mathrm{B})|^2 $], must enter the spin~current with a positive~(negative)~sign, and vice~versa for the holelike~parts~[$ |g(\mathbf{k}_\parallel ; E_\mathrm{B})|^2 $ and~$ |h(\mathbf{k}_\parallel ; E_\mathrm{B})|^2 $], which describe states that effectively propagate along the opposite transverse~directions, we obtain
    \begin{align}
        I_{\eta,\hat{z}}^\mathrm{s} &= \frac{\hbar}{2} \sum_{E_\mathrm{B}} \frac{|\Delta_\mathrm{S}(0)| \tanh \left( 1.74 \sqrt{T_\mathrm{C}/T-1} \right)}{2E_\mathrm{B}} \nonumber \\
        &\hspace{10 pt} \times \frac{A}{(2\pi)^2} \int \mathrm{d}^2 \mathbf{k}_\parallel \, \frac{\hbar k_\eta}{m} \left[ \big| e(\mathbf{k}_\parallel ; E_\mathrm{B}) \big|^2 - \big| f(\mathbf{k}_\parallel ; E_\mathrm{B}) \big|^2 \right. \nonumber \\
        &\hspace{20 pt} \left. - \big| g(\mathbf{k}_\parallel ; E_\mathrm{B}) \big|^2 + \big| h(\mathbf{k}_\parallel ; E_\mathrm{B}) \big|^2 \right] \times \tanh \left( \frac{E_\mathrm{B}}{2k_\mathrm{B}T} \right) ;
    \end{align}
    The two equivalent spin~current~formulas were given as~Eqs.~\eqref{EqSpinCurrentFurusakiTsukada} and~\eqref{EqSpinCurrentBoundStates} in~Sec.~\ref{SecJSCurrents}.

    \bibliography{paper}

    \onecolumngrid
    \newpage
    

\section*{Supplementary material}

# SUPPLEMENTAL MATERIAL

## Anomalous Josephson Hall effect charge and transverse spin currents in superconductor/ferromagnetic insulator/superconductor junctions

Andreas Costa[1, *] and Jaroslav Fabian[1]

[1]*Institute for Theoretical Physics, University of Regensburg, 93040 Regensburg, Germany*

In this Supplemental Material, we present the technical details not included into the manuscript, and additional analyses of the bound state spectra of our system, which may be particularly helpful to follow the connections drawn in Sec. V of our manuscript. If not otherwise stated, we use the abbreviations declared in the manuscript.

## I. SYSTEM PARAMETERS

The strengths of *scalar* and *magnetic* tunneling through the F-I barrier of our system, as well as the interfacial SOC, are classified by the dimensionless parameters summarized in Tab. S1. The *superconducting phase difference* and the *in-plane magnetization angle* are tuned from zero to $2\pi$ in our calculations to effectively obtain the current-phase and current-magnetization direction relations, respectively.

TABLE S1. Dimensionless system parameters; $q_{\mathrm{F}} = \sqrt{2m\mu}/\hbar$ is the Fermi wave vector.

| | |
|---|---|
| $\bar{\lambda}_{\mathrm{SC}} = \dfrac{2m\lambda_{\mathrm{SC}}}{\hbar^2 q_{\mathrm{F}}}$ | effective *scalar* tunneling strength |
| $\bar{\lambda}_{\mathrm{MA}} = \dfrac{2m\lambda_{\mathrm{MA}}}{\hbar^2 q_{\mathrm{F}}}$ | effective *magnetic* tunneling strength |
| $\bar{\lambda}_{\mathrm{R}} = \dfrac{2m\alpha}{\hbar^2}$ | effective *Rashba SOC* strength |
| $\bar{\lambda}_{\mathrm{D}} = \dfrac{2m\beta}{\hbar^2}$ | effective *linearized Dresselhaus SOC* strength |
| $\phi_{\mathrm{S}} \in [0; 2\pi]$ | *superconducting phase difference* |
| $\Phi \in [0; 2\pi]$ | *in-plane magnetization angle* in the F-I |

To stress that the parameters, assumed for all discussed calculations, refer to realistic configurations, we want to give some concrete examples. The scalar tunneling strength $\bar{\lambda}_{\mathrm{SC}} = 1$ would, for instance, correspond to a barrier height of $0.75\,\mathrm{eV}$ and a barrier width of about $0.40\,\mathrm{nm}$ (assuming $q_{\mathrm{F}} \approx 8 \times 10^7\,\mathrm{cm}^{-1}$ [S1] as a realistic value for metals' Fermi wave vector). The linearized Dresselhaus SOC strength for thin tunneling barriers is approximated by $\beta \approx \bar{\lambda}_{\mathrm{SC}} q_{\mathrm{F}} \gamma$ [S2, S3], with $\gamma$ being the material's cubic Dresselhaus SOC parameter. Specifically for our example, $\bar{\lambda}_{\mathrm{D}} = 0.5$ suggests $\beta \approx 1.9\,\mathrm{eV}\,\mathring{\mathrm{A}}^2$ (an AlP barrier would have $\beta \approx 1.7\,\mathrm{eV}\,\mathring{\mathrm{A}}^2$ [S2]). Effective Rashba SOC strengths of $\bar{\lambda}_{\mathrm{R}} = 0.5, 1.0, 2.0$, and $4.0$ resemble the Rashba SOC parameters $\alpha \approx 1.9\,\mathrm{eV}\,\mathring{\mathrm{A}}^2$, $\alpha \approx 3.8\,\mathrm{eV}\,\mathring{\mathrm{A}}^2$, $\alpha \approx 7.6\,\mathrm{eV}\,\mathring{\mathrm{A}}^2$, and $\alpha \approx 15.2\,\mathrm{eV}\,\mathring{\mathrm{A}}^2$; those values lie all well within the experimentally accessible regime [S3, S4]. Moreover, the regarded magnetic tunneling strengths, $\bar{\lambda}_{\mathrm{MA}} \approx 10^{-3}$, refer to ferromagnetic exchange gaps in the meV-range, which are much smaller than typical exchange gaps in Fs (typically about $1\,\mathrm{eV}$). Therefore, it is justified to term our system *weakly ferromagnetic*. We limit our considerations to thus small values of $\bar{\lambda}_{\mathrm{MA}}$ since we want to demonstrate that weak ferromagnetism is already sufficient to create *sizable AJHE charge and transverse spin currents*. This comes along with the great advantage that weak ferromagnetism can efficiently coexist with superconductivity in one single junction, without a dramatic suppression of the system's superconducting properties.

---

* E-Mail: andreas.costa@physik.uni-regensburg.de

## II. GENERALIZED FURUSAKI–TSUKADA METHOD

As introduced in the manuscript, we consider a three-dimensional Josephson junction composed of two semi-infinite S regions, spanning $z < 0$ and $z > 0$ half-spaces. Both electrodes are separated by an ultrathin F-I tunneling barrier, which simultaneously breaks space inversion symmetry and lets interfacial Rashba [S5] and Dresselhaus [S6] SOC [S2, S7] emerge. We include that layer into our model in terms of a deltalike barrier [S8–S13], containing scalar and magnetic tunneling potentials, as well as both types of interfacial SOC. In electron–hole Nambu space, formed by the basis set $\Psi = [\psi^\uparrow, \psi^\downarrow, (\psi^\downarrow)^\dagger, (-\psi^\uparrow)^\dagger]^\top$, the Josephson junction's stationary BdG equation [S14], describing quasiparticle excitations, takes the form

$$\begin{bmatrix} \hat{\mathcal{H}}_e & \hat{\Delta}_S(z) \\ \hat{\Delta}_S^\dagger(z) & \hat{\mathcal{H}}_h \end{bmatrix} \Psi(\mathbf{r}) = E\,\Psi(\mathbf{r}). \tag{S1}$$

The single-particle Hamiltonians for electrons and holes, $\hat{\mathcal{H}}_e$ and $\hat{\mathcal{H}}_h$, as well as the ($s$-wave) superconducting pairing potential, $\hat{\Delta}_S(z)$, got defined in the manuscript. Assuming translational invariance (scattering of particles only happens along the $\hat{z}$-direction), the transverse wave vector (parallel to the F-I interface), $\mathbf{k}_\parallel = [k_x, k_y, 0]^\top$, needs to be conserved. Therefore, we substitute the general ansatz

$$\Psi(\mathbf{r}) = \psi(z)\mathrm{e}^{\mathrm{i}(\mathbf{k}_\parallel \cdot \mathbf{r}_\parallel)}, \tag{S2}$$

with $\mathbf{r}_\parallel = [x, y, 0]^\top$ being the vector of transverse spatial coordinates, into the BdG equation, Eq. (S1), to effectively reduce the problem to finding the one-dimensional scattering states along $\hat{z}$, $\psi(z)$, once in the left and once in the right S.

Generally speaking, we need to distinguish between *four* possible scenarios of incident quasiparticles from the left S that are scattered at the interface: (1) an incoming up-spin electronlike quasiparticle, (2) an incoming down-spin electronlike quasiparticle, (3) an incoming up-spin holelike quasiparticle, and (4) an incoming down-spin holelike quasiparticle. The latter two processes are necessary to account for the possibility of incoming electronlike quasiparticles from the right S. To compute the total current flow in the end, we need to determine the imbalance between the electronlike quasiparticles moving to the right and those moving to the left, and thus cannot simply neglect electronlike quasiparticles incident from the right S. The most general solution for the ($\hat{z}$-projected) scattering states in the left S ($z < 0$), accounting for all mentioned situations, reads then

$$\psi^{(i)}(z < 0) = \psi^{(i)}_{\text{incoming}}(z < 0) + \mathcal{A}^{(i)} \begin{bmatrix} u \\ 0 \\ 0 \\ v \end{bmatrix} \mathrm{e}^{-\mathrm{i}q_{z,e}z} + \mathcal{B}^{(i)} \begin{bmatrix} 0 \\ u \\ v \\ 0 \end{bmatrix} \mathrm{e}^{-\mathrm{i}q_{z,e}z} + C^{(i)} \begin{bmatrix} v \\ 0 \\ u \\ 0 \end{bmatrix} \mathrm{e}^{\mathrm{i}q_{z,h}z} + \mathcal{D}^{(i)} \begin{bmatrix} 0 \\ v \\ 0 \\ u \end{bmatrix} \mathrm{e}^{\mathrm{i}q_{z,h}z}, \tag{S3}$$

while we obtain

$$\psi^{(i)}(z > 0) = \mathcal{E}^{(i)} \begin{bmatrix} u\mathrm{e}^{\mathrm{i}\phi_S} \\ 0 \\ v \\ 0 \end{bmatrix} \mathrm{e}^{\mathrm{i}q_{z,e}z} + \mathcal{F}^{(i)} \begin{bmatrix} 0 \\ u\mathrm{e}^{\mathrm{i}\phi_S} \\ 0 \\ v \end{bmatrix} \mathrm{e}^{\mathrm{i}q_{z,e}z} + \mathcal{G}^{(i)} \begin{bmatrix} v\mathrm{e}^{\mathrm{i}\phi_S} \\ 0 \\ u \\ 0 \end{bmatrix} \mathrm{e}^{-\mathrm{i}q_{z,h}z} + \mathcal{H}^{(i)} \begin{bmatrix} 0 \\ v\mathrm{e}^{\mathrm{i}\phi_S} \\ 0 \\ u \end{bmatrix} \mathrm{e}^{-\mathrm{i}q_{z,h}z} \tag{S4}$$

in the right S ($z > 0$). The superscript $i$, with $i \in \{1; 2; 3; 4\}$, refers to the four possible quasiparticle injection scenarios ordered in the same way as stated above; the related scattering states differ in the incoming waves and the respective scattering coefficients. The incoming waves are given by

$$\psi^{(1)}_{\text{incoming}}(z < 0) = \begin{bmatrix} u \\ 0 \\ v \\ 0 \end{bmatrix} \mathrm{e}^{\mathrm{i}q_{z,e}z} \tag{S5}$$

for an incident up-spin electronlike quasiparticle,

$$\psi^{(2)}_{\text{incoming}}(z < 0) = \begin{bmatrix} 0 \\ u \\ 0 \\ v \end{bmatrix} \mathrm{e}^{\mathrm{i}q_{z,e}z} \tag{S6}$$

for an incident down-spin electronlike quasiparticle,

$$\psi^{(3)}_{\text{incoming}}(z < 0) = \begin{bmatrix} v \\ 0 \\ u \\ 0 \end{bmatrix} \mathrm{e}^{-\mathrm{i}q_{z,h}z} \tag{S7}$$

for an incident up-spin holelike quasiparticle, and likewise

$$\psi^{(4)}_{\text{incoming}}(z < 0) = \begin{bmatrix} 0 \\ v \\ 0 \\ u \end{bmatrix} e^{-iq_{z,h}z} \tag{S8}$$

for an incident down-spin holelike quasiparticle. The physical meaning of the scattering coefficients can be unraveled if we consider, for instance, an incoming up-spin electronlike quasiparticle [process (1)]. That quasiparticle can either be reflected back as an electronlike quasiparticle, which we call *SR*, or undergo *AR*, getting basically reflected as a holelike quasiparticle. The related *spin-resolved* coefficients (owing to the SOC, each process can either be accompanied by a spin flip or not) are denoted by $\mathcal{A}^{(1)}$ and $\mathcal{B}^{(1)}$, as well as by $C^{(1)}$ and $\mathcal{D}^{(1)}$. Analogously, $\mathcal{E}^{(1)}$, $\mathcal{F}^{(1)}$, $\mathcal{G}^{(1)}$, and $\mathcal{H}^{(1)}$ indicate electronlike and holelike transmissions into the right S. For the three remaining injection processes, the coefficients (which are of course different as we explicitly emphasized by the differing superscripts) can be interpreted in a similar way.

The $\hat{z}$-projections of the electronlike and holelike quasiparticles' wave vectors are given by

$$q_{z,e} = q_{z,e}(\mathbf{k}_\parallel; E) = \sqrt{\frac{2m}{\hbar^2}\left[\mu + \sqrt{E^2 - |\Delta_S|^2}\right] - \mathbf{k}_\parallel^2} \quad \text{and} \quad q_{z,h} = q_{z,h}(\mathbf{k}_\parallel; E) = \sqrt{\frac{2m}{\hbar^2}\left[\mu - \sqrt{E^2 - |\Delta_S|^2}\right] - \mathbf{k}_\parallel^2}. \tag{S9}$$

The quasiparticle excitation energies, $E$, as well as the superconducting energy gap, $|\Delta_S|$, are both typically much smaller than the chemical potential, $\mu$, i.e., $\mu \ll E$ and $|\Delta_S| \ll \mu$. Therefore, one may use the commonly approximated wave vectors, $q_{z,e} \approx q_{z,h} \approx q_z \equiv \sqrt{q_F^2 - \mathbf{k}_\parallel^2}$ to simplify the further theoretical treatment; $q_F = \sqrt{2m\mu}/\hbar$ is the Fermi wave vector. The factors $u = u(E)$ and $v = v(E)$ are the usual Bardeen–Cooper–Schrieffer coherence factors, satisfying

$$u(E) = \sqrt{\frac{1}{2}\left(1 + \sqrt{1 - \frac{|\Delta_S|^2}{E^2}}\right)} = \sqrt{1 - v^2(E)}. \tag{S10}$$

To attain the unknown scattering coefficients, the states are required to fulfill the interfacial ($z = 0$) boundary conditions

$$\psi(z)\big|_{z=0_-} = \psi(z)\big|_{z=0_+}, \tag{S11}$$

as well as

$$\left\{\left[\frac{\hbar^2}{2m}\frac{\mathrm{d}}{\mathrm{d}z} + \lambda_{\text{SC}}\right]\boldsymbol{\eta} + \lambda_{\text{MA}}\boldsymbol{\omega}\right\}\psi(z)\big|_{z=0_-} + \begin{bmatrix} \boldsymbol{\Omega}\cdot\hat{\boldsymbol{\sigma}} & \mathbf{0} \\ \mathbf{0} & -(\boldsymbol{\Omega}\cdot\hat{\boldsymbol{\sigma}}) \end{bmatrix}\psi(z)\big|_{z=0_-} = \frac{\hbar^2}{2m}\frac{\mathrm{d}}{\mathrm{d}z}\boldsymbol{\eta}\psi(z)\big|_{z=0_+}, \tag{S12}$$

where

$$\boldsymbol{\eta} = \begin{bmatrix} 1 & 0 & 0 & 0 \\ 0 & 1 & 0 & 0 \\ 0 & 0 & -1 & 0 \\ 0 & 0 & 0 & -1 \end{bmatrix} \quad \text{and} \quad \boldsymbol{\omega} = \begin{bmatrix} 0 & e^{-i\Phi} & 0 & 0 \\ e^{i\Phi} & 0 & 0 & 0 \\ 0 & 0 & 0 & e^{-i\Phi} \\ 0 & 0 & e^{i\Phi} & 0 \end{bmatrix}; \tag{S13}$$

$\boldsymbol{\Omega} = [(\alpha - \beta)k_y, -(\alpha + \beta)k_x, 0]$ contains the Rashba and Dresselhaus SOC. For each of the four quasiparticle injection processes, Eqs. (S11)–(S12) represent a system of eight linear equations for eight unknown scattering coefficients. Due to the equations' complexity, we do not give analytical expressions for the coefficients, but rather solve for them numerically.

Once all scattering coefficients are determined, the transverse *AJHE current* components, $I_\eta$ ($\eta \in \{x; y\}$), can be evaluated from an extended Green's function-based [S15] Furusaki–Tsukada method [S16]. Close to the barrier (i.e., at $z = 0$), one obtains

$$I_\eta \approx \frac{ek_BT}{2\hbar}|\Delta_S|\frac{A_\eta}{(2\pi)^2}\int \mathrm{d}^2\mathbf{k}_\parallel \sum_{\omega_n} \frac{k_\eta}{\sqrt{q_F^2 - \mathbf{k}_\parallel^2}}\left[\frac{C^{(1)}(i\omega_n) + \mathcal{D}^{(2)}(i\omega_n) + \mathcal{A}^{(3)}(i\omega_n) + \mathcal{B}^{(4)}(i\omega_n)}{\sqrt{\omega_n^2 + |\Delta_S|^2}}\right]; \tag{S14}$$

$e$ is the (positive) elementary charge, $k_B$ stands for Boltzmann's constant, and $\omega_n = (2n + 1)\pi k_BT$, with integer $n$, represents the fermionic Matsubara frequencies (at temperature $T$ and given in units of $1/\hbar$). The cross-sectional area through which the current $I_\eta$ flows is denoted by $A_\eta$. For simplicity, we assume that the tunneling and Hall contact areas are equal, i.e., $A_x = A_y = A_z \equiv A$. The AR coefficients entering Eq. (S14) are obtained by applying the boundary conditions given by Eqs. (S11)–(S12) to the generic scattering states, Eqs. (S3)–(S4), (numerically) solving the resulting system of equations, and analytically continuing the excitation energies, $E \to i\omega_n$. Within our calculations, we can then treat the tunneling strengths,

$\overline{\lambda}_{SC}$ and $\overline{\lambda}_{MA}$, the SOC strengths, $\overline{\lambda}_{R}$ and $\overline{\lambda}_{D}$, the superconducting phase difference, $\phi_S$, as well as the magnetization angle in the F-I, $\Phi$, as tunable parameters. For the AJHE currents presented in Fig. 3 of our manuscript, we numerically extracted the (analytically continued) AR coefficients and substituted them into Eq. (S14); the summations over $\mathbf{k}_\parallel$ and $\omega_n$ were also performed fully numerically. If we compute the currents at *finite temperature*, we additionally need to account for the correct temperature dependence of the superconducting energy gap. We employed the usual Bardeen–Cooper–Schrieffer-like scaling for that, $|\Delta_S(T \neq 0)| = |\Delta_S(0)| \tanh(1.74 \sqrt{T_C/T - 1})$, with $T_C$ being the superconductors' critical temperature. For our concrete examples in the manuscript, we used $|\Delta_S(0)| \approx 2.5$ meV and $T_C \approx 16$ K as representative values for $s$-wave superconductors [S17]. The AJHE currents in the manuscript got normalized according to $(I_\eta e)/[G_S \pi |\Delta_S(0)|]$, where $G_S = (Ae^2 q_F^2)/(4\pi^2\hbar)$ is Sharvin's conductance of a three-dimensional point contact with interfacial cross-section area $A$.

## III. SKEW AR—EFFECTIVE BARRIER MODEL

We mentioned in the manuscript that the dominant contributions to the AJHE currents originate from the singlet channel [S13], i.e., from spin-conserving skew ARs of incident electronlike quasiparticles. Owing to the interfacial SOC, these quasiparticles are not only exposed to the usual interfacial scalar and magnetic potential (containing scalar and magnetic tunneling terms), but also to an additional transverse momentum- and spin-dependent potential; see Eq. (3) in the manuscript for the limiting case of $\beta = 0$, $k_x = 0$, and $\Phi = 0$. Following Eq. (S14), each quasiparticle undergoing skew AR contributes to the AJHE currents proportionally to the product of its transverse velocity along $\tilde{\eta}$ ($\sim k_\eta$) and the corresponding AR coefficient. The latter can be extracted from a simplified BdG scattering description, assuming that the interfacial SOC is weak and spin-flip scattering gets negligible. The Nambu scattering states [see Eqs. (S3)–(S4)] are then decomposed into their spin-resolved blocks [$\sigma = +(-)1$ for spin up (spin down)],

$$\psi^\sigma(z < 0) = \begin{bmatrix} u \\ v \end{bmatrix} e^{iq_{z,e}z} + \mathcal{A}^\sigma \begin{bmatrix} u \\ v \end{bmatrix} e^{-iq_{z,e}z} + C^\sigma \begin{bmatrix} v \\ u \end{bmatrix} e^{iq_{z,h}z} \tag{S15}$$

and

$$\psi^\sigma(z > 0) = \mathcal{E}^\sigma \begin{bmatrix} u e^{i\phi_S} \\ v \end{bmatrix} e^{iq_{z,e}z} + \mathcal{G}^\sigma \begin{bmatrix} v e^{i\phi_S} \\ u \end{bmatrix} e^{-iq_{z,h}z}; \tag{S16}$$

$\mathcal{A}^\sigma$ and $C^\sigma$ are the SR and AR coefficients, while $\mathcal{E}^\sigma$ and $\mathcal{G}^\sigma$ refer to transmissions. For simplicity, we just deal with an incoming electronlike quasiparticle at the moment. The states need to fulfill the interfacial ($z = 0$) boundary conditions

$$\psi^\sigma(z)\big|_{z=0_-} = \psi^\sigma(z)\big|_{z=0_+}, \tag{S17}$$

as well as

$$\left[\frac{\hbar^2}{2m}\frac{\mathrm{d}}{\mathrm{d}z} + V^\sigma_{\mathrm{eff}}\right]\psi^\sigma(z)\big|_{z=0_-} = \frac{\hbar^2}{2m}\frac{\mathrm{d}}{\mathrm{d}z}\psi^\sigma(z)\big|_{z=0_+}, \tag{S18}$$

where $V^\sigma_{\mathrm{eff}} = \lambda_{SC} + \sigma\lambda_{MA} + \sigma\alpha k_y$ is the effective interfacial scattering potential [see Eq. (3) in the manuscript]. To give simple analytical results, we focus on zero energy [as $u(E = 0) = 1/\sqrt{2}$ and $v(E = 0) = -\mathrm{i}/\sqrt{2}$], $k_y = \pm q_F/2$, and keep the superconducting phase difference at $\phi_S = \pi/2$. The AR coefficient's amplitude reads then

$$|C^\sigma| = \frac{3}{3 + 2\left(Z^\sigma_{\mathrm{eff}}\right)^2}, \tag{S19}$$

with the dimensionless scattering potential parameter $Z = (2mV^\sigma_{\mathrm{eff}})/(\hbar^2 q_F) = \overline{\lambda}_{SC} + \sigma\overline{\lambda}_{MA} \pm \sigma\overline{\lambda}_R/2$. In Fig. 2 of the manuscript, we show $|C^\sigma|$ as a function of $Z^\sigma_{\mathrm{eff}}$ and exemplarily illustrate the skew AR process for $\overline{\lambda}_{SC} = 2$, $\overline{\lambda}_{MA} = 0.25$, and $\overline{\lambda}_R = 1$. Incoming up-spin electronlike quasiparticles are then predominantly Andreev reflected at $k_y < 0$ (referring to $-\sigma\overline{\lambda}_R/2|_{\sigma=1}$) and their down-spin counterparts at $k_y > 0$ (referring to $+\sigma\overline{\lambda}_R/2|_{\sigma=-1}$), generating AJHE currents along $\hat{y}$ and $-\hat{y}$, respectively. Due to the finite $\overline{\lambda}_{MA}$, the second process happens more likely and an effective small AJHE current along $-\hat{y}$ builds up. As we discussed in the manuscript, incoming holelike quasiparticles model electronlike quasiparticles impinging on the F-I from the right S and contribute to the overall AJHE currents in the same way.

## IV. BOUND STATE TECHNIQUE

A characteristic spectroscopic fingerprint of Josephson junctions is the formation of subgap bound states, well-localized around the weak link connecting the two superconducting electrodes. Particularly in the additional presence of ferromagnetic components, like the F-I layer in our junction, those states split into two branches [S12, S18]: a conventional *ABS branch* [S19, S20] and a *YSR branch* [S21–S24]. Within this section, we formulate an alternative theoretical framework to compute AJHE currents from the junctions' bound state spectra. The final results serve as an essential cross-check for the Green's function technique and eventually allow us to draw distinct connections between the bound state characteristics and the AJHE.

### A. Bound state energies

To extract our junction's bound state spectrum from the BdG equation, Eq. (S1), we construct the general ansatz for $\psi(z)$ *without considering an incoming wave*. For *positive* energies, $E > 0$, the ansatz in the left S ($z < 0$) reads

$$\psi(z < 0; \mathbf{k}_\parallel; E) = a(\mathbf{k}_\parallel; E) \begin{bmatrix} u(E) \\ 0 \\ v(E) \\ 0 \end{bmatrix} e^{-iq_{z,e}(\mathbf{k}_\parallel; E)z} + b(\mathbf{k}_\parallel; E) \begin{bmatrix} 0 \\ u(E) \\ 0 \\ v(E) \end{bmatrix} e^{-iq_{z,e}(\mathbf{k}_\parallel; E)z}$$

$$+ c(\mathbf{k}_\parallel; E) \begin{bmatrix} v(E) \\ 0 \\ u(E) \\ 0 \end{bmatrix} e^{iq_{z,h}(\mathbf{k}_\parallel; E)z} + d(\mathbf{k}_\parallel; E) \begin{bmatrix} 0 \\ v(E) \\ 0 \\ u(E) \end{bmatrix} e^{iq_{z,h}(\mathbf{k}_\parallel; E)z} \quad \text{(S20)}$$

and the corresponding one in the right S ($z > 0$)

$$\psi(z > 0; \mathbf{k}_\parallel; E) = e(\mathbf{k}_\parallel; E) \begin{bmatrix} u(E)e^{i\phi_S} \\ 0 \\ v(E) \\ 0 \end{bmatrix} e^{iq_{z,e}(\mathbf{k}_\parallel; E)z} + f(\mathbf{k}_\parallel; E) \begin{bmatrix} 0 \\ u(E)e^{i\phi_S} \\ 0 \\ v(E) \end{bmatrix} e^{iq_{z,e}(\mathbf{k}_\parallel; E)z}$$

$$+ g(\mathbf{k}_\parallel; E) \begin{bmatrix} v(E)e^{i\phi_S} \\ 0 \\ u(E) \\ 0 \end{bmatrix} e^{-iq_{z,h}(\mathbf{k}_\parallel; E)z} + h(\mathbf{k}_\parallel; E) \begin{bmatrix} 0 \\ v(E)e^{i\phi_S} \\ 0 \\ u(E) \end{bmatrix} e^{-iq_{z,h}(\mathbf{k}_\parallel; E)z}. \quad \text{(S21)}$$

The wave vectors and coherence factors got defined in Sec. II; see Eqs. (S9)–(S10). The full wave functions for the three-dimensional problem are then

$$\Psi(x, y, z; \mathbf{k}_\parallel; E) = \psi(z; \mathbf{k}_\parallel; E) e^{i(\mathbf{k}_\parallel \cdot \mathbf{r}_\parallel)}. \quad \text{(S22)}$$

Although we are now interested in the bound state spectrum instead of the real scattering problem (thus, we also did not consider incoming waves), there is still a certain similarity between both approaches so that one might still identify the coefficients as the SR, AR, and transmission coefficients introduced in Sec. II. Formally, there would always be a second, functionally independent, solution, $\bar{\psi}(z)$, belonging to *negative* energies. This solution could be obtained from $\psi(z)$ by interchanging $u \longmapsto -v^*$ and $v \longmapsto u^*$ in the Nambu spinors. In the following, we will neglect the second (formal) solution and just keep in mind that to each positive bound state energy, $E_B$, there will always be symmetrically another bound state at energy $-E_B$. Applying the boundary conditions in Eqs. (S11)–(S12) to the states in Eqs. (S20)–(S21) results in a homogeneous eight-dimensional system of equations. To obtain a nontrivial solution (for fixed $\mathbf{k}_\parallel$), the determinant of the system's coefficient matrix must be zero, yielding a secular equation we can (numerically) solve to extract the bound state energies, $E = E_B$.

Once *all* bound state energies are determined, we can proceed and find the (still unknown) coefficients in the bound state wave functions in Eqs. (S20)–(S21). It is important to notice that all those coefficients are not only functions of the bound state energies, $E_B$, but also of the transverse wave vector, $\mathbf{k}_\parallel$ (explicitly and implicitly since the bound state energies themselves also depend on $\mathbf{k}_\parallel$). We strengthened that fact already in the ansatz for the bound state wave functions, Eqs. (S20)–(S21), by regarding the coefficients as functions of $\mathbf{k}_\parallel$ and $E = E_B$. To attain analytical expressions for the coefficients, we exploit the interfacial boundary conditions, given by Eqs. (S11) and (S12), for another time. Applying Eq. (S11), ensuring continuity of the wave function at the interface, allows us to rewrite the wave function coefficients in the region left of the F-I interface, $a(\mathbf{k}_\parallel; E_B)$, $b(\mathbf{k}_\parallel; E_B)$, $c(\mathbf{k}_\parallel; E_B)$, and $d(\mathbf{k}_\parallel; E_B)$, in terms of the remaining four coefficients, which initially appeared in the

wave function in the right half-junction. To be more specific, we obtain

$$a(\mathbf{k}_\parallel; E_B) = \varepsilon_1(E_B) \cdot e(\mathbf{k}_\parallel; E_B) + \gamma_1(E_B) \cdot g(\mathbf{k}_\parallel; E_B), \tag{S23}$$

$$b(\mathbf{k}_\parallel; E_B) = \varphi_1(E_B) \cdot f(\mathbf{k}_\parallel; E_B) + \eta_1(E_B) \cdot h(\mathbf{k}_\parallel; E_B), \tag{S24}$$

$$c(\mathbf{k}_\parallel; E_B) = \varepsilon_2(E_B) \cdot e(\mathbf{k}_\parallel; E_B) + \gamma_2(E_B) \cdot g(\mathbf{k}_\parallel; E_B), \tag{S25}$$

and lastly,

$$d(\mathbf{k}_\parallel; E_B) = \varphi_2(E_B) \cdot f(\mathbf{k}_\parallel; E_B) + \eta_2(E_B) \cdot h(\mathbf{k}_\parallel; E_B), \tag{S26}$$

where we introduced, for the sake of compactness, the "weighting factors"

$$\varepsilon_1(E_B) = \frac{e^{i\phi_S} - v^2(E_B)/u^2(E_B)}{1 - v^2(E_B)/u^2(E_B)}, \qquad\qquad \varepsilon_2(E_B) = \frac{(1 - e^{i\phi_S}) \cdot v(E_B)/u(E_B)}{1 - v^2(E_B)/u^2(E_B)}, \tag{S27}$$

$$\varphi_1(E_B) = \varepsilon_1(E_B), \qquad\qquad \varphi_2(E_B) = \varepsilon_2(E_B), \tag{S28}$$

$$\gamma_1(E_B) = \frac{(e^{i\phi_S} - 1) \cdot v(E_B)/u(E_B)}{1 - v^2(E_B)/u^2(E_B)}, \qquad\qquad \gamma_2(E_B) = \frac{1 - e^{i\phi_S} v^2(E_B)/u^2(E_B)}{1 - v^2(E_B)/u^2(E_B)}, \tag{S29}$$

$$\eta_1(E_B) = \gamma_1(E_B), \qquad\qquad \eta_2(E_B) = \gamma_2(E_B), \tag{S30}$$

defining which fraction of wave function coefficients in the right part is needed to fully express those in the left part.

In the end, all coefficients in the bound state wave function need to be chosen such that the wave function is properly normalized. Using the fact that $\mathrm{Im}[q_{z,e}(\mathbf{k}_\parallel; E_B)] = -\mathrm{Im}[q_{z,h}(\mathbf{k}_\parallel; E_B)]$, which can be directly extracted from the wave vectors in Eq. (S9) by substituting $0 \le E = E_B \le |\Delta_S|$, the normalization condition can be recast as

$$\frac{|u(E_B)|^2 + |v(E_B)|^2}{2\mathrm{Im}[q_{z,e}(\mathbf{k}_\parallel; E_B)]} \Big[ \big|a(\mathbf{k}_\parallel; E_B)\big|^2 + \big|b(\mathbf{k}_\parallel; E_B)\big|^2 + \big|c(\mathbf{k}_\parallel; E_B)\big|^2 + \big|d(\mathbf{k}_\parallel; E_B)\big|^2$$
$$+ \big|e(\mathbf{k}_\parallel; E_B)\big|^2 + \big|f(\mathbf{k}_\parallel; E_B)\big|^2 + \big|g(\mathbf{k}_\parallel; E_B)\big|^2 + \big|h(\mathbf{k}_\parallel; E_B)\big|^2 \Big] = 1. \tag{S31}$$

By means of Eqs. (S23)–(S26), together with Eqs. (S27)–(S30), we can bring the normalization condition, Eq. (S31), into a form in which only the wave function coefficients in the right half-junction need to be known. To further simplify the description, we take advantage of the second boundary condition in Eq. (S12), accounting for the jump-like discontinuity in the wave function's first derivative. Applying this condition to the bound state wave function ansatz, Eqs. (S20)–(S21), yields four equations for four unknown coefficients, after substituting Eqs. (S23)–(S26). Therefore, we can always eliminate two different coefficients from the equations and express consecutively $f(\mathbf{k}_\parallel; E_B)$, $g(\mathbf{k}_\parallel; E_B)$, and $h(\mathbf{k}_\parallel; E_B)$ solely by $e(\mathbf{k}_\parallel; E_B)$. To become concrete, we extract the relations

$$f(\mathbf{k}_\parallel; E_B) = \Sigma_1(\mathbf{k}_\parallel; E_B) \cdot e(\mathbf{k}_\parallel; E_B), \tag{S32}$$

$$g(\mathbf{k}_\parallel; E_B) = \Sigma_2(\mathbf{k}_\parallel; E_B) \cdot e(\mathbf{k}_\parallel; E_B), \tag{S33}$$

as well as

$$h(\mathbf{k}_\parallel; E_B) = \Sigma_3(\mathbf{k}_\parallel; E_B) \cdot e(\mathbf{k}_\parallel; E_B), \tag{S34}$$

with the "weighting factors"

$$\Sigma_1(\mathbf{k}_\parallel; E_B) = \Big[ \overline{\varepsilon}_3(\mathbf{k}_\parallel; E_B)\overline{\eta}_2(\mathbf{k}_\parallel; E_B)\overline{\gamma}_1(\mathbf{k}_\parallel; E_B) - \overline{\varepsilon}_2(\mathbf{k}_\parallel; E_B)\overline{\eta}_3(\mathbf{k}_\parallel; E_B)\overline{\gamma}_1(\mathbf{k}_\parallel; E_B) - \overline{\varepsilon}_3(\mathbf{k}_\parallel; E_B)\overline{\eta}_1(\mathbf{k}_\parallel; E_B)\overline{\gamma}_2(\mathbf{k}_\parallel; E_B)$$
$$+ \overline{\varepsilon}_1(\mathbf{k}_\parallel; E_B)\overline{\eta}_3(\mathbf{k}_\parallel; E_B)\overline{\gamma}_2(\mathbf{k}_\parallel; E_B) + \overline{\varepsilon}_2(\mathbf{k}_\parallel; E_B)\overline{\eta}_1(\mathbf{k}_\parallel; E_B)\overline{\gamma}_3(\mathbf{k}_\parallel; E_B) - \overline{\varepsilon}_1(\mathbf{k}_\parallel; E_B)\overline{\eta}_2(\mathbf{k}_\parallel; E_B)\overline{\gamma}_3(\mathbf{k}_\parallel; E_B) \Big] \Big/ \Lambda(\mathbf{k}_\parallel; E_B), \tag{S35}$$

$$\Sigma_2(\mathbf{k}_\parallel; E_B) = \Big[ -\overline{\varepsilon}_3(\mathbf{k}_\parallel; E_B)\overline{\eta}_2(\mathbf{k}_\parallel; E_B)\overline{\varphi}_1(\mathbf{k}_\parallel; E_B) + \overline{\varepsilon}_2(\mathbf{k}_\parallel; E_B)\overline{\eta}_3(\mathbf{k}_\parallel; E_B)\overline{\varphi}_1(\mathbf{k}_\parallel; E_B) + \overline{\varepsilon}_3(\mathbf{k}_\parallel; E_B)\overline{\eta}_1(\mathbf{k}_\parallel; E_B)\overline{\varphi}_2(\mathbf{k}_\parallel; E_B)$$
$$- \overline{\varepsilon}_1(\mathbf{k}_\parallel; E_B)\overline{\eta}_3(\mathbf{k}_\parallel; E_B)\overline{\varphi}_2(\mathbf{k}_\parallel; E_B) - \overline{\varepsilon}_2(\mathbf{k}_\parallel; E_B)\overline{\eta}_1(\mathbf{k}_\parallel; E_B)\overline{\varphi}_3(\mathbf{k}_\parallel; E_B) + \overline{\varepsilon}_1(\mathbf{k}_\parallel; E_B)\overline{\eta}_2(\mathbf{k}_\parallel; E_B)\overline{\varphi}_3(\mathbf{k}_\parallel; E_B) \Big] \Big/ \Lambda(\mathbf{k}_\parallel; E_B), \tag{S36}$$

and

$$\Sigma_3(\mathbf{k}_\parallel; E_B) = \Big[ \overline{\varepsilon}_3(\mathbf{k}_\parallel; E_B) \overline{\gamma}_2(\mathbf{k}_\parallel; E_B) \overline{\varphi}_1(\mathbf{k}_\parallel; E_B) - \overline{\varepsilon}_2(\mathbf{k}_\parallel; E_B) \overline{\gamma}_3(\mathbf{k}_\parallel; E_B) \overline{\varphi}_1(\mathbf{k}_\parallel; E_B) - \overline{\varepsilon}_3(\mathbf{k}_\parallel; E_B) \overline{\gamma}_1(\mathbf{k}_\parallel; E_B) \overline{\varphi}_2(\mathbf{k}_\parallel; E_B)$$
$$+ \overline{\varepsilon}_1(\mathbf{k}_\parallel; E_B) \overline{\gamma}_3(\mathbf{k}_\parallel; E_B) \overline{\varphi}_2(\mathbf{k}_\parallel; E_B) + \overline{\varepsilon}_2(\mathbf{k}_\parallel; E_B) \overline{\gamma}_1(\mathbf{k}_\parallel; E_B) \overline{\varphi}_3(\mathbf{k}_\parallel; E_B) - \overline{\varepsilon}_1(\mathbf{k}_\parallel; E_B) \overline{\gamma}_2(\mathbf{k}_\parallel; E_B) \overline{\varphi}_3(\mathbf{k}_\parallel; E_B) \Big] \Big/ \Lambda(\mathbf{k}_\parallel; E_B). \tag{S37}$$

The "overline" coefficients read

$$\overline{\varepsilon}_1(\mathbf{k}_\parallel; E_B) = -\frac{q_{z,e}(\mathbf{k}_\parallel; E_B)}{q_F} u(E_B) e^{i\phi_S} - \left[ \frac{q_{z,e}(\mathbf{k}_\parallel; E_B)}{q_F} + i\overline{\lambda}_{SC} \right] u(E_B) \varepsilon_1(E_B) - \left[ -\frac{q_{z,h}(\mathbf{k}_\parallel; E_B)}{q_F} + i\overline{\lambda}_{SC} \right] v(E_B) \varepsilon_2(E_B), \tag{S38}$$

$$\overline{\varphi}_1(\mathbf{k}_\parallel; E_B) = -i \left[ \overline{\lambda}_{MA} e^{-i\Phi} + \overline{\Omega}(\mathbf{k}_\parallel) \right] u(E_B) \varepsilon_1(E_B) - i \left[ \overline{\lambda}_{MA} e^{-i\Phi} + \overline{\Omega}(\mathbf{k}_\parallel) \right] v(E_B) \varepsilon_2(E_B), \tag{S39}$$

$$\overline{\gamma}_1(\mathbf{k}_\parallel; E_B) = \frac{q_{z,h}(\mathbf{k}_\parallel; E_B)}{q_F} v(E_B) e^{i\phi_S} - \left[ \frac{q_{z,e}(\mathbf{k}_\parallel; E_B)}{q_F} + i\overline{\lambda}_{SC} \right] u(E_B) \gamma_1(E_B) - \left[ -\frac{q_{z,h}(\mathbf{k}_\parallel; E_B)}{q_F} + i\overline{\lambda}_{SC} \right] v(E_B) \gamma_2(E_B), \tag{S40}$$

$$\overline{\eta}_1(\mathbf{k}_\parallel; E_B) = -i \left[ \overline{\lambda}_{MA} e^{-i\Phi} + \overline{\Omega}(\mathbf{k}_\parallel) \right] u(E_B) \gamma_1(E_B) - i \left[ \overline{\lambda}_{MA} e^{-i\Phi} + \overline{\Omega}(\mathbf{k}_\parallel) \right] v(E_B) \gamma_2(E_B), \tag{S41}$$

$$\overline{\varepsilon}_2(\mathbf{k}_\parallel; E_B) = -i \left[ \overline{\lambda}_{MA} e^{i\Phi} + \overline{\Omega}^*(\mathbf{k}_\parallel) \right] u(E_B) \varepsilon_1(E_B) - i \left[ \overline{\lambda}_{MA} e^{i\Phi} + \overline{\Omega}^*(\mathbf{k}_\parallel) \right] v(E_B) \varepsilon_2(E_B), \tag{S42}$$

$$\overline{\varphi}_2(\mathbf{k}_\parallel; E_B) = -\frac{q_{z,e}(\mathbf{k}_\parallel; E_B)}{q_F} u(E_B) e^{i\phi_S} - \left[ \frac{q_{z,e}(\mathbf{k}_\parallel; E_B)}{q_F} + i\overline{\lambda}_{SC} \right] u(E_B) \varepsilon_1(E_B) - \left[ -\frac{q_{z,h}(\mathbf{k}_\parallel; E_B)}{q_F} + i\overline{\lambda}_{SC} \right] v(E_B) \varepsilon_2(E_B), \tag{S43}$$

$$\overline{\gamma}_2(\mathbf{k}_\parallel; E_B) = -i \left[ \overline{\lambda}_{MA} e^{i\Phi} + \overline{\Omega}^*(\mathbf{k}_\parallel) \right] u(E_B) \gamma_1(E_B) - i \left[ \overline{\lambda}_{MA} e^{i\Phi} + \overline{\Omega}^*(\mathbf{k}_\parallel) \right] v(E_B) \gamma_2(E_B), \tag{S44}$$

$$\overline{\eta}_2(\mathbf{k}_\parallel; E_B) = \frac{q_{z,h}(\mathbf{k}_\parallel; E_B)}{q_F} v(E_B) e^{i\phi_S} - \left[ \frac{q_{z,e}(\mathbf{k}_\parallel; E_B)}{q_F} + i\overline{\lambda}_{SC} \right] u(E_B) \gamma_1(E_B) - \left[ -\frac{q_{z,h}(\mathbf{k}_\parallel; E_B)}{q_F} + i\overline{\lambda}_{SC} \right] v(E_B) \gamma_2(E_B), \tag{S45}$$

$$\overline{\varepsilon}_3(\mathbf{k}_\parallel; E_B) = \frac{q_{z,e}(\mathbf{k}_\parallel; E_B)}{q_F} v(E_B) + \left[ \frac{q_{z,e}(\mathbf{k}_\parallel; E_B)}{q_F} + i\overline{\lambda}_{SC} \right] v(E_B) \varepsilon_1(E_B) + \left[ -\frac{q_{z,h}(\mathbf{k}_\parallel; E_B)}{q_F} + i\overline{\lambda}_{SC} \right] u(E_B) \varepsilon_2(E_B), \tag{S46}$$

$$\overline{\varphi}_3(\mathbf{k}_\parallel; E_B) = i \left[ -\overline{\lambda}_{MA} e^{-i\Phi} + \overline{\Omega}(\mathbf{k}_\parallel) \right] v(E_B) \varepsilon_1(E_B) + i \left[ -\overline{\lambda}_{MA} e^{-i\Phi} + \overline{\Omega}(\mathbf{k}_\parallel) \right] u(E_B) \varepsilon_2(E_B), \tag{S47}$$

$$\overline{\gamma}_3(\mathbf{k}_\parallel; E_B) = -\frac{q_{z,h}(\mathbf{k}_\parallel; E_B)}{q_F} u(E_B) + \left[ \frac{q_{z,e}(\mathbf{k}_\parallel; E_B)}{q_F} + i\overline{\lambda}_{SC} \right] v(E_B) \gamma_1(E_B) + \left[ -\frac{q_{z,h}(\mathbf{k}_\parallel; E_B)}{q_F} + i\overline{\lambda}_{SC} \right] u(E_B) \gamma_2(E_B), \tag{S48}$$

$$\overline{\eta}_3(\mathbf{k}_\parallel; E_B) = i \left[ -\overline{\lambda}_{MA} e^{-i\Phi} + \overline{\Omega}(\mathbf{k}_\parallel) \right] v(E_B) \gamma_1(E_B) + i \left[ -\overline{\lambda}_{MA} e^{-i\Phi} + \overline{\Omega}(\mathbf{k}_\parallel) \right] u(E_B) \gamma_2(E_B), \tag{S49}$$

and finally,

$$\overline{\varepsilon}_4(\mathbf{k}_\parallel; E_B) = i \left[ -\overline{\lambda}_{MA} e^{i\Phi} + \overline{\Omega}^*(\mathbf{k}_\parallel) \right] v(E_B) \varepsilon_1(E_B) + i \left[ -\overline{\lambda}_{MA} e^{i\Phi} + \overline{\Omega}^*(\mathbf{k}_\parallel) \right] u(E_B) \varepsilon_2(E_B), \tag{S50}$$

$$\overline{\varphi}_4(\mathbf{k}_\parallel; E_B) = \frac{q_{z,e}(\mathbf{k}_\parallel; E_B)}{q_F} v(E_B) + \left[ \frac{q_{z,e}(\mathbf{k}_\parallel; E_B)}{q_F} + i\overline{\lambda}_{SC} \right] v(E_B) \varepsilon_1(E_B) + \left[ -\frac{q_{z,h}(\mathbf{k}_\parallel; E_B)}{q_F} + i\overline{\lambda}_{SC} \right] u(E_B) \varepsilon_2(E_B), \tag{S51}$$

$$\overline{\gamma}_4(\mathbf{k}_\parallel; E_B) = i \left[ -\overline{\lambda}_{MA} e^{i\Phi} + \overline{\Omega}^*(\mathbf{k}_\parallel) \right] v(E_B) \gamma_1(E_B) + i \left[ -\overline{\lambda}_{MA} e^{i\Phi} + \overline{\Omega}^*(\mathbf{k}_\parallel) \right] u(E_B) \gamma_2(E_B), \tag{S52}$$

as well as,

$$\overline{\eta}_4(\mathbf{k}_\parallel; E_B) = -\frac{q_{z,h}(\mathbf{k}_\parallel; E_B)}{q_F} u(E_B) + \left[ \frac{q_{z,e}(\mathbf{k}_\parallel; E_B)}{q_F} + i\overline{\lambda}_{SC} \right] v(E_B) \gamma_1(E_B) + \left[ -\frac{q_{z,h}(\mathbf{k}_\parallel; E_B)}{q_F} + i\overline{\lambda}_{SC} \right] u(E_B) \gamma_2(E_B); \tag{S53}$$

the common denominator in Eqs. (S35)–(S37) can be written as

$$\Lambda(\mathbf{k}_\parallel; E_B) = -\overline{\eta}_3(\mathbf{k}_\parallel; E_B) \overline{\gamma}_2(\mathbf{k}_\parallel; E_B) \overline{\varphi}_1(\mathbf{k}_\parallel; E_B) + \overline{\eta}_2(\mathbf{k}_\parallel; E_B) \overline{\gamma}_3(\mathbf{k}_\parallel; E_B) \overline{\varphi}_1(\mathbf{k}_\parallel; E_B) + \overline{\eta}_3(\mathbf{k}_\parallel; E_B) \overline{\gamma}_1(\mathbf{k}_\parallel; E_B) \overline{\varphi}_2(\mathbf{k}_\parallel; E_B)$$
$$- \overline{\eta}_1(\mathbf{k}_\parallel; E_B) \overline{\gamma}_3(\mathbf{k}_\parallel; E_B) \overline{\varphi}_2(\mathbf{k}_\parallel; E_B) - \overline{\eta}_2(\mathbf{k}_\parallel; E_B) \overline{\gamma}_1(\mathbf{k}_\parallel; E_B) \overline{\varphi}_3(\mathbf{k}_\parallel; E_B) + \overline{\eta}_1(\mathbf{k}_\parallel; E_B) \overline{\gamma}_2(\mathbf{k}_\parallel; E_B) \overline{\varphi}_3(\mathbf{k}_\parallel; E_B). \tag{S54}$$

In all preceding equations, we used the effective (dimensionless) parameters $\overline{\lambda}_{SC}$ and $\overline{\lambda}_{MA}$, representing the scalar and magnetic tunneling strengths, respectively; see Tab. S1 for the definition of these quantities. Moreover, the influence of SOC is captured by the dimensionless SOC "matrix element"

$$\overline{\Omega}(\mathbf{k}_\parallel) = \left( \overline{\lambda}_R - \overline{\lambda}_D \right) k_y / q_F + i \left( \overline{\lambda}_R + \overline{\lambda}_D \right) k_x / q_F, \tag{S55}$$

where the effective Rashba and Dresselhaus SOC parameters, $\overline{\lambda}_R$ and $\overline{\lambda}_D$, can again be looked up in Tab. S1.

Eventually, we can put Eqs. (S23)–(S26), as well as Eqs. (S32)–(S34) all together into the normalization condition, Eq. (S31), to end up with

$$\frac{|u(E_B)|^2 + |v(E_B)|^2}{2\,\mathrm{Im}[q_{z,e}(\mathbf{k}_\parallel; E_B)]} \Big[ |\varepsilon_1(E_B) + \gamma_1(E_B)\Sigma_2(\mathbf{k}_\parallel; E_B)|^2 + |\varepsilon_1(E_B)\Sigma_1(\mathbf{k}_\parallel; E_B) + \gamma_1(E_B)\Sigma_3(\mathbf{k}_\parallel; E_B)|^2$$
$$+ |\varepsilon_2(E_B) + \gamma_2(E_B)\Sigma_2(\mathbf{k}_\parallel; E_B)|^2 + |\varepsilon_2(E_B)\Sigma_1(\mathbf{k}_\parallel; E_B) + \gamma_2(E_B)\Sigma_3(\mathbf{k}_\parallel; E_B)|^2$$
$$+ 1 + |\Sigma_1(\mathbf{k}_\parallel; E_B)|^2 + |\Sigma_2(\mathbf{k}_\parallel; E_B)|^2 + |\Sigma_3(\mathbf{k}_\parallel; E_B)|^2 \Big] \cdot |e(\mathbf{k}_\parallel; E_B)|^2 = 1. \quad (S56)$$

Equation (S56) can be inverted to initially identify $|e(\mathbf{k}_\parallel; E_B)|^2$. All remaining (absolute squares of the) bound state wave function coefficients are afterwards uniquely determined by going back with the given equations.

### B. Current calculation

As soon as we know the specific form of the bound state wave functions (including all appearing coefficients), we are fully equipped to compute the AJHE current flows in our system. To avoid dealing with Cooper pairs in the superconductors, we evaluate the current inside the F-I region. Electrons initially forming Cooper pairs in one of the superconductors essentially tunnel through the F-I barrier via the available bound states and pair again in the second S. The effectively resulting Cooper pair exchange can generate net Josephson currents. Owing to the interfacial SOC, the transferred Cooper pairs are additionally subject to the interfacial skew tunneling, discussed in detail in our manuscript, which gives eventually rise to transverse AJHE (super)currents. The great advantage of the bound state-based picture is that all electrical current inside the F-I is carried by *single* electrons occupying the bound states; the latter can be fully characterized within the methodology introduced in Sec. IV A. To extract the net (balanced) current flows, we need to properly account for Cooper pair electrons tunneling from the left into the right S *and* for Cooper pair electrons simultaneously tunneling along the opposite direction. Most commonly, one exploits the BdG description's electron–hole symmetry and considers electrons *and* holes that are transferred from left to right; the hole contribution includes then automatically the current originating from electrons incident from the right S. Our bound state wave function ansatz, Eqs. (S20)–(S21), indeed contains superpositions of electronlike and holelike parts, and thus already captures all relevant information for computing electrical currents in the correct way.

To quantify the current carried by *each occupied bound state*, we apply the electrical current density operator to the obtained bound state wave function at the F-I interface. Note that the latter is supposed to be ultrathin and the wave function itself must be continuous at the interface [required by the boundary condition in Eq. (S11)]. We can therefore rely either on $\Psi(x, y, z < 0; \mathbf{k}_\parallel; E_B)$ or on $\Psi(x, y, z > 0; \mathbf{k}_\parallel; E_B)$, and simply take the limit $z \to 0_+$ in the end. Throughout our considerations, we adopt the convention that *tunneling currents* flowing *from the left to the right* are counted *negatively* (as electrons tunneling from left to right possess *positive* velocities, but *negative* charge, $-e$). The average contribution of the state governed by the bound state wave function $\Psi(x, y, z; \mathbf{k}_\parallel; E_B)$ to the electrical current density along the $\hat{\eta}$-direction ($\hat{\eta} \in \{\hat{x}, \hat{y}\}$) is then determined by

$$j_\eta(\mathbf{k}_\parallel; E_B) = \lim_{z \to 0_+} \left\langle \langle \hat{j}_\eta \rangle \Big|_{\Psi(x,y,z>0;\mathbf{k}_\parallel;E_B)} \right\rangle = \lim_{z \to 0_+} \left\{ \left\langle \Psi(x, y, z > 0; \mathbf{k}_\parallel; E_B) \Big| \hat{j}_\eta \Big| \Psi(x, y, z > 0; \mathbf{k}_\parallel; E_B) \right\rangle \tanh\left(\frac{E_B}{2 k_B T}\right) \right\}, \quad (S57)$$

where

$$\hat{j}_\eta = -e \begin{bmatrix} -\mathrm{i}\frac{\hbar}{m}\frac{\partial}{\partial \eta} & 0 & 0 & 0 \\ 0 & -\mathrm{i}\frac{\hbar}{m}\frac{\partial}{\partial \eta} & 0 & 0 \\ 0 & 0 & -\mathrm{i}\frac{\hbar}{m}\frac{\partial}{\partial \eta} & 0 \\ 0 & 0 & 0 & -\mathrm{i}\frac{\hbar}{m}\frac{\partial}{\partial \eta} \end{bmatrix} \quad (S58)$$

represents the electrical current density operator. Its general form is physically justified. The current density operator of single particles is, in general, given by the product of the particles' charge and their velocities. In our case, the current density operator must be replaced by a matrix-like formulation (matching the Nambu BdG Hamiltonian). Its first two components act on the electron block of the bound state wave function. They are indeed identified as the product of the electron charge, $-e$, and the electrons' transverse velocities along $\hat{\eta}$, $\hat{v}_{\eta,e} = (\partial \hat{\mathcal{H}}_e) / (\hbar \partial k_\eta)$. This relationship is actually the same we expect from elementary single-particle quantum mechanics. Nevertheless, one should be aware that, by formally calculating $\hat{v}_{\eta,e}$ inside the F-I, one would obtain additional off-diagonal terms in Eq. (S58) when considering $\hat{v}_{x,e}$ or $\hat{v}_{y,e}$ (i.e., directions parallel to the interface). Those terms originate from the transverse momentum-dependent SOC contribution to the single-particle Hamiltonian and effectively couple the up-spin and down-spin components of the bound state wave functions. Since the wave functions in our case simultaneously contain only pure up-spin or down-spin parts and no mixture of both (this would change in the presence of SOC or

Zeeman fields directly inside the superconductors), we could neglect the off-diagonal terms in the current density operator. Since holes enter the BdG modeling as time-reversed electrons, i.e., with opposite transverse velocities *and* charge, the hole-block of the current operator is formally identical to the electron one. The analogy becomes plausible, for instance, by explicitly deriving the current's continuity equation from the *time-dependent* BdG equation in the considered Nambu representation. Furthermore, the *thermal occupation factor*, $\tanh[E_B/(2k_BT)]$, entering Eq. (S57), ensures that the regarded bound state got indeed occupied.

To compute the *total AJHE currents*, $I_\eta(E_B)$, originating from *one* bound state with energy $E_B$, we need to average the current density, Eq. (S57), over all transverse channels,

$$
\begin{aligned}
I_\eta(E_B) &= \frac{A}{(2\pi)^2} \int d^2\mathbf{k}_\parallel \, j_\eta(\mathbf{k}_\parallel; E_B) \\
&= -2eu(E_B)v(E_B)\frac{A}{(2\pi)^2} \int d^2\mathbf{k}_\parallel \frac{\hbar k_\eta}{m} \left[ \left| e(\mathbf{k}_\parallel; E_B) \right|^2 + \left| f(\mathbf{k}_\parallel; E_B) \right|^2 + \left| g(\mathbf{k}_\parallel; E_B) \right|^2 + \left| h(\mathbf{k}_\parallel; E_B) \right|^2 \right] \tanh\left( \frac{E_B}{2k_BT} \right) \\
&= -e \frac{|\Delta_S|}{E_B} \frac{A}{(2\pi)^2} \int d^2\mathbf{k}_\parallel \frac{\hbar k_\eta}{m} \left[ \left| e(\mathbf{k}_\parallel; E_B) \right|^2 + \left| f(\mathbf{k}_\parallel; E_B) \right|^2 + \left| g(\mathbf{k}_\parallel; E_B) \right|^2 + \left| h(\mathbf{k}_\parallel; E_B) \right|^2 \right] \tanh\left( \frac{E_B}{2k_BT} \right).
\end{aligned}
\tag{S59}
$$

In the second step, we inserted the explicit form of the bound state wave function in the right S [see Eq. (S21)] and used that $u(E_B)v(E_B) = |\Delta_S|/(2E_B)$ for $0 \leq E_B \leq |\Delta_S|$. The specific form of Eq. (S59) is quite rewarding as it reminds us a lot of the usually established quantum mechanical current formulas. Electrical current is a measure for spatial charge separation. Therefore, the total current flow along the $\hat{\eta}$-direction can be generically evaluated by multiplying the charge of single electrons, $-e$, by their transverse velocities along $\hat{\eta}$ [$v_\eta = (\hbar k_\eta)/m$], and by an additional "weighting factor" that represents the density of electrons propagating along that direction (recall that we evaluate the current inside the F-I, in which no Cooper pairs are formed). Particularly in our case, this "weighting factor" originates from the wave function matching, as one realizes when inspecting the analogy between the elementary quantum mechanical foundation and Eq. (S59). Note that the influence of the interfacial SOC is hidden in both the bound state energies and the (absolute squares of the) wave function coefficients; see the mathematical framework described above.

In nonmagnetic Josephson junctions, the whole bound state spectrum is dominated by ABS and the resulting current flow would already be correctly incorporated in Eq. (S59). However, magnetic barriers lead to a splitting of the bound state spectrum into ABS on the one and YSR states on the other hand [S12, S18]. In order to extract a well-balanced current, we then need to average over both branches of bound states so that the expression for the total (interfacial) AJHE currents eventually has the form

$$
I_\eta = -e \sum_{E_B} \frac{|\Delta_S|}{2E_B} \frac{A}{(2\pi)^2} \int d^2\mathbf{k}_\parallel \frac{\hbar k_\eta}{m} \left[ \left| e(\mathbf{k}_\parallel; E_B) \right|^2 + \left| f(\mathbf{k}_\parallel; E_B) \right|^2 + \left| g(\mathbf{k}_\parallel; E_B) \right|^2 + \left| h(\mathbf{k}_\parallel; E_B) \right|^2 \right] \tanh\left( \frac{E_B}{2k_BT} \right).
\tag{S60}
$$

The same formulation is, in principle, also possible to quantify the (tunneling) Josephson current. One again starts from Eq. (S57), together with Eq. (S58), now for $\eta = z$ and substitutes the bound state wave function in Eq. (S21). The final expression for the total Josephson current flow, $I_J$, reads

$$
I_J \approx -e \sum_{E_B} \frac{|\Delta_S|}{2E_B} \frac{A}{(2\pi)^2} \int d^2\mathbf{k}_\parallel \frac{\hbar q_z}{m} \left[ \left| e(\mathbf{k}_\parallel; E_B) \right|^2 + \left| f(\mathbf{k}_\parallel; E_B) \right|^2 - \left| g(\mathbf{k}_\parallel; E_B) \right|^2 - \left| h(\mathbf{k}_\parallel; E_B) \right|^2 \right] \tanh\left( \frac{E_B}{2k_BT} \right);
\tag{S61}
$$

we approximated, for simplicity, $q_{z,e} \approx q_{z,h} \approx q_z$ as mentioned in Sec. II. Despite the similarity of the equations' generic forms, we recognize two fundamental differences when comparing Eq. (S61) to the formula for the AJHE current components, Eq. (S60). To get from the transverse AJHE currents to the (longitudinal) Josephson current, we need to replace the electrons' transverse velocities by the ones along the longitudinal $\hat{z}$-direction. More surprisingly, the (absolute squares of the) wave function coefficients belonging to holelike quasiparticle transmissions, $|g(\mathbf{k}_\parallel; E_B)|^2$ and $|h(\mathbf{k}_\parallel; E_B)|^2$, enter the Josephson current formula with a negative sign, whereas they were counted positively for the AJHE currents in Eq. (S60). To understand the reason behind this observation, we need to think about the orientations of the transmitted electronlike *and* holelike quasiparticles' wave vectors (basically indicating their velocity vectors). While their transverse components, $\mathbf{k}_\parallel = [k_x, k_y, 0]^\top$, are typically aligned along opposite directions, the longitudinal parts point along the same direction [since both types of quasiparticles tunnel (propagate) from left to right]. The opposite charge of electrons and holes compensates the differing signs in their transverse velocities, while introducing a relative opposite sign in their longitudinal velocities. Therefore, the electronlike and holelike velocity prefactors in the current formulas, Eqs. (S60) and (S61), enter with the same signs in the first (transverse AJHE currents) and with opposite signs in the second (Josephson current) case. Similar relations appear when modeling transverse electrical transport in ferromagnet/S junctions within a generalized Blonder–Tinkham–Klapwijk approach [S13, S25].

To complete this section, we want to stress that Eq. (S61) generalizes the simple relation, $I_J \sim \partial[E_B(\phi_S)]/\partial\phi_S$, which is typically employed to extract Josephson currents from the bound state spectrum. Nonetheless, the presented approach works more practically for complicated junctions, e.g., if SOC in the bulk superconductors would be present. The latter gives rise to

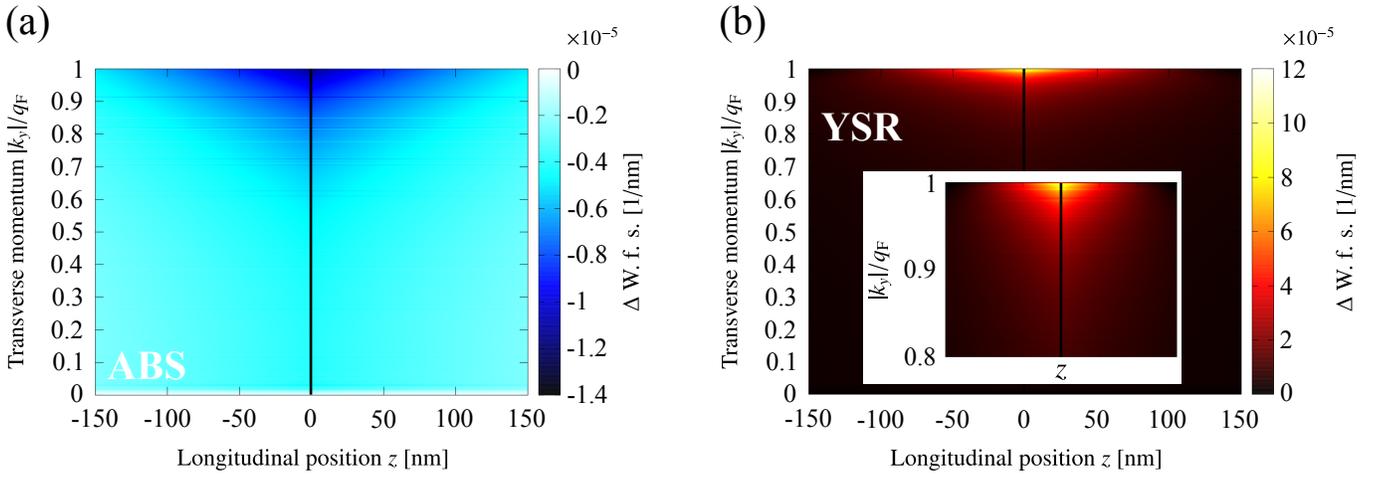

FIG. S1. (a) Calculated spatial dependence of the ABS wave functions' absolute square *differences*, $\Delta$ W. f. s., for all allowed transverse momenta, $0 \leq |k_y| \leq q_F$ (assuming $k_x = 0$). Positive $\Delta$ W. f. s. indicates that the bound state wave functions at $k_y \geq 0$ exceed those at the respective $-k_y$ (and vice versa). All other junction parameters are the same as in Fig. 5 of our manuscript. (b) Same calculation as in (a) for the YSR states. The inset shows a zoom into the plot at $0.8q_F \leq |k_y| \leq q_F$.

additional (SOC-governed) current contributions which are not included in the simple formula. Moreover, our final result for the Josephson current in Eq. (S61) is fully consistent with the formula derived by Furusaki [S26]. The case of transverse currents, which is the particularly interesting one in our work, has not been treated there. For the data presented in Fig. 5(b) of our manuscript, we numerically implemented the strategy described in Secs. IV A and IV B. After solving the system of equations— originating from matching the bound state wave functions at the interface—for the bound state energies, we determined all wave function coefficients and substituted everything into Eq. (S60) to obtain the AJHE current components. We resolved the contributions of the ABS and the YSR states in order to unravel their individual impact.

## C. Bound state wave function squares

To illustrate the influence of SOC on the bound state spectrum, we discuss the absolute squares of the bound state wave functions in Fig. 5(a) of our manuscript. For completeness, we briefly summarize the underlying formulas. Following the generic bound state wave function ansatz in Eqs. (S20) and (S21), the wave functions' absolute squares can be expressed as

$$\left|\psi(z<0; \mathbf{k}_\parallel; E_B)\right| = 2u(E_B)v(E_B)\left[\left|a(\mathbf{k}_\parallel; E_B)\right|^2 + \left|b(\mathbf{k}_\parallel; E_B)\right|^2 + \left|c(\mathbf{k}_\parallel; E_B)\right|^2 + \left|d(\mathbf{k}_\parallel; E_B)\right|^2\right] e^{2\mathrm{Im}[q_{z,e}(\mathbf{k}_\parallel; E_B)]z} \tag{S62}$$

in the left S ($z < 0$) and

$$\left|\psi(z>0; \mathbf{k}_\parallel; E_B)\right| = 2u(E_B)v(E_B)\left[\left|e(\mathbf{k}_\parallel; E_B)\right|^2 + \left|f(\mathbf{k}_\parallel; E_B)\right|^2 + \left|g(\mathbf{k}_\parallel; E_B)\right|^2 + \left|h(\mathbf{k}_\parallel; E_B)\right|^2\right] e^{-2\mathrm{Im}[q_{z,e}(\mathbf{k}_\parallel; E_B)]z} \tag{S63}$$

in the right S ($z > 0$), respectively. The plane wave propagation parallel to the F-I interface (i.e., along $\hat{x}$ and $\hat{y}$) does not contribute to the *bound state* wave function squares.

The results presented in Fig. 5(a) of our manuscript were obtained by numerically solving for the ABS and YSR bound state energies, $E_B$, and extracting the coefficients entering the bound state wave functions according to the methodology described in Sec. IV A, before finally evaluating Eqs. (S62) and (S63). The inset of Fig. 5(a) depicts the spatial dependence of the bound state wave functions' absolute square *differences*,

$$\Delta \text{ W. f. s.} = \left|\psi(z; k_x, k_y; E_B)\right|^2 - \left|\psi(z; -k_x, -k_y; E_B)\right|^2, \tag{S64}$$

for $k_y = \pm 0.99q_F$. To provide a more thorough characterization, the color plots in Fig. S1 illustrate the spatial dependence of $\Delta$ W. f. s. for all possible values of $|k_y| \in [0; q_F]$ (assuming $k_x = 0$), once for the ABS and once for the YSR states. The data shown in the inset of Fig. 5(a) in our manuscript essentially indicates a cut through the color plots along $|k_y| = 0.99q_F$. Since the bound states' individual contributions to the AJHE currents increase proportionally with $\Delta$ W. f. s. [recall, e.g., Eq. (5) in the manuscript], the overall AJHE current amplitudes are mostly determined by those states for which $|\mathbf{k}_\parallel| \rightarrow q_F$ so that the

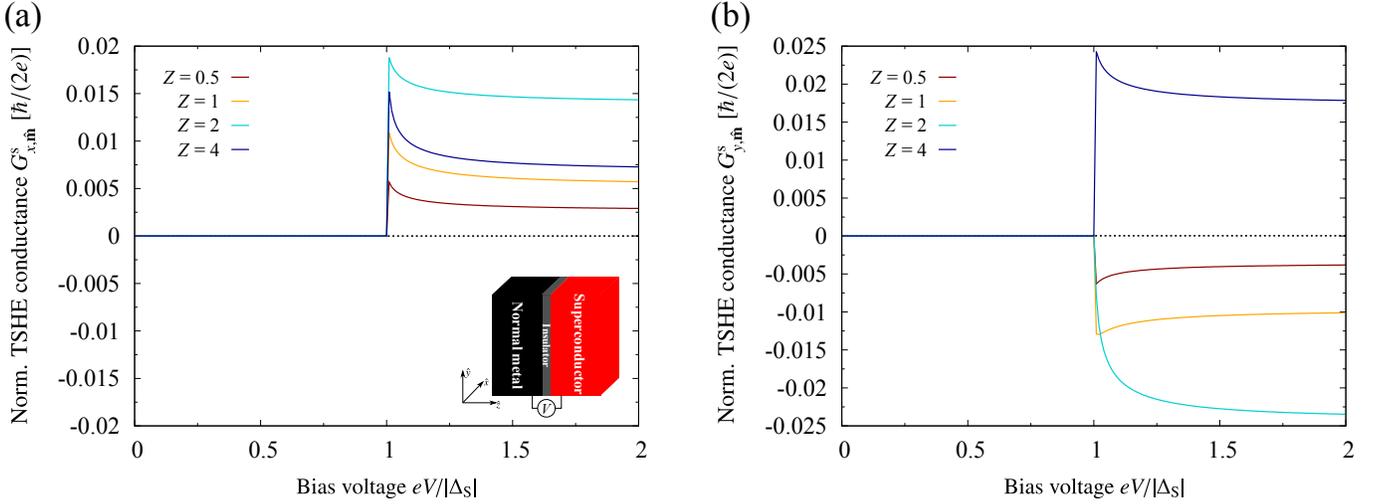

FIG. S2. Calculated maximal TSHE conductances, (a) $G^s_{x,th}$ and (b) $G^s_{y,th}$, normalized to $G_S$, as a function of the applied bias voltage ($eV/|\Delta_S|$) and for different indicated effective barrier strengths, $Z$; consult Ref. [S13] for the choice of the parameters. The inset shows a schematic sketch of the considered junction.

SOC-induced asymmetries, we discuss in Sec. VI, and simultaneously $\Delta$ W. f. s. get most distinct. This is actually also the reason why we chose $|k_y| = 0.99 q_F$ as an example for the plot in our manuscript.

All features, we extracted for $|k_y| = 0.99 q_F$, are universal and apply likewise to other values of $|k_y|$. Specifically, the current contributions originating from the YSR part of the bound state spectrum always remarkably overcome the ones stemming from the ABS (revealed by the larger $\Delta$ W. f. s. amplitudes for YSR states), while both flow along reversed directions (i.e., the respective $\Delta$ W. f. s. differ in sign). Moreover, the exponential damping of the bound state wave functions with increasing distance from the interface, we briefly addressed in the manuscript, is clearly visible (at least for larger $|k_y|$; otherwise, $\Delta$ W. f. s. becomes thus small that we cannot properly resolve its spatial dependence on the chosen scale of the color plots).

## V. TRANSVERSE SPIN CURRENTS

Besides the AJHE charge current flows, also their *transverse spin current counterparts* offer an interesting subject for deeper studies. While the conventional spin Hall effect [S27, S28] evolved into a well-established phenomenon in several normal-conducting systems, it turns out that the situation becomes much more subtle when turning the system superconducting. In particular, it is the fundamental time-reversal (electron–hole) symmetry of superconductors which may dramatically suppress (superconducting) spin Hall effects, as we will demonstrate in the following.

### A. Tunneling spin Hall effects in normal metal/superconductor tunnel junctions

A recent work [S29] concluded that skew tunneling through the insulating spin-active interfaces (including Rashba and Dresselhaus SOC) of normal metal/normal metal tunnel junctions gives rise to sizable transverse *tunneling spin Hall effect (TSHE) spin currents*. Note that the skew tunneling mechanism is actually very similar to the one in our Josephson junction, except that one solely needs to deal with single electrons (instead of Cooper pairs) in the normal system. Therefore, the TAHE charge current components indeed vanish (since the system is not ferromagnetic) and only the TSHE spin currents are nonzero.

Replacing the right metallic electrode by a ($s$-wave) S results in an intriguing interplay of *skew SRs* and *skew ARs* [S13], which, for instance, remarkably enhances the TAHE charge current flows once the system becomes ferromagnetic (e.g., if the left electrode consists of a ferromagnet). To unravel their impact on the TSHE spin currents, we generalize the BTK tunneling conductance formula [S25], initially applied to describe electrical transport through metal/S junctions, to the TSHE spin conductances. The considered normal metal/S junction (see the inset of Fig. S2 for a sketch) is modeled within the usual BdG approach; consult Ref. [S13] for details. The transverse spin current parts are basically obtained from the TAHE charge current formula [S13] by replacing the electron charge, $-e$, by $\hbar/(2e)$, and properly accounting for the incident and scattered particles' spins. To avoid dealing with quasiparticles inside the S, we evaluate the TSHE currents in the normal metal (close to the interfacial barrier). As stated in the manuscript, we are interested in *particle* spin currents, meaning that we only differentiate between spin up and spin down, but do not care about electrons' and holes' charge (contrary to the TAHE charge currents). To be specific,

up-spin electrons (holes) propagating along $\hat{\eta} \in \{\hat{x}; \hat{y}\}$ contribute *positively* to the spin conductances, while down-spin electrons' (holes') contributions are weighted *negatively* (and vice versa for electrons and holes that move along $-\hat{\eta}$). An overall positive sign of the TSHE spin conductances indicates then that the skew tunneling mechanism predominantly separates up-spin carriers at $\eta > 0$ and down-spin carriers at $\eta < 0$, resulting in the finite TSHE spin current flows. After some calculations, the TSHE spin conductance components, normalized to Sharvin's conductance ($G_S$) of a three-dimensional point contact, are given by

$$\frac{G^s_{\eta,\hat{\mathbf{m}}}}{G_S} = -\frac{\hbar}{2e} \sum_{\sigma=\pm 1} \frac{1}{2\pi} \int d^2\mathbf{k}_\parallel \frac{k_\eta}{k_z} \sigma \left\{ \left[ \left| r^{\sigma,\sigma}_e(eV) \right|^2 - \left| r^{\sigma,-\sigma}_e(eV) \right|^2 \right] - \left[ \left| r^{\sigma,\sigma}_h(-eV) \right|^2 - \left| r^{\sigma,-\sigma}_h(-eV) \right|^2 \right] \right\}; \tag{S65}$$

for convenience, we projected the individual scattering processes' spin conductance contributions on the unit vector $\hat{\mathbf{m}} = [\cos\Phi, \sin\Phi, 0]^\top$, which points along the magnetization orientation once the metal would become (weakly) ferromagnetic. In sharp contrast to the TAHE charge current formula [S13], SRs (coefficients $r^{\sigma,\pm\sigma}_e$) and ARs (coefficients $r^{\sigma,\pm\sigma}_h$) enter the TSHE spin conductance with *opposite* signs. The puzzling competition of skew SRs and skew ARs can hence really noticeably damp the TSHE conductances as we will additionally emphasize in the next paragraph. Besides, $V$ denotes the bias voltage applied to the junction and $k_z$ electrons' (holes') longitudinal wave vector (see Ref. [S13]).

Figure S2 illustrates the calculated TSHE spin conductance components as a function of the applied bias voltage, $eV$, and for various interfacial barrier strengths. Most remarkably, the TSHE conductances get not only heavily suppressed by the intricate interplay of skew SRs and skew ARs, but even vanish in the subgap region (i.e., at $eV < |\Delta_S|$). In fact, the complete absence of the TSHE in this regime can be traced back to a qualitative skew reflection picture of incident electrons at the metal/S interface, simultaneously invoking SRs and ARs. We discuss this physical picture in Fig. S3 and its caption. To generate nonzero TSHE conductance, $eV$ needs to be increased to values exceeding $|\Delta_S|$. Effectively, the TSHE conductances are then additionally governed by skew tunnelings and not only by skew reflections. SRs and ARs do no longer fully compensate each other and sizable TSHE conductances are expected to appear. Once the junction eventually approaches its normal-conducting state at $eV \gg |\Delta_S|$, ARs do no longer have any influence on the TSHE conductances. Even in the tunneling limit ($Z \gg 1$), when all the physics emerging in the system is ruled by skew SRs, the TSHE conductances remain still finite (which is clear as the AR contributions initially canceling the SR parts are now negligibly small), as predicted in the aforementioned work [S29]. We can therefore conclude that the unique correlations between (skew) SRs and (skew) ARs [being one manifestation of the junction's time-reversal (electron–hole) symmetry] in normal metal/S contacts usually acts against the TSHE and makes its realization in superconducting systems extremely challenging.

### B. Transverse spin currents in S/F-I/S junctions

To access the *transverse $\hat{\sigma}_z$-spin (super)current components*, $I^s_{\eta,\hat{z}}$, in our S/F-I/S Josephson junction, we can either extend the Furusaki–Tsukada technique [S30] or rely on our bound state description. Note that the $\hat{\sigma}_x$- and $\hat{\sigma}_y$-spin currents obviously vanish when inspecting the generic form of the scattering states inside the superconductors. Close to the interface ($z = 0$), the Furusaki–Tsukada approach yields

$$
\begin{aligned}
I^s_{\eta,\hat{z}} &\approx -\frac{k_B T}{4} |\Delta_S| \frac{A}{(2\pi)^2} \int d^2\mathbf{k}_\parallel \sum_{\omega_n} \frac{k_\eta}{\sqrt{q^2_F - \mathbf{k}^2_\parallel}} \left[ \frac{-\mathcal{C}^{(1)}(i\omega_n) + \mathcal{D}^{(2)}(i\omega_n) + \mathcal{A}^{(3)}(i\omega_n) - \mathcal{B}^{(4)}(i\omega_n)}{\sqrt{\omega^2_n + |\Delta_S|^2}} \right] \\
&= \frac{k_B T}{4} |\Delta_S| \frac{A}{(2\pi)^2} \int d^2\mathbf{k}_\parallel \sum_{\omega_n} \frac{k_\eta}{\sqrt{q^2_F - \mathbf{k}^2_\parallel}} \left[ \frac{\mathcal{C}^{(1)}(i\omega_n) - \mathcal{D}^{(2)}(i\omega_n) - \mathcal{A}^{(3)}(i\omega_n) + \mathcal{B}^{(4)}(i\omega_n)}{\sqrt{\omega^2_n + |\Delta_S|^2}} \right].
\end{aligned} \tag{S66}
$$

Simply speaking, Eq. (S66) follows from the AJHE charge current formula, Eq. (S14), by replacing the electron charge, $-e$, in the prefactor by $\hbar/(2e)$ and weighting the contributions of the individual quasiparticle scattering processes with reasonable signs. To give one example, let us focus on $\mathcal{C}^{(1)}$, essentially describing an incident up-spin electronlike quasiparticle (transversely propagating along $\hat{\eta}$) which gets Andreev reflected as an up-spin holelike quasiparticle. Since the retro-reflected hole moves along the opposite transverse direction compared to the incoming electronlike quasiparticle (but has still the same spin), we count its contribution to the *particle* spin current *negatively*. Given the first sign, the ones of the remaining three processes need to be chosen in a consistent manner, i.e., as in Eq. (S66). The spin current results presented in Figs. 7 and 8 of our manuscript were obtained by numerically evaluating Eq. (S66) for the indicated parameters.

Alternatively, we could extract $I^s_{\eta,\hat{z}}$ from the bound state wave functions we constructed in Sec. IV. Analogously to the AJHE charge current formula [see Eq. (S60)], we end up with

$$I^s_{\eta,\hat{z}} = \frac{\hbar}{2} \sum_{E_B} \frac{|\Delta_S|}{2E_B} \frac{A}{(2\pi)^2} \int d^2\mathbf{k}_\parallel \frac{\hbar k_\eta}{m} \left[ \left| e(\mathbf{k}_\parallel; E_B) \right|^2 - \left| f(\mathbf{k}_\parallel; E_B) \right|^2 - \left| g(\mathbf{k}_\parallel; E_B) \right|^2 + \left| h(\mathbf{k}_\parallel; E_B) \right|^2 \right] \tanh\left( \frac{E_B}{2k_B T} \right), \tag{S67}$$

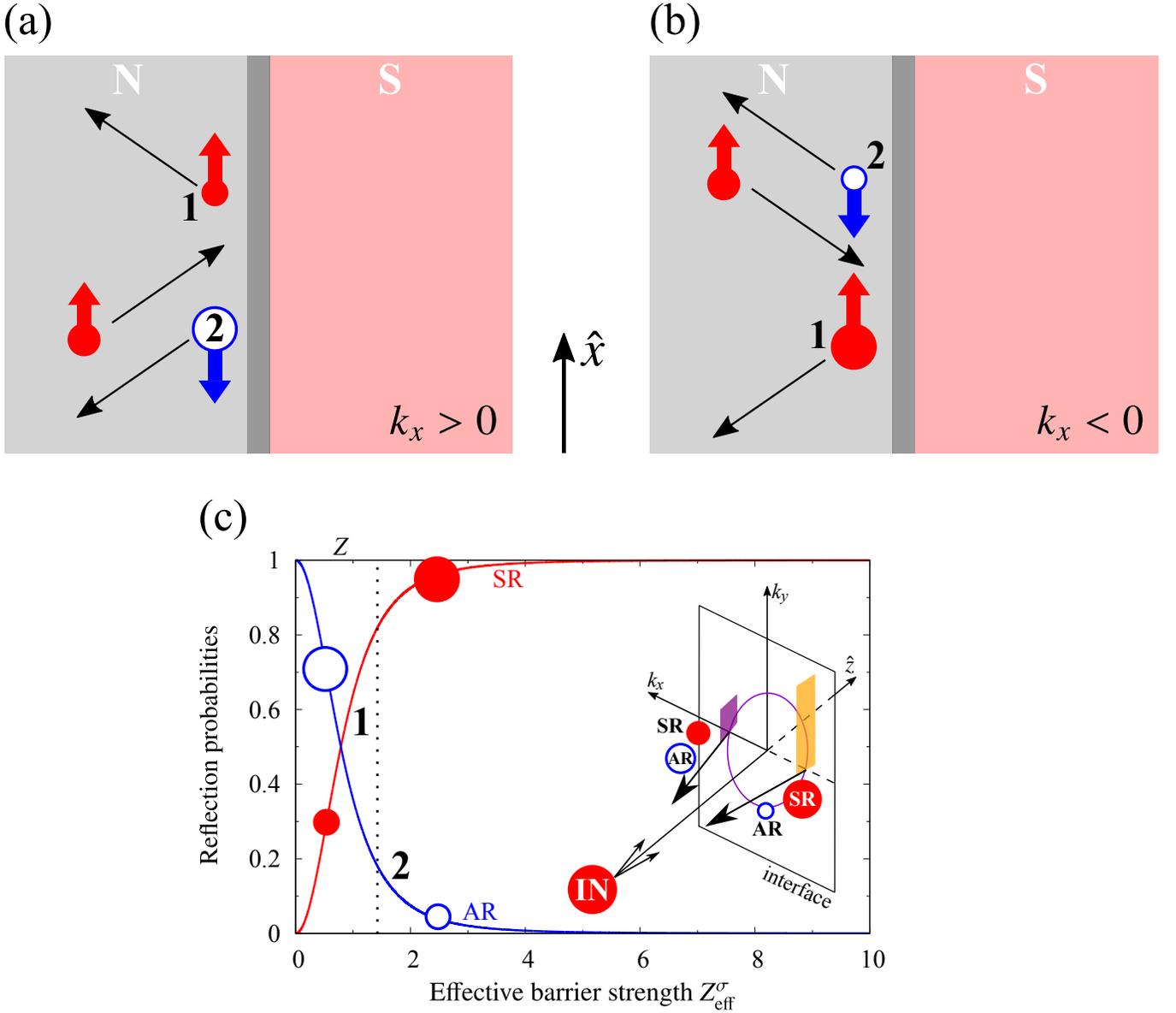

FIG. S3. (a) Illustration of SR (1) and AR (2) of an incoming up-spin electron with transverse momentum $k_x > 0$ ($k_y = 0$) at the normal metal (N)/S interface. Electrons are colored red, holes blue; red and blue arrows indicate the spins and black arrows the carriers' propagation directions. Spin-flip reflections do not need to be included for N/S junctions since interfacial spin flips convert, on average, equal numbers of up-spins to down-spins and vice versa, and do therefore not contribute to the TSHE spin conductances. (b) Same illustration as in (a) for an incident $k_x < 0$-spin up electron. (c) Scattering at the junction's spin-active interface can be understood in terms of skew SRs and skew ARs at an effective barrier, $Z_{eff}^\sigma$, similarly to the quasiparticle description we established for Josephson junctions in our manuscript. Process (a) essentially causes a *positive* TSHE spin current (conductance) since spin up carriers accumulate at $x > 0$ and spin down carriers at $x < 0$, respectively, and vice versa for process (b). The overall SR and AR contributions to the TSHE spin conductance are proportional to the *differences* of the skew SR and skew AR probabilities at $k_x > 0$ and $k_x < 0$ (since we need to integrate over all transverse momenta in the end). Since both skew reflection probability *differences* become equal in nonmagnetic junctions (basically, subtracting the probabilities indicated by the large electrons and holes from the ones highlighted by the small electrons and holes, and taking their absolute amplitudes, yields the same), processes (a) and (b) must generate the *same amounts* of TSHE spin currents, flowing along *opposite* directions, and the total TSHE spin current (conductance) must vanish. Finite TSHE spin conductances require therefore $eV > |\Delta_S|$ so that additionally skew transmissions come into play and skew SRs and skew ARs do no longer completely cancel.

substituting $(-e) \longmapsto \hbar/(2e)$ and recognizing that the electronlike parts of up-spin bound states $[\sim |e(\mathbf{k}_\|; E_B)|^2]$ and down-spin states $[\sim |f(\mathbf{k}_\|; E_B)|^2]$ must enter the spin current formula with *opposite* signs. Moreover, to each occupied electronlike bound state propagating along $\hat{\eta}$, the related spin-resolved holelike parts $[\sim |g(\mathbf{k}_\|; E_B)|^2$ and $\sim |h(\mathbf{k}_\|; E_B)|^2]$ describe states moving along $-\hat{\eta}$ and therefore need to be included with another overall negative sign. In Fig. 9 of the manuscript, we demonstrate the persuading coincidence between the spin currents computed from the Furusaki–Tsukada and the bound state techniques.

## VI. SOC ASYMMETRIES IN THE BOUND STATE SPECTRA

In our manuscript, we propose a distinct correlation between the AJHE and the underlying bound states forming at the junction interface. To become more familiar with the characteristics of these states, we briefly investigate the bound state spectra of some representative junctions.

First, let us consider an effectively *one-dimensional* junction in the absence of interfacial SOC. Figure S4 illustrates the bound state energies of this junction as a function of the magnetic tunneling strength, $\overline{\lambda}_{MA}$, once in the absence of interfacial scalar tunneling, $\overline{\lambda}_{SC} = 0$, and once in the presence of moderate interfacial scalar tunneling, $\overline{\lambda}_{SC} = 2$; one may compare the chosen parameters to the ones outlined in Sec. I. The bound state energies generally depend in a characteristic manner on the superconducting phase difference between the two superconductors. For simplicity, we set the phase difference to $\phi_S = 0$. A detailed investigation of the phase difference dependence could be found, e.g., in our earlier work [S12]. As long as SOC is absent, no magnetoanisotropic effects come into play [S11, S31] and the specific orientation of the magnetization vector inside the F-I has no influence on the energies of the bound states. The shown results were obtained by implementing the technique described in Sec. IV and (numerically) solving for nontrivial bound state solutions (for the indicated parameter sets). Apparently, if there is *no* magnetic tunneling, $\overline{\lambda}_{MA} = 0$, we only recover degenerate bound states at the energies $E_B = \pm|\Delta_S|$, independently of the actual strength of the scalar tunneling. In fact, those states correspond to the well-known ABS [S32, S33], suggesting bound state energies

$$E_B = \pm|\Delta_S| \sqrt{\frac{\overline{\lambda}_{SC}^2 + 4\cos^2(\phi_S/2)}{\overline{\lambda}_{SC}^2 + 4}}, \tag{S68}$$

which simply merge into $E_B = \pm|\Delta_S|$ at $\phi_S = 0$.

More striking is the second case in which additional magnetic tunneling is taken into account, $\overline{\lambda}_{MA} \neq 0$. Our calculations in Fig. S4 clearly reveal that each bound state band splits into two distinct bands (in total, there are four bands then as all bound states symmetrically appear also at negative energies due to the electron–hole symmetry of our system) as soon as $\overline{\lambda}_{MA}$ becomes finite. While one band reflects the properties of ABS ($E_B = \pm|\Delta_S|$ independently of the chosen $\overline{\lambda}_{MA}$) and we therefore still refer to these states as ABS, the second band's properties strongly differ. The bound state energies of that band are initially shifted more and more towards the center of the superconducting gap (with increasing $\overline{\lambda}_{MA}$), cross zero energy at a critical $\overline{\lambda}_{MA}^{crit.}$ [S12], and finally approach again the gap edge in the dirty limit (very strong magnetic tunnelings, $\overline{\lambda}_{MA} \gg 1$). Owing to the full analogy between the descried features of those states and the initially in bulk superconductors (in the presence of magnetic impurities) appearing YSR states [S21–S24], we call them YSR states. Both the ABS and the YSR states are typically symmetric with respect to $\overline{\lambda}_{MA}$'s sign, covering antiferromagnetic and ferromagnetic coupling scenarios, respectively. The existence of a $\overline{\lambda}_{MA}^{crit.}$, together with the YSR states' unexpected spin pattern (not discussed here), turned out to be the responsible physical mechanism behind the existence of 0-$\pi$ transitions in magnetic Josephson junctions [S12, S18].

The analysis of the bound state spectrum gets more subtle when considering the *three-dimensional* case, in which the bound state energies additionally depend in a nontrivial manner on the (sign of the) transverse wave vector, $\mathbf{k}_\|$; see, e.g., the general ansatz for the bound state wave functions in Eq. (S22) that involves not only the evanescent solutions along $\hat{z}$, but also the plane wave parts along the transverse directions. The bound state energies additionally depend on the interfacial Rashba and Dresselhaus SOC terms, which scale *linearly* with $\mathbf{k}_\|$ and therefore flip their signs when reversing $\mathbf{k}_\|$. This might eventually cause pronounced SOC asymmetries in the bound state spectra.

In order to understand the exact behavior, we shall once again go back to the comprehensive analytical description formulated in Sec. IV. Let us focus on the bound state energies, $E_B$, for one moment. These energies are extracted by matching the generic bound state wave function ansatz at the F-I interface and looking for nontrivial solutions. The latter precisely refer to the $E_B$'s we want to determine. The equations obtained from the boundary conditions comprise both transverse wave vector components, $k_x$ and $k_y$, once by substituting the explicit forms of the quasiparticle wave vectors, see Eqs. (S9), and once via the SOC part as we mentioned in the previous paragraph. The first is not important for us since $k_x$ and $k_y$ simply enter the quasiparticle wave vectors in Eq. (S9) quadratically. Reversing the sign of $\mathbf{k}_\|$ leaves the quasiparticle wave vectors untouched. We shall rather focus on the SOC "matrix element", Eq. (S55), which indeed contains $k_x$ and $k_y$ in linear terms. Reversing one of the momenta modifies the SOC "matrix element" and hence also the bound state energies. To confirm this speculation, we exemplarily show the bound state

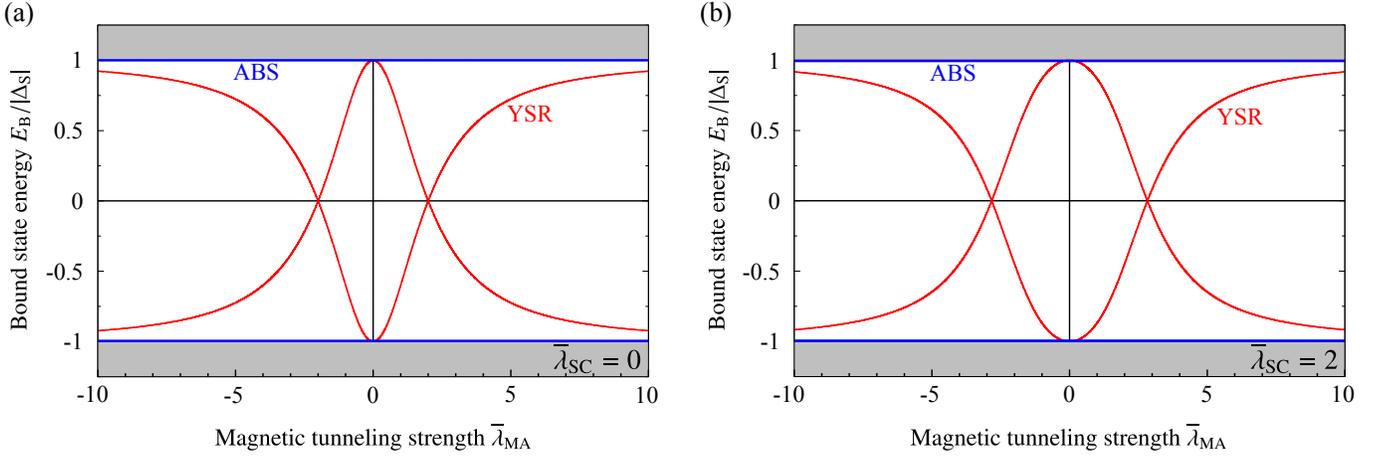

FIG. S4. (a) Generic spectrum of the ABS (blue) and YSR states (red) in *one-dimensional* S/F-I/S junctions as a function of the magnetic tunneling strength, $\bar{\lambda}_{\mathrm{MA}}$, and at zero superconducting phase difference ($\phi_{\mathrm{S}} = 0$). Scalar tunneling and interfacial SOC are absent, $\bar{\lambda}_{\mathrm{SC}} = \bar{\lambda}_{\mathrm{R}} = \bar{\lambda}_{\mathrm{D}} = 0$. (b) Same bound state spectrum, but in the presence of moderate interfacial scalar tunneling, $\bar{\lambda}_{\mathrm{SC}} = 2$.

energies as a function of the transverse momentum $k_y$ in Fig. S5; for simplicity, we set $k_x = 0$. The scalar tunneling strength got fixed to a weak value of $\bar{\lambda}_{\mathrm{SC}} = 1$, whereas the regarded magnetic tunneling strength, $\bar{\lambda}_{\mathrm{MA}} = 2$, needs to be quite large to see clear effects (otherwise, the effects would still be there, but barely visible). The superconducting phase difference is $\phi_{\mathrm{S}} = \pi/2$ and the magnetization in the F-I barrier is aligned along $\hat{x}$ ($\Phi = 0$). While SOC is fully absent in Fig. S5(a), we assume finite Rashba and Dresselhaus SOC in Fig. S5(b). For clarity, the strength of the Rashba SOC gets even more enhanced in Figs. S5(c)–(d). As deduced, the energies of both the ABS and the YSR states are not altered with a reversal of $k_y$'s sign as long as SOC is not there. The only $\mathbf{k}_{\parallel}$-dependent part in our model is the wave vector with the quadratic scaling in $\mathbf{k}_{\parallel}$ so that interchanging $\mathbf{k}_{\parallel} \longmapsto -\mathbf{k}_{\parallel}$ does not alter the equations determining the bound state energies. This does no longer hold when SOC starts to increase. Particularly in the situation we are concerned with, the SOC "matrix element", Eq. (S55), changes its overall sign (recall that we set $k_x = 0$). It is exactly this overall sign change which consecutively needs to be accounted for in all subsequent equations (when applying the boundary conditions to the bound state wave functions), and that eventually modifies the bound state energies when reversing $k_y$'s sign. Consequently, the bound state energy bands become asymmetric with respect to $k_y$'s sign. The asymmetry becomes more distinct with large magnetic tunneling strength and large SOC; see, for instance, Figs. S5(c)–(d), where we additionally highlighted the range of strongest asymmetries by red arrows as a guide for eyes. Just the case of equal Rashba and Dresselhaus SOC strengths, $\bar{\lambda}_{\mathrm{R}} = \bar{\lambda}_{\mathrm{D}}$, behaves in an extraordinary way and the bound state asymmetries disappear.

There are two more unexpected and important observations. First, the $\mathbf{k}_{\parallel}$-asymmetry is much more apparent in the YSR states than in the ABS. The reason for this is that the ABS are usually located close to the gap edge and vary only slightly with modulating $\mathbf{k}_{\parallel}$. Contrarily, the YSR states typically lie more inside the superconducting gap—see also our discussions of the one-dimensional system in Fig. S4—, and are hence much more sensitive to a change of any system parameters, i.e., also to a sign change of $\mathbf{k}_{\parallel}$. Second, comparing Fig. S5(b) with Fig. S5(d) reveals furthermore that an increase of the Rashba SOC strength might be sufficient to shift the region of stronger $k_y$-asymmetry from $k_y < 0$ to $k_y > 0$ (highlighted by the red arrows). The result of that will become clear within the next paragraph.

The most natural question we need to answer from the experimental point of view is: *Are the predicted $\mathbf{k}_{\parallel}$-dependent asymmetries of the bound state energies directly extractable from transport measurements?* Recalling our theoretical framework worked out in Sec. IV, electrical current (tunneling and transverse AJHE parts) is carried by electrons tunneling from one S into the second one via the bound states at the F-I interface. This means in particular for the transverse AJHE current components that each *occupied* bound state with energy $E_{\mathrm{B}}$ and the transverse (electron) velocity $v_\eta = (\hbar k_\eta)/m$ (along the $\hat{\eta}$-direction) carries an amount of current given by the product of this transverse velocity, the electron charge, and a "weighting factor". Precisely this argumentation justified our reasoning to obtain Eq. (S60). The "weighting factor" (playing the role of an effective charge density factor in usual quantum mechanical approaches) depends on $\mathbf{k}_{\parallel}$ via the bound state energies, $E_{\mathrm{B}}$. Physically, a transverse AJHE current starts to flow if there are more occupied states (at a given $E_{\mathrm{B}}$) propagating along the $\hat{\eta}$-direction than occupied counterpropagating states. If we assume that SOC is absent, we concluded that the $E_{\mathrm{B}}$'s of the propagating and the respective counterpropagating state are the same, owing to the peculiar form of the equations derived from the boundary conditions. Therefore, there is *always* a counterpropagating state to each propagating one, having exactly the same energy and carrying the same amount of charge current, just along the opposite transverse direction. The net AJHE current must inevitably be zero. The situation changes if SOC is present, creating the discussed asymmetry in the bound state energies. To each state propagating along the $\hat{\eta}$-direction, we now find a counterpropagating state with a *(slightly) different* bound state en-

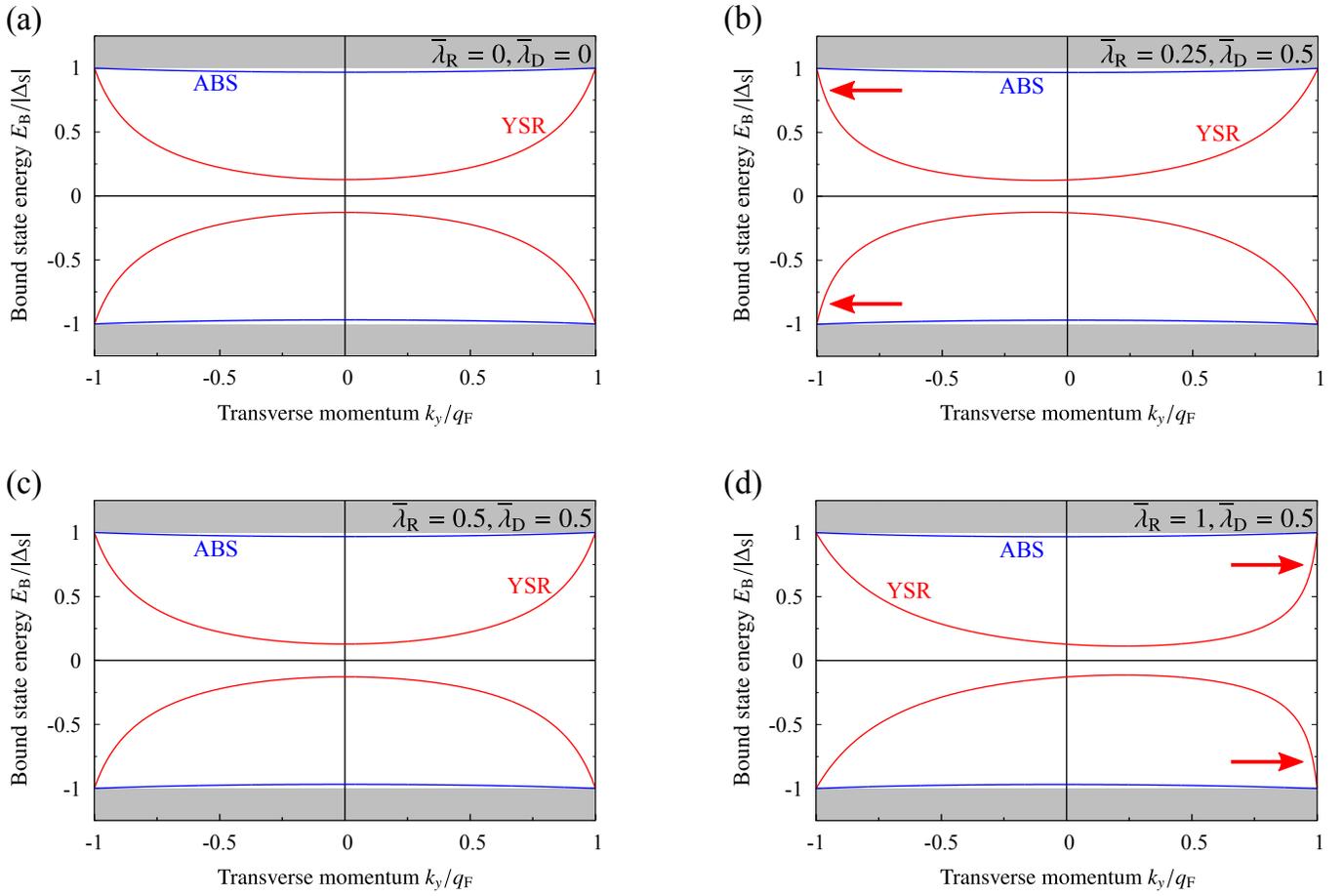

FIG. S5. (a) Spectrum of the ABS (blue) and YSR states (red) in *three-dimensional* S/F-I/S junctions as a function of the transverse momentum $k_y$ (normalized to the Fermi wave vector, $q_F$) and in the presence of weak interfacial scalar tunneling, $\overline{\lambda}_{SC} = 1$, setting $k_x = 0$ and $\phi_S = \pi/2$. The magnetization is aligned along $\hat{x}$ (i.e., $\Phi = 0$) and both Rashba and Dresselhaus SOC are absent, $\overline{\lambda}_R = \overline{\lambda}_D = 0$. (b) Same bound state spectrum in the presence of Rashba SOC with strength $\overline{\lambda}_R = 0.25$ and Dresselhaus SOC with strength $\overline{\lambda}_D = 0.5$. The red arrows indicate the position of the mostly pronounced $k_y$-asymmetry of the YSR branch. (c) Same bound state spectrum in the presence of Rashba SOC with strength $\overline{\lambda}_R = 0.5$ and Dresselhaus SOC with strength $\overline{\lambda}_D = 0.5$. (d) Same bound state spectrum in the presence of Rashba SOC with strength $\overline{\lambda}_R = 1$ and Dresselhaus SOC with strength $\overline{\lambda}_D = 0.5$. The red arrows indicate the position of the mostly pronounced $k_y$-asymmetry of the YSR-like branch.

ergy (simply due to the fact that reversing $\mathbf{k}_\parallel$'s sign modifies the bound state energies in the presence of SOC). The asymmetry in $\mathbf{k}_\parallel$ directly manifests itself in differing "weighting factors" for the propagating and counterpropagating states, which finally triggers a finite AJHE current. The current's overall amplitude is hence directly linked to the strength of the asymmetry in the bound state spectrum; large magnetic tunneling strength and large SOC could, in principle, be expected to significantly enhance the current. However, one should be aware that the increase always competes with a decrease stemming from the additional disorder brought into the system so that the current cannot become arbitrarily huge.

Coming back to our previous statements, two further conclusions become possible. Since the $\mathbf{k}_\parallel$-asymmetry is much more pronounced in the YSR part of the spectrum, also these states should predominantly control the magnitudes of the AJHE current. This is indeed the case as we discuss in more details in Fig. 5(b) of our manuscript. Moreover, we already realized above that increasing the Rashba SOC strength might shift the regime of strongest $\mathbf{k}_\parallel$-asymmetry from negative to positive transverse momenta, affecting the AJHE current by suddenly reversing its direction when increasing the Rashba SOC strength. The latter can be seen, for example, in Fig. 3 of our manuscript; one should nonetheless notice that we used different parameters there. The reversal of AJHE currents may be regarded as the transverse analog to the well-known 0-$\pi$ transitions [S34], usually emerging in the (tunneling) Josephson current.

After our generic analyses of the asymmetric bound state spectra in the considered S/F-I/S junctions, we shortly want to comment on the magnetization orientation's role. Owing to the striking competition between the ferromagnetic barrier on the one and the interfacial SOC on the other hand, the AJHE charge current components must scale according to $I_x \sim \sin \Phi$ and $I_y \sim \cos \Phi$, where the angle $\Phi$ measures the orientation of the magnetization in the F-I with respect to the $\hat{x}$-axis; see our calculations

presented in Fig. 3 of the manuscript. Therefore, we expect $I_x = 0$ and only $I_y \neq 0$ in our case (recall that we suppose $\Phi = 0$). The finite AJHE current along $\hat{y}$ is indeed justified by the $k_y$-dependent asymmetry in the bound state spectra (apart from the case $\bar{\lambda}_R = \bar{\lambda}_D$, in which both the $k_y$-asymmetry and the AJHE current vanish, as expected from the absent skew tunneling mechanism we proposed in the manuscript for these parameters). Repeating similar calculations to obtain the bound state energies as a function of $k_x$ shows that there is in fact no such asymmetry with respect to $k_x$, in agreement with the expected $I_x = 0$. The whole situation changes when considering $\Phi = \pi/2$; the asymmetry in the energies of the bound states appears now when reversing the sign of $k_x$ and disappears when the sign of $k_y$ changes, effectively leading to $I_x \neq 0$ and $I_y = 0$, just as expected. To keep our work compact, we do not show all those additional bound state calculations. This is a distinct manifestation that all the emerging asymmetries (and thus the existence of net AJHE current flows themselves) must indeed be traced back to the intriguing interplay of the magnetization in the F-I barrier and the interfacial SOC. Looking up the concrete magnetization part of the junction's Hamiltonian (see, e.g., the modeling in our manuscript), we assert that fully aligning the magnetization in the F-I along $\hat{x}$ [$\Phi = 0 \pmod{2\pi}$] involves the $\hat{\sigma}_x$-Pauli matrix. Since the same Pauli matrix appears in the SOC Hamiltonian in connection with $k_y$, the magnetic field oriented fully along $\hat{x}$ couples to the $k_y$-dependent part of the SOC and hence eventually gives rise to the strongly pronounced $k_y$-asymmetry of the bound state energies and maximal $I_y$. Via the Pauli matrix $\hat{\sigma}_y$, magnetizations purely along $\hat{y}$ [$\Phi = \pi/2 \pmod{2\pi}$] couple to $k_x$ and generate maximal $I_x$. For intermediate magnetization orientations, we obtain a mixture of both couplings and consequently, $I_x \neq 0$ and $I_y \neq 0$; there are (smaller) asymmetries of the bound state energies with respect to *both* $k_x$ and $k_y$. We decided for the superconducting phase difference of $\phi_S = \pi/2$ as the underlying physics (especially the Josephson physics) becomes more evident there than in the special case of zero phase difference and we wanted to stress that the discussed effects do not require trivial phase differences.

---



\end{document}